\documentclass[fleqn,usenatbib]{mnras}

\usepackage{newtxtext,newtxmath}

\usepackage[T1]{fontenc}

\DeclareRobustCommand{\VAN}[3]{#2}
\let\VANthebibliography\thebibliography
\def\thebibliography{\DeclareRobustCommand{\VAN}[3]{##3}\VANthebibliography}

\usepackage{booktabs}
\usepackage{graphicx}	
\usepackage{amsmath}	
\usepackage{layouts}
\usepackage{gensymb}
\usepackage{hyperref}
\usepackage[encapsulated]{CJK}

\newcommand{\mrm}{\mathrm}
\newcommand{\mbf}{\mathbf}
\newcommand{\mcal}{\mathcal}
\newcommand{\Msun}{\mathrm{M_{\odot}}}

\title[Unified AGN model for galaxy formation]{A unified accretion disc model for supermassive black holes in galaxy formation simulations: method and implementation}

\author[S. Koudmani et al.]{
Sophie Koudmani,$^{1,2}$\thanks{E-mail: skoudmani@flatironinstitute.org}
Rachel S. Somerville,$^{1}$
Debora Sijacki,$^{3}$
Martin A. Bourne,$^{3}$
\begin{CJK*}{UTF8}{gbsn}
Yan-Fei Jiang (姜燕飞)$^{1}$
\end{CJK*}
\newauthor
\ and Kasar Profit$^{1,4}$
\\
$^{1}$ Center for Computational Astrophysics, Flatiron Institute, 162 5$^{th}$ Avenue, New York, NY 10010, USA\\
$^{2}$ St Catharine's College, University of Cambridge, Trumpington Street, Cambridge CB2 1RL, UK\\
$^{3}$ Institute of Astronomy and Kavli Institute for Cosmology, University of Cambridge, Madingley Road, Cambridge, CB3 0HA, UK\\
$^{4}$ Colgate University, 13 Oak Drive, Hamilton, NY 13346, USA
}

\date{Accepted 2024 May 28. Received 2024 May 8; in original form 2023 December 13}

\pubyear{2024}

\begin{document}
\label{firstpage}
\pagerange{\pageref{firstpage}--\pageref{lastpage}}
\maketitle

\begin{abstract}
It is well established that supermassive black hole (SMBH) feedback is crucial for regulating the evolution of massive, if not all, galaxies. However, modelling the interplay between SMBHs and their host galaxies is challenging due to the vast dynamic range. Previous simulations have utilized simple subgrid models for SMBH accretion, while recent advancements track the properties of the unresolved accretion disc, usually based on the thin $\alpha$-disc model. However, this neglects accretion in the radiatively inefficient regime, expected to occur through a thick disc for a significant portion of an SMBH's lifetime. To address this, we present a novel `unified' accretion disc model for SMBHs, harnessing results from the analytical advection-dominated inflow-outflow solution (ADIOS) model and state-of-the-art general relativistic (radiation-)magnetohydrodynamics (GR(R)MHD) simulations. Going from low to high Eddington ratios, our model transitions from an ADIOS flow to a thin $\alpha$-disc via a truncated disc, incorporating self-consistently SMBH spin evolution due to Lense-Thirring precession. Utilizing the moving mesh code \textsc{arepo}, we perform simulations of single and binary SMBHs within gaseous discs to validate our model and assess its impact. The disc state significantly affects observable luminosities, and we predict markedly different electromagnetic counterparts in SMBH binaries. Crucially, the assumed disc model shapes SMBH spin magnitudes and orientations, parameters that gravitational wave observatories like LISA and IPTA are poised to constrain. Our simulations emphasize the importance of accurately modelling SMBH accretion discs and spin evolution, as they modulate the available accretion power, profoundly shaping the interaction between SMBHs and their host galaxies.
\end{abstract}




\section{Introduction}
Galaxy formation in the $\Lambda$CDM Universe is a hierarchical process whereby primordial density fluctuations collapse into dark matter haloes with cosmic time and grow via inflows and mergers. The gravitational pull of the dark matter leads the gas to collapse within these haloes, where it can radiatively cool and form stars. To prevent excessive star formation and reproduce the observed galaxy stellar mass function, theoretical models need to include feedback processes \citep[see e.g.][for recent reviews]{somerville_physical_2015,vogelsberger_cosmological_2020}. Supernova feedback is commonly invoked for the regulation of the baryon cycle in low-mass galaxies, whilst the more energetic active galactic nuclei (AGN) feedback from supermassive black holes (SMBHs) is required to suppress star formation in massive galaxies. 

Recent tantalizing observations suggest that AGN feedback could also play a role at the low-mass end of the galaxy population, with observational hints at AGN-driven outflows in nearby dwarfs \citep[e.g.][]{penny_sdss-iv_2018,manzano-king_agn-driven_2019,liu_integral_2020,davis_radio_2022,aravindan_comparison_2023} as well as indications of significant AGN activity from overmassive BHs in high-redshift low-mass galaxies \citep[e.g.][]{harikane_jwstnirspec_2023,maiolino_small_2024,maiolino_jades_2023,ubler_ga-nifs_2024}.

SMBHs are likely to reside in the centre of the majority of massive galaxies \citep[see e.g. review by][]{kormendy_coevolution_2013} and intermediate-mass black holes (IMBHs) may reside in the centres of the majority of massive dwarfs \citep{greene_intermediate-mass_2020}. As these galaxies are expected to undergo several major mergers with similarly sized galaxies over their cosmic lifetime, IMBHs and SMBHs also grow via mergers in addition to gas accretion. The merging process of stellar-mass BHs has been recently directly observed with the Advanced LiGO and VIRGO detectors which have opened a gravitational wave window on our Universe, even likely pushing into the IMBH regime \citep{abbott_gw190521_2020}. While these ground-based detectors are not sensitive to the range of gravitational waves emitted by merging SMBHs, IPTA and LISA in future will be able to probe them. This potential has been spectacularly demonstrated by the detection of a signal consistent with a gravitational wave background from SMBH mergers by members of IPTA \citep{verbiest_international_2016,agazie_comparing_2024}, NANOGrav \citep{agazie_nanograv_2023}, Parkes PTA \citep{reardon_search_2023} and joint results from European and Indian PTAs \citep{antoniadis_second_2023} marking the onset of the multi-messenger era for galaxy formation.

SMBH feedback in galaxy formation is an inherently multi-scale process spanning 14 orders of magnitude from the event horizon ($\sim 10^{-6}$~pc for Sgr A*) to the cosmic web ($\sim 10^{8}$~pc). This renders an ab-initio treatment computationally infeasible and hence cosmological simulations have to resort to including SMBH physics via so-called subgrid models \citep[also see discussion in][]{curtis_resolving_2015}. The modelling of the gas accretion onto the AGN is essential for obtaining realistic AGN feedback as the luminosity is directly tied to the accretion rate. What is more, accretion disc physics is crucial for accurately modelling the SMBH spin evolution which dictates the radiative efficiency and jet powers as well as influences gravitational recoils.

Most cosmological simulations \citep[e.g.][]{dubois_black_2014,schaye_eagle_2015,sijacki_illustris_2015,pillepich_simulating_2018} have adopted the `Bondi-Hoyle-Lyttleton' model \citep{hoyle_effect_1939,bondi_mechanism_1944,bondi_spherically_1952} for black hole (BH) accretion, which allows to infer SMBH growth rates based on the gas properties at the Bondi radius ($\sim5 - 5000$~pc for SMBHs), significantly reducing the resolution requirements. Though note that many cosmological simulations do not even resolve the Bondi radius, especially for low-mass SMBHs, though additional refinement in the central region can circumvent these issues \citep[e.g.][]{curtis_powerful_2016}. However, the Bondi model is very simplistic, assuming radial symmetry, neglecting angular momentum transfer and cannot be self-consistently used to predict luminosities of quasars. Moreover, due to the strong dependency of the Bondi accretion rate on SMBH mass, the Bondi model makes it very difficult to grow light seeds or reproduce observations of bright AGN in dwarfs \citep[e.g.][]{koudmani_little_2021,koudmani_two_2022,haidar_black_2022}, whilst likely overestimating the gas accretion rate for elliptical galaxies and galaxy clusters \citep[e.g.][]{russell_inside_2015,bambic_agn_2023,guo_toward_2023}. To address these limitations more advanced approaches have emerged to model AGN accretion processes following two main strategies. 

In the first approach, the range of spatial or temporal scales is reduced to make the problem computationally tractable. Recent examples include the hydrodynamic simulations of an elliptical galaxy by \citet{guo_toward_2023}, resolving SMBH accretion from galaxy scales all the way down to scales similar to the black hole horizon, and recent general relativistic magnetohydrodynamics (GRMHD) simulations bridging the scales from the event horizon to the Bondi radius and beyond \citep{lalakos_bridging_2022,cho_bridging_2023-1}. Bridging from large scales inwards, the cosmological-zoom in simulation of a high-redshift quasar by \citet{hopkins_forged_2024,hopkins_forged_2024-1} follow the gas flows from the cosmological environment to the scale of the SMBH's accretion disc ($\sim 10^{-4}$~pc) for $\sim 10^{4}$~yr. While these attempts can reach an impressive dynamical range, they are computationally prohibitively expensive to allow for a study of a representative SMBH population.

The second approach is to improve the physical realism of the subgrid models. For example, extensions to the Bondi model have been developed to take the gas angular momentum into account in hydrodynamical simulations, e.g. \citet{krumholz_bondi_2005}, \citet{rosas-guevara_impact_2015} or \citet{curtis_resolving_2016}. Furthermore, the torque-based accretion model infers the gas accretion rates motivated by analytical calculations and numerical simulations of angular momentum transport and gas inflow in galaxies, from scales of $\sim$~kpc to deep inside the potential of the central SMBH \citep{hopkins_analytic_2011,angles-alcazar_black_2013,angles-alcazar_gravitational_2017}, recently extended by \citet{rennehan_three_2023} to include AGN feedback in different physical accretion regimes. However, the torque-based model does not track small-scale angular momentum flows which are crucial to model the SMBH spin evolution, which modulates the radiative efficiency and jet power.

As a compromise between these two main approaches, the accretion disc particle method has been introduced by \citet{power_accretion_2011} and further developed by a range of groups \citep[e.g.][]{dubois_black_2014,fiacconi_galactic_2018,yuan_active_2018,beckmann_dense_2019,bustamante_spin_2019,cenci_black_2021,tartenas_improving_2022,husko_spin-driven_2022,husko_winds_2024,bollati_connection_2023,massonneau_how_2023}, including self-consistent evolution of the SMBH spin. The common approach here is to significantly increase the resolution around the SMBH to directly measure the mass and angular momentum flows onto the accretion disc, typically requiring to resolve scales of $\sim10^{-2}$~pc corresponding to the outer radius of the accretion disc. The accretion disc and SMBH properties are then evolved in a subgrid fashion. These models build on the rich body of work which has focused on developing AGN accretion disc prescriptions for semi-analytical models of galaxy formation \citep[e.g.][]{volonteri_distribution_2005,volonteri_evolution_2013,berti_cosmological_2008,lagos_black_2009,fanidakis_grand_2011,barausse_evolution_2012,dotti_orientation_2013,sesana_linking_2014,gaspari_unifying_2017,lagos_quenching_2024}.

However, these accretion disc models mostly focus on the thin disc case (the high-Eddington-ratio regime) for the steady state as this regime can be described analytically with the Shakura-Sunyaev disc model \citep{shakura_black_1973}, whilst there is no equivalent global analytical disc model for the low-Eddington-ratio regime \citep[though see][]{fanidakis_grand_2011,yuan_active_2018}. Recent cosmological simulations projects have started incorporating thick discs into their frameworks \citep[e.g.][]{dubois_introducing_2021,husko_spin-driven_2022,husko_winds_2024}, however, these generally have to make some approximations based on the Bondi rate as the mass and angular momentum flows onto the disc are not directly resolved.

Observations of X-ray binaries and AGN indicate that at low Eddington ratios ($f_\mathrm{Edd} \lesssim 0.01$) accretion occurs via a different mode than the standard thin disc, e.g. observations show hard X-ray spectra instead of soft black body-like spectra and low radiative efficiencies. The ADAF (Advection Dominated Accretion Flow) solution has been shown to have the properties required to provide a physical description of the observations \citep[e.g.][]{narayan_advection-dominated_2008}. Moreover, most X-ray binaries in the hard and quiescent state are deduced to accrete via so-called truncated discs with an outer thin disc and an inner ADAF component \citep[e.g.][]{yuan_hot_2014}. There is also some observational evidence that this phenomenon could transfer over to AGN \citep[e.g.][]{trump_accretion_2011,yu_origin_2011,nemmen_spectral_2014}. 

Theoretical arguments suggest that ADAFs should produce strong winds and this outflow behaviour can be modelled with the general advection-dominated inflow -- outflow solution (ADIOS) for radiatively inefficient accretion flows \citep{blandford_fate_1999}. 

Building on these observations and theoretical frameworks, as well as results from general relativistic (radiation-)magnetohydrodynamics (GR(R)MHD) simulations, we have developed a unified accretion disc subgrid model that proceeds via the thin disc model from \citet{fiacconi_galactic_2018} at high Eddington ratios and via the ADIOS flow model at low Eddington ratios. For intermediate Eddington ratios, we smoothly model the transition between these two modes via a truncated disc. A unified model for accretion through a truncated thin disc and inner ADIOS flow does not currently exist for cosmological simulations, so our new method is an important advancement for modelling accretion onto AGN in general.

In this paper, we present the numerical method and implementation as well as a suite of validation simulations, including single SMBHs and SMBH binaries in gaseous discs. The latter crucially demonstrates the scientific relevance of our model as electromagnetic counterpart and spin evolution predictions are hugely sensitive to the disc state. Due to significant observational challenges there are currently only a handful of confirmed binary detections \citep{rodriguez_compact_2006,bansal_constraining_2017,kharb_candidate_2017}. High-resolution simulations of SMBH binaries \citep{dittmann_survey_2022,dittmann_evolution_2024,franchini_resolving_2022,siwek_orbital_2023,siwek_preferential_2023,bollati_dynamical_2023,bourne_dynamics_2023} are crucial as next-generation facilities are set to significantly extend these samples with dedicated surveys across the electromagnetic spectrum \citep{dorazio_observational_2023}.

The remainder of this paper is structured as follows. Section~\ref{sec:ADIOSMethods} presents the theoretical background as well as the main equations governing the unified accretion disc model. The simulation suite is presented in Section~\ref{sec:sims} and the results, with a special focus on a comparison between the thin disc model and unified disc model predictions, are presented in Section~\ref{sec:results}. We discuss the observational implications and possible extensions of our model in Section~\ref{sec:discussion} and conclude in Section~\ref{sec:conclusions}.

\section{Methodology} \label{sec:ADIOSMethods}
\begin{figure*}
    \centering
    \includegraphics[width=0.95\textwidth]{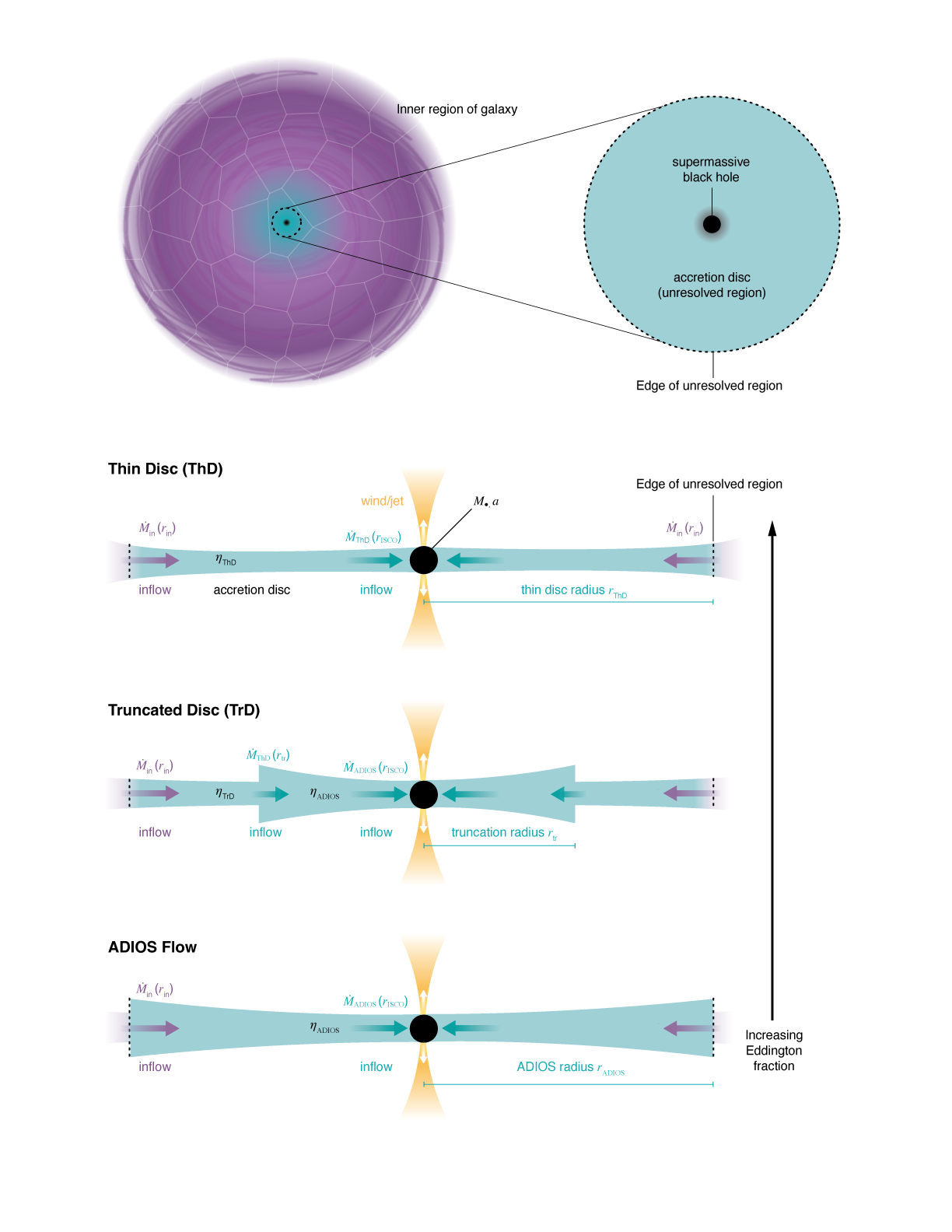}
    \vspace{-1.2cm}
    \caption{Overview of the unified accretion disc model. The top panel illustrates how the subgrid model is connected to the simulation via the boundary conditions at the outer edge of the accretion disc. The resolution in the region around the central SMBH (purple-shaded) is increased down to the scale of the accretion disc (turquoise-shaded), where the inflow rates are measured from the hydro solver and added to the SMBH -- accretion disc particle, which represents the properties (mass and angular momentum) of both the SMBH and the disc. With our unified accretion disc particle method we track both the radiatively efficient regime (thin disc) and radiatively inefficient regime (thick disc), as illustrated in the lower panel. The inflows from the hydro solver at the outer edge of the disc are evolved via our unified model based on the latest high-resolution GR(R)MHD simulations, allowing us to track the mass and spin evolution of the SMBH. As the Eddington ratio increases we smoothly transition from an advection-dominated ADIOS flow (optically thin, geometrically thick) to a Shakura-Sunyaev thin $\alpha$-disc via a truncated disc configuration with an inner ADIOS flow and outer thin disc. Crucially, in a galaxy formation context, this model can be used to inject wind and jet feedback based on the properties of the accretion flow.}
    \label{fig:truncated_disc_sketch}
\end{figure*}

In this Section, we provide the theoretical background as well as derivations of the relevant equations for our unified accretion disc model. Readers who would just like to have an overview of the methods without delving into the fine details are advised to just read Sections~\ref{subsec:NumericalImplOverview}~and~\ref{subsec:ModelSummary}, in conjunction with Figure~\ref{fig:truncated_disc_sketch} and Table~\ref{tab:AccDiscEquations}.

\subsection{Numerical implementation overview} \label{subsec:NumericalImplOverview}

Our unified accretion disc model evolves the SMBH mass, $M_{\bullet}$, and angular momentum, $\mathbf{J}_{\bullet}$, as well as the accretion disc mass, $M_\mrm{d}$, and angular momentum, $\mathbf{J}_\mrm{d}$, in a subgrid fashion, where the SMBH and its accretion disc are modelled as a composite SMBH --accretion disc particle (see Figure~\ref{fig:truncated_disc_sketch}, upper panel). This builds on previous accretion disc based SMBH growth models in galaxy formation \citep[e.g.][]{power_accretion_2011,dubois_black_2014,fiacconi_galactic_2018,bustamante_spin_2019,cenci_black_2021,tartenas_improving_2022}. In the majority of these cases, the thin $\alpha$-disc model is used to evolve the SMBH and accretion disc properties where the disc is assumed to be in local thermal equilibrium, and can radiate its heat efficiently \citep{shakura_black_1973}. The kinematic viscosity $\nu$ is parametrized as $\nu = \alpha c_\mathrm{s} H$, where $\alpha$ is a dimensionless parameter representing the efficiency of viscous transport, $c_\mathrm{s}$ is the sound speed and $H$ is the scale height. As the thin $\alpha-$disc offers a global, steady-state analytical solution and allows for the self-consistent calculation of rest-mass-energy and angular momentum transfer at the innermost stable circular orbit (ISCO), it is the commonly chosen for subgrid accretion disc prescriptions.

However, the thin $\alpha$-disc model makes several simplifying assumptions about the nature of the accretion flow -- in particular, it is only valid in the radiatively-efficient regime. At low Eddington ratios, in the radiatively-inefficient regime, the accretion flow is expected to transition to a geometrically thick disc following the ADAF solution. \citet{narayan_advection-dominated_1995} noted that ADAFs will likely be associated with strong winds and \citet{blandford_fate_1999} derived a family of self-similar solutions, advection-dominated inflow-outflow solutions `ADIOS', with a wide range of outflow efficiencies. Analytical models on their own cannot predict the wind loss in the advection-dominated regime and hence the wind parameters have to be informed by numerical simulations. In this work, we aim to extend the thin $\alpha$-disc model implementation presented in \citet{fiacconi_galactic_2018} to the radiatively inefficient regime based on the analytical ADIOS model and results from high-resolution GR(R)MHD simulations, with the transition between the two disc states modelled as a truncated accretion disc.

With our unified accretion disc model, the gas mass and angular momentum inflows at the scale of the outer accretion disc are directly measured from the hydro solver and added appropriately to the subgrid accretion disc. The properties of the SMBH and its surrounding accretion disc are then evolved by the subgrid model (see Figure~\ref{fig:truncated_disc_sketch}).

Firstly, we summarise the relevant equations for the mass and spin evolution of the accretion disc particle system. The SMBH mass, $M_{\bullet}$, evolves as:
\begin{equation}\label{eq:bh_mass_evol}
\frac{{\rm d} M_{\bullet}}{{\rm d}t} = (1 - \eta) \dot{M}_\mrm{BH,0}\,,
\end{equation}
where $\dot{M}_\mrm{BH,0}$ is the rest mass flux onto the SMBH and $\eta$ is the radiative efficiency. The disc mass, $M_{\rm d}$, evolution is described by:
\begin{equation}\label{eq:disc_mass_evol}
\frac{{\rm d} M_{\rm d}}{{\rm d}t} = -\dot{M} + \dot{M}_{\rm in}\,,
\end{equation}
where $\dot{M}$ is the steady state rest mass flux through the disc and $\dot{M}_{\rm in}$ is the gas mass inflow onto the disc (see Section~\ref{subsubsec:InflowQuants} for details on how this quantity is calculated by the hydro solver). Note that if there is no mass loss via a jet or a wind then $\dot{M} = \dot{M}_\mrm{BH,0}$. Here we distinguish between these two quantities because we model mass loss through a wind in the radiatively inefficient regime, following the ADIOS model.

The angular momentum of the SMBH, $\mathbf{J}_{\bullet}$, evolves both due to accretion and due to Lense-Thirring precession (if misaligned from the disc angular momentum vector $\mathbf{J}_{\rm d}$):
\begin{equation}\label{eq_bh_am_evol}
\begin{aligned}
\frac{{\rm d}\mathbf{J}_{\bullet}}{{\rm d}t} = \dot{M}_\mrm{BH,0} L_{\rm inner}~{\rm sign}(\mathbf{j}_{\bullet} \cdot \mathbf{j}_{\rm d})~\mathbf{j}_{\bullet} + \left.\frac{\mrm{d}\mbf{J}_\mrm{\bullet}}{\mrm{d}t}\right|_{\substack{\mrm{LT}}}\,,
\end{aligned}
\end{equation}
where $L_{\rm inner}$ is the (spin-dependent) specific angular momentum at the inner boundary of the accretion flow and $\mathbf{j}_{\bullet}$ and $\mathbf{j}_{\rm d}$ are the angular momentum unit vectors (versors). The angular momentum of the disc then evolves as:
\begin{equation}\label{eq_disc_am_evol}
\frac{{\rm d}\mathbf{J}_{\rm d}}{{\rm d}t} = - \dot{M} L_{\rm inner}~{\rm sign}(\mathbf{j}_{\bullet} \cdot \mathbf{j}_{\rm d})~\mathbf{j}_{\bullet} - \left.\frac{\mrm{d}\mbf{J}_\mrm{\bullet}}{\mrm{d}t}\right|_{\substack{\mrm{LT}}} + \dot{\mathbf{J}}_{\rm in}\,,
\end{equation}
where $\dot{\mathbf{J}}_{\rm in}$ is the gas angular momentum flow onto the disc, see Section~\ref{subsubsec:InflowQuants} for a more detailed definition.

Here we assume that the outflows are occurring in the innermost region so that the winds carry away the specific angular momentum associated with the inner boundary of the accretion flow, which represents a lower limit for the angular momentum loss. Furthermore, we note that in the truncated disc state, there would be an additional torque between the inner ADIOS flow and outer thin disc. Fully accounting for this effect is beyond the scope of this work, however, we include the impact of the outer thin disc on the precession frequency of the inner flow following \citet{bollimpalli_effect_2023} (see Section~\ref{subsubsec:AngMomLT} for details).

With the equations above, the evolution of the SMBH -- disc particle system can be fully specified given $\eta$, $\dot{M}_\mrm{BH,0}$, $\dot{M}$, $L_{\rm inner}$ and $\left.\frac{\mrm{d}\mbf{J}_\mrm{\bullet}}{\mrm{d}t}\right|_{\substack{\mrm{LT}}}$.

For the thin disc model, all of these quantities can be obtained from global analytical models \citep[see][for details]{fiacconi_galactic_2018}, however, for the ADAF model only self-similar analytical solutions exist which are not valid at the inner or outer flow boundaries due to their scale-free nature. In particular $\eta$ and $L_{\rm inner}$ cannot be calculated analytically in the ADAF regime. Instead the disc equations have to be evaluated numerically or, alternatively, these characteristic quantities can be extracted from GR(R)MHD simulations.

The remainder of the methodology section is structured as follows. Firstly, we provide details on the numerical implementation, including parameter selection and the connection between the subgrid model and the hydrodynamical simulation, in Section~\ref{subsec:NumericalImplDeets}. Subsequently, we outline the process by which the disc state is determined within our subgrid model in Section~\ref{subsec:DiscModeDeterm}. We then elaborate on our analytical modelling and use of the latest GR(R)MHD simulations to track mass and angular momentum transfer for the ADAF and truncated disc states, respectively, which are detailed in Sections~\ref{subsec:MassTransferADIOS} and \ref{subsec:AngTransferADIOS}. Finally, we present a summary of our unified accretion disc model in Section~\ref{subsec:ModelSummary}.

\subsection{Numerical implementation details} \label{subsec:NumericalImplDeets}

\subsubsection{Conventions}
In this paper, we adopt the convention whereby lower-case $r$ is a radius expressed in units of the Schwarzschild radius $R_\mrm{s}=\frac{2\mrm{G}M_{\bullet}}{\mrm{c}^{2}}$, where $\mrm{G}$ is Newton's gravitational constant, $M_{\bullet}$ is the SMBH mass, and $\mrm{c}$ is the speed of light.

\subsubsection{Accretion disc parameters and calibrations}
It is now widely accepted that angular momentum transport in ionized accretion flows occurs via the magnetorotational instability \citep[][]{balbus_magnetothermal_1991, balbus_instability_1998}. MHD simulations \citep[e.g.][]{mckinney_general_2012,tchekhovskoy_general_2012,sadowski_energy_2013,ryan_radiative_2017,liska_formation_2018} model this self-consistently but for our hydro-only simulations we need to employ an $\alpha$-like prescription for the viscous stress. For a fully ionized accretion disc, we would expect relatively high viscosities \citep[e.g.][]{martin_physical_2019}. Here we need to work with an effective viscosity prescription and we set $\alpha$ to the canonical value of $\alpha=0.1$ for the thin disc state.

In the thick disc regime, the disc scale height obeys $H/R > \alpha$. As the disc is unresolved in our simulations, we fix this to a value which matches the high-resolution simulations that we use as inputs for our subgrid model, setting $H/R = 0.3$ \citep[see][]{tchekhovskoy_general_2012}. We then calibrate $\alpha$ as a constrained parameter to ensure a smooth transition between the truncated and pure ADAF/ADIOS regime (see Section~\ref{subsubsec:MassFlowPureADAF}).

In the thin disc regime, we have $H/R < \alpha$ and we use the truncated scale height as a constrained parameter to calibrate the viscous timescale of the disc to the timescales obtained from the accretion model by \citet{fiacconi_galactic_2018}, see also Section~\ref{subsubsec:MassFlowTruncDisc}.

\subsubsection{Inflow quantities} \label{subsubsec:InflowQuants}
The resolved inflow quantities are calculated in analogy to \citet{fiacconi_galactic_2018}. These fluxes are obtained from the hydro solver and connect our subgrid model with the hydrodynamical simulation. In brief, the inflow radius, $r_\mrm{in}$, is defined as the kernel-weighted average size of the gas cell neighbours of the SMBH within the smoothing length. The mass inflow rate, $\dot{M}_\mrm{in}$, is defined as the mass flux on to the SMBH particle, which is calculated via a kernel-weighted average of the mass flux provided by the neighbours within the smoothing length. Similarly $\mbf{L}_\mrm{in}$ is the kernel-weighted average specific angular momentum. Following \citet{bourne_dynamics_2023}, we modify the inflow calculation for the angular momentum as this approach tends to overestimate the magnitude of $\mbf{L}_\mrm{in}$. Instead, we only take the direction of $\mbf{L}_\mrm{in}$ and set the magnitude of the inflow specific angular momentum to the specific angular momentum of the subgrid accretion disc. In practice, we typically find reasonable agreement between the specific angular momentum of the subgrid disc and the specific angular momentum of gas on the scale of the subgrid disc.

For a given timestep\footnote{Note that, to ensure that we temporally resolve the physical timescales introduced by our model, we introduce a time step limiter so that the maximum timestep corresponds to at most ten per cent of the accretion and/or spin alignment timescale, following \citet{fiacconi_galactic_2018}.} $\Delta t$, the disc mass $M_\mrm{d}$ and disc angular momentum $\mbf{J}_\mrm{d}$ then get updated as:
\begin{equation}
    M_\mrm{d} \rightarrow M_\mrm{d} + \dot{M}_\mrm{in} \Delta t\,,
\end{equation}
\begin{equation}
    \mbf{J}_\mrm{d} \rightarrow \mbf{J}_\mrm{d} + \dot{M}_\mrm{in} \mbf{L}_\mrm{in} \Delta t\,.
\end{equation}
\subsection{Disc mode} \label{subsec:DiscModeDeterm}

There are three disc modes: the thin disc (ThD), truncated disc (TrD) consisting of an inner ADIOS flow and outer thin disc, and a pure ADIOS flow. 

Initially, one has to make an educated guess with regards to which state the system should be in. To make this guess effectively, we calculate an auxiliary Eddington fraction, $f_\mrm{Edd,aux}$, at the beginning of every timestep, based on the expected Eddington fraction if the system were in the thin disc state\footnote{Indeed, the advantage of the steady-state thin $\alpha$-disc solution is that it allows for the unique determination of the mass flow rate through the disc given the four state variables -- BH and disc mass as well as BH and disc angular momentum. Hence, once these variables are initialised, one can evolve the system by measuring the mass and angular momentum inflow rates onto the disc.} which is derived in \citet{fiacconi_galactic_2018} following the thin $\alpha$-disc model from \citet{shakura_black_1973}.
\begin{equation}
    f_\mrm{Edd,aux} = f_\mrm{Edd,ThD}=\frac{\dot{M}}{\dot{M}_\mrm{Edd}}\,, 
\end{equation}
with
\begin{equation} \label{eq:fEddThD}
    f_\mrm{Edd,ThD} = 0.76 \left( \frac{M_\mrm{d}}{10^{4} \ \Msun}\right)^{5}
    \left( \frac{M_{\bullet}}{10^{6} \ \Msun}\right)^{-47/7} \left( \frac{a J_\mrm{d}/J_\mrm{\bullet}}{3}\right)^{-25/7}\, ,
\end{equation}
and the Eddington rate defined as
\begin{equation}
    \dot{M}_\mathrm{Edd} = \frac{4 \pi \mathrm{G} M_\mathrm{BH} \mathrm{m_{p}}}{\eta_\mathrm{ref} \mathrm{\sigma_{T} \mathrm{c}}},
\end{equation}
where $\mathrm{m_{p}}$ is the proton mass, $\eta_\mathrm{ref}=0.1$ is the reference radiative efficiency, $\mathrm{\sigma_{T}}$ is the Thomson cross section and $\mathrm{c}$ is the speed of light.
We then use $f_\mrm{Edd,aux}$ to calculate the characteristic radii of the disc system and set the disc state accordingly. To determine the accretion mode of the disc system, we compare the radius of the thin disc innermost stable circular orbit (ISCO), $r_\mrm{ISCO}$, the truncation radius, $r_\mrm{tr}$, and the outer thin disc radius, $r_\mrm{ThD}$. 

At intermediate Eddington ratios, the outer thin disc is expected to be truncated at a characteristic radius $r_\mrm{tr}$, transitioning into a hot accretion flow. The physical reason for this transition is not fully understood and there are several possible mechanisms such as the disc evaporation model, the turbulent diffusion model and the large viscosity model -- all of which predict that the truncation radius $r_\mrm{tr}$ should decrease as $f_\mrm{Edd}$ increases in agreement with observations of black hole binaries and low-luminosity AGN \citep[see][and references therein]{yuan_hot_2014}. Also see \citet{liu_application_2009} for a discussion on the application of disc evaporation models to AGN and \citet{cho_analytical_2022} for a generalised disc evaporation and transition model applicable to both X-ray binaries and AGN.

Firstly, we estimate the radius that the thin disc would have in the absence of disc truncation using the expression from \citet{fiacconi_galactic_2018}:
\begin{equation}
    r_\mrm{ThD} = 40 \left( \frac{M_\mrm{d}}{\Msun} \right)^{4/5} \left( \frac{\alpha}{0.1}\right)^{16/25} \left( \frac{M_{\bullet}}{10^{6}\ \Msun} \right)^{-44/25} f_\mrm{Edd,aux}^{-14/25}\,,
\end{equation}
and following  \citet{yuan_active_2018}, we estimate the truncation radius as:
\begin{equation}
    r_\mrm{tr} = 3 \left( \frac{2 \times 10^{-2}}{f_\mrm{Edd,aux}} \right)^{2}\,.
\end{equation}

This expression is based on disc truncation by diffusion where the energy from the interior of the disc heats the gas above the virial temperature forming an inner hot flow and truncating the outer thin disc \citep[see][]{honma_global_1996}. 

However, note that alternative truncation mechanisms give the same qualitative behaviour. In the case of the disc evaporation model, the specific evaporation rate generally decreases as a function of radius. Hence, as the Eddington ratio decreases, the truncation radius (where evaporation and gas inflow are balanced) increases. For magnetically truncated discs, we would expect similarly for the truncation radius to increase as the Eddington fraction decreases \citep[whilst higher magnetic flux at a fixed Eddington fraction may increase the truncation radius, however, this is beyond the scope of this work, see][for details]{liska_formation_2022}. More holistically, global radiation MHD simulations around SMBHs have demonstrated the link between optical depth trends and disc truncation, where the optical depth increases with larger radii and higher Eddington ratios. Hence the truncation radius, the radius where the optical depth falls below one, decreases as the Eddington ratio increases \citep[see][]{jiang_global_2019}.

Finally, $r_\mrm{ISCO}$ is given by
\begin{equation}
    r_\mrm{ISCO} = \frac{\Lambda(a)}{2}\,,
\end{equation}
where $\Lambda(a)$ is a function of the SMBH spin $a$ \citep[see e.g. equation B2 in][for an expression for $\Lambda(a)$]{fiacconi_galactic_2018}. $\Lambda(a)$ is largest for retrograde spinning BHs (orbiting against SMBH spin), whilst highly-spinning prograde BHs have the smallest ISCO.

\begin{figure}
    \centering
    \includegraphics[width=\columnwidth]{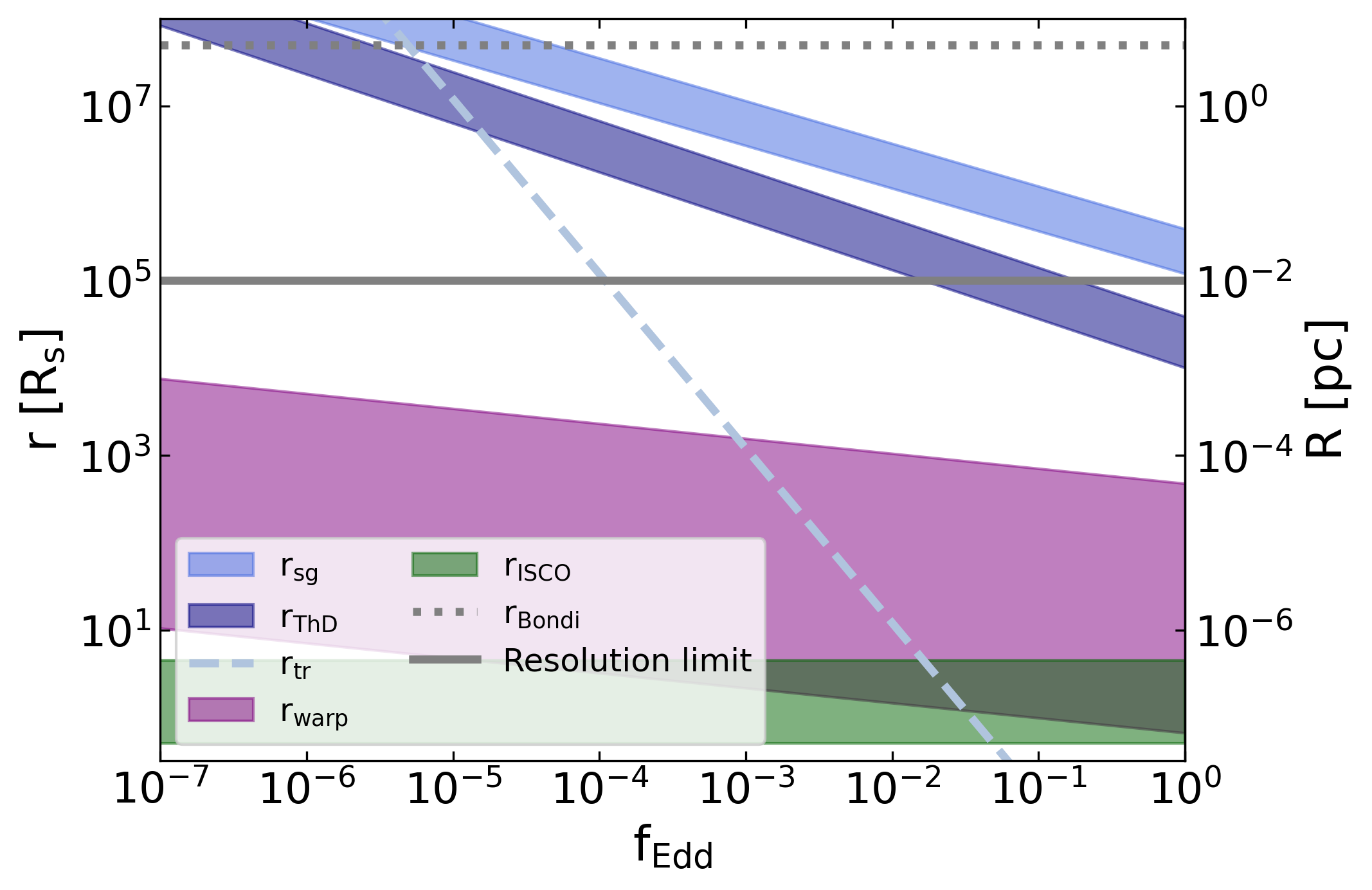}
    \caption{Characteristic disc radii as a function of Eddington fraction. The self-gravity radius ($r_\mathrm{sg}$, light-blue-shaded region), outer thin disc radius ($r_\mathrm{ThD}$, dark-blue-shaded region) and warp radius ($r_\mathrm{warp}$, purple-shaded region) are calculated for SMBH mass $M_{\bullet}=10^{6}\ \mathrm{M_\mathrm{\odot}}$ and disc mass $M_\mathrm{d}=10^{3}\ \mathrm{M_\mathrm{\odot}}$, applicable to the simulations presented in Sections~\ref{sec:sims}~and~\ref{sec:results}. We also plot the truncation radius ($r_\mathrm{tr}$, light-blue dashed line) and ISCO ($r_\mathrm{ISCO}$, green-shaded region).  The shaded regions indicate the parameter space covered by the possible SMBH spin values. For comparison, we also indicate the resolution limit of the simulations presented in this paper as a solid grey line and the Bondi radius $r_\mathrm{Bondi}$ as a dotted grey line.}
    \label{fig:radii_adios}
\end{figure}

These characteristic disc radii are shown as a function of Eddington fraction in Figure~\ref{fig:radii_adios}. For reference, we also show the warp radius (crucial for determining the Lense-Thirring precession regime, see Section~\ref{subsubsec:AngMomLT}), the self-gravity radius beyond which the disc could fragment\footnote{Note that our model does not yet include the regime where the self-gravity of the disc becomes significant. This is left for future work. Instead, following \citet{fiacconi_galactic_2018}, we restrict the mass inflows onto the subgrid accretion disc such that it does not grow beyond the self gravity radius of the $\alpha$-disc in the thin and truncated regimes. In the pure ADIOS regime, we restrict the disc mass to $M_\mrm{d} \leq \frac{H}{R} M_\mrm{\bullet}$ \citep[see e.g.][]{king_evolution_2008,fanidakis_grand_2011}.} \citep[see][for details]{fiacconi_galactic_2018}, and the Bondi radius, as well as the typical resolution scale for the simulations presented in this work.

The ratio between the ISCO, truncation radius and thin disc radius, then determines our disc state at each timestep, with the thin disc at high Eddington ratios, the truncated disc emerging at intermediate Eddington ratios and the pure ADIOS flow at low Eddington ratios (see Figure~\ref{fig:truncated_disc_sketch} for an illustration of the three main disc states):
\begin{itemize}
    \item \textit{Thin Disc}: If $r_\mrm{tr} \leq r_\mrm{ISCO}$, the ADIOS flow component is within the ISCO of the thin disc, so we proceed with the standard thin disc model from \citet{fiacconi_galactic_2018} if the Eddington fraction is sufficiently high.
    \item \textit{Truncated Disc}: If $r_\mrm{ISCO} < r_\mrm{tr} < r_\mrm{ThD}$ and $\sqrt{\mrm{G}M_{\bullet}R_\mrm{tr}} \leq J_\mrm{d}/M_\mrm{d}$, we evolve the disc system as a truncated disc.
    \item \textit{ADIOS Flow}: If $r_\mrm{tr} \geq r_\mrm{ThD}$ (i.e. the size of the ADIOS flow is larger than the size of the thin disc) or the gas does not have sufficient angular momentum to circularise at the truncation radius ($\sqrt{\mrm{G}M_{\bullet}R_\mrm{tr}} > J_\mrm{d}/M_\mrm{d}$), then we proceed with the pure ADIOS flow model. In this case, the size of the flow, $r_\mrm{ADIOS}$, may increase to larger sizes than the typical inflow scale, the Bondi radius. To avoid this complication, we limit the size of the ADIOS flow to $r_\mathrm{Bondi}$, which also ensures that we do not expand the flow beyond the so-called `strong ADAF radius', which provides a theoretical upper limit $r_\mrm{max}$ beyond which the ADAF solution becomes invalid \citep[see][]{narayan_advection-dominated_1995}.
\end{itemize}
Having developed a mechanism for determining the state of the accretion disc system, we are now ready to calculate the mass transfer, luminosity and angular momentum transfer for each of the accretion disc configurations. For the thin disc case, we simply follow the equations from \citet{fiacconi_galactic_2018} to calculate the mass and spin evolution as well as SMBH luminosities. In Sections~\ref{subsec:MassTransferADIOS}~and~\ref{subsec:AngTransferADIOS}, we outline the equations governing the mass transfer and angular momentum transfer if we have a pure ADIOS flow or a truncated disc with an inner ADIOS flow.

\subsection{Mass transfer} \label{subsec:MassTransferADIOS}

\begin{figure}
    \centering
    \includegraphics[width=\columnwidth]{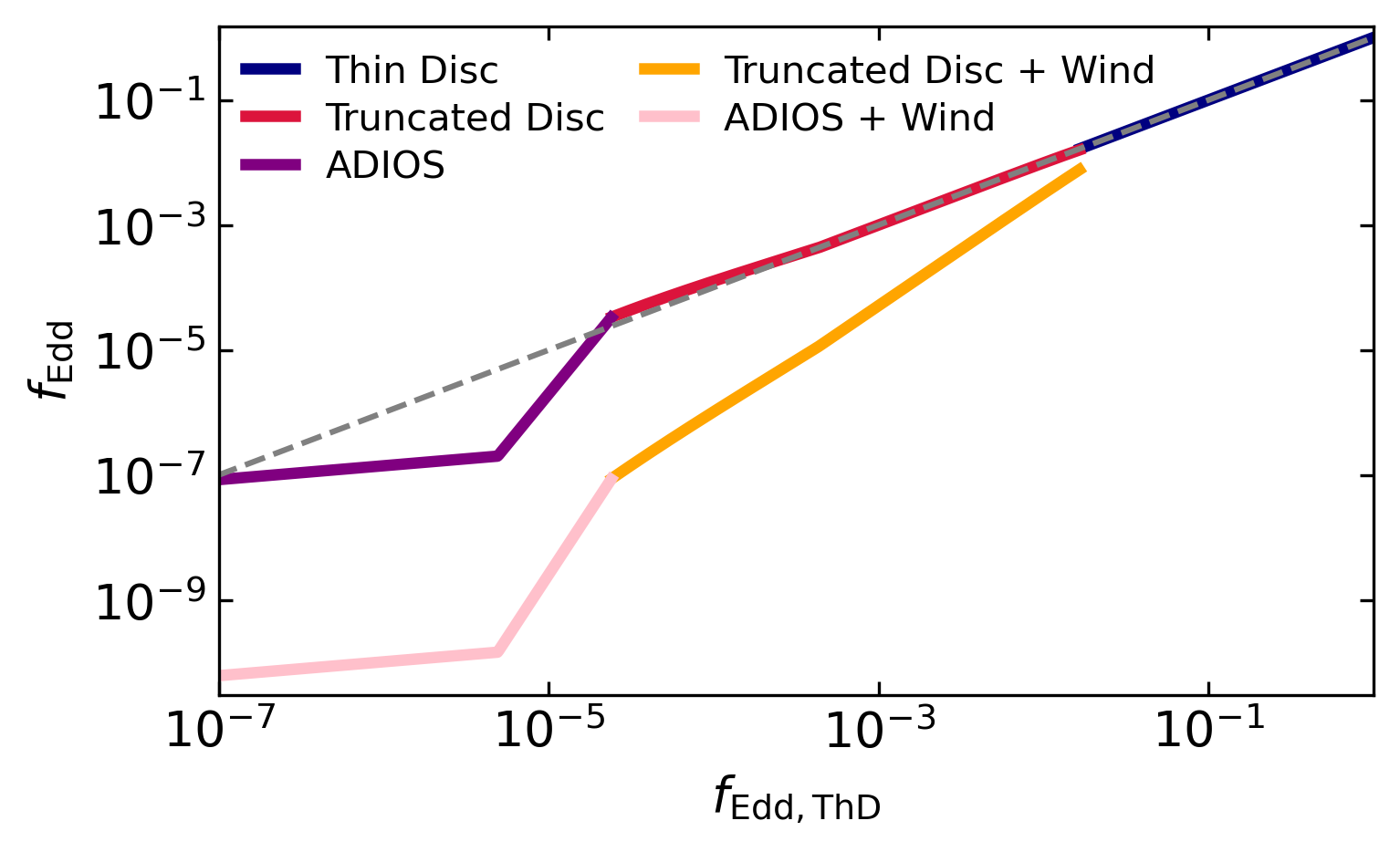}
    \caption{Eddington fractions in the unified accretion disc model as a function of Eddington fraction assuming the thin-disc-only model from \citet{fiacconi_galactic_2018} for a SMBH of mass $M_\mathrm{\bullet}=10^{6} \ \Msun$. The thin disc regime is plotted as the blue line, whilst the truncated disc regime is plotted as the red (no wind loss) and orange (with wind loss) lines. The pure ADIOS flow regime is indicated by the dark purple (no wind loss) and pink lines (with wind loss). To guide the eye, the grey-dashed line indicates the Eddington fraction if we extended the thin disc only model from \citet{fiacconi_galactic_2018} to low Eddington ratios. In the truncated disc regime, the accretion rate without wind loss follows the thin disc prediction as the outer thin disc sets the feeding rate of the inner hot flow (though there may be variability that is not captured by our model). Note the change in the slope at $f_\mathrm{Edd,ThD} \lesssim 10^{-5}$ in the pure ADIOS regime occurs as the size of the thick disc becomes limited by the Bondi radius.}
    \label{fig:fedd_disc_states}
\end{figure}
In the ADIOS model \citep{blandford_fate_1999} the accretion rate depends on the radius due to the wind loss. Therefore the definition of the Eddington fraction becomes more subtle as this will also change with radius. Here we follow \citet{xie_radiative_2012} and define the Eddington fraction of the ADIOS flows with respect to the mass accretion rate at the (outer) event horizon, $r_\mathrm{H}(a)= \frac{1 + \sqrt{1-a^{2}}}{2}$, which represents the rest mass flux onto the SMBH, $\dot{M}_\mrm{BH,0}$. The Eddington fraction is then given by:
\begin{equation}
    f_\mrm{Edd} = f_\mrm{Edd,ADIOS} = \frac{\dot{M}_\mrm{BH,0}}{\dot{M}_\mrm{Edd}}\, ,
\end{equation}
Once the state of the disc has been decided, we can then drop $f_\mathrm{Edd,aux}$ and use the actual Eddington fraction of the ADIOS flow to estimate the radiative efficiency, $\eta_\mrm{ADIOS}$. We model the radiative efficiency $\eta_\mrm{ADIOS}$, which is dependent on the ADIOS Eddington fraction $f_\mrm{Edd}$, using the analytical fitting function obtained by numerically solving the dynamical equations of a two-temperature plasma from \citet{xie_radiative_2012} corrected by the GR-R-MHD simulation results from \citet{ryan_radiative_2017} as compiled in \citet{inayoshi_transition_2019}:
\begin{equation}
\log \eta_\mrm{ADIOS} = \begin{cases}
-1.0 -(0.0162/f_\mrm{Edd})^4 & {\rm for}~~0.023\leq f_\mrm{Edd},
\vspace{1mm}\\
\sum_{n} a_n (\log{f_\mrm{Edd}})^n & {\rm for}~~~10^{-4}<f_\mrm{Edd}<0.023,
\vspace{1mm}\\
\sum_{n} b_n (\log{f_\mrm{Edd}})^n & {\rm for}~~~10^{-8}<f_\mrm{Edd}\leq 10^{-4},
\end{cases}
\end{equation}
The fitted values are $a_0=-0.807$, $a_1=0.27$, $a_n=0$ ($n\geq 2$), $b_0=-1.749$, $b_1=-0.267$, $b_2=-0.07492$ and $b_n=0$ ($n\geq3$), see \citet{inayoshi_transition_2019} for details.

Note that the radiative efficiency will likely also depend on the magnetic field threading the disc \citep[see][]{liska_magnetic_2024}, however, capturing these effects is beyond the scope of this work. In the future, we also plan to explore different radiative efficiency prescriptions based on the weak magnetic field `SANE' state and saturated magnetic field `MAD' state.

Using the radiative efficiency, we can then calculate the luminosity of the ADIOS flow as
\begin{equation} \label{eq:PureADAFlum}
\dot{E}_\mrm{ADIOS} = \eta_\mrm{ADIOS}(f_\mrm{Edd}) \dot{M}_\mrm{BH,0} \mrm{c}^2\,,
\end{equation}
and the SMBH growth rate as:
\begin{equation}
    \dot{M}_{\bullet} = (1-\eta_\mrm{ADIOS}) \dot{M}_\mrm{BH,0}\,.
\end{equation}
The rest mass flux onto the SMBH, $\dot{M}_\mrm{BH,0}$, will depend on the disc model. In particular, we obtain different rest mass fluxes for the inner ADIOS flow of the truncated disc model and the `pure' ADIOS state. This is mainly due to different accretion rates at the outer edge of the ADIOS flow (in the truncated disc case this is set by the feeding rate at the inner edge of the thin disc component) and different wind mass losses (in the truncated disc case these tend to be lower due to the smaller size of the ADIOS flow). In Sections~\ref{subsubsec:MassFlowPureADAF} and \ref{subsubsec:MassFlowTruncDisc}, we outline how $\dot{M}_\mrm{BH,0}$ is calculated for the pure ADIOS flow model and the truncated disc model, respectively. Note that in this paper, we include the possibility of wind loss according to the ADIOS model, however, we do not inject this wind mass\footnote{This means that mass is not strictly conserved within the accretion-only ADIOS subgrid model as the wind mass is being drained from the subgrid disc but not being re-injected into the simulation domain.}.
This makes the set-up well-defined and allows us to better isolate the effect of the different disc states as live wind injection will complicate the thermodynamics. Coupling our unified accretion model with wind and jet feedback is beyond the scope of this paper and will be addressed in future work.

\subsubsection{Pure ADIOS flow model} \label{subsubsec:MassFlowPureADAF}
To estimate the mass accretion rate through the pure ADIOS flow, we need to estimate the viscous timescale in the advection-dominated regime, integrating the inverse of the radial velocity from the inner to the outer edge of the ADIOS flow. As we limit the size of the accretion flow to the Bondi radius $r_\mrm{Bondi}$, we obtain $r_\mrm{ADIOS}$ as:
\begin{equation}
    r_\mrm{ADIOS} = \min(r_\mrm{tr}, r_\mrm{Bondi})\,.
\end{equation}
We then estimate the mass flow rate at $r_\mrm{ADIOS}$ as:
\begin{equation}
    \dot{M}(r_\mrm{ADIOS}) = {M}_\mrm{d}/\tau_\mrm{visc,ADIOS}\,, 
\end{equation}
where $\tau_\mrm{visc,ADIOS}$ is the viscous timescale of the ADIOS flow. Since the disc mass $M_\mrm{d}$ is updated as $M_\mrm{d} \rightarrow M_\mrm{d} + \dot{M}_\mrm{in} \Delta t$, this mass accretion rate at $r_\mrm{ADIOS}$ is simply the inflow rate rescaled by the viscous timescale and corrected for disc draining.

The viscous timescale may be estimated from the radial velocity of the ADIOS flow $v_\mrm{r,ADIOS}$ as:
\begin{equation}
    \tau_\mrm{visc,ADIOS} = \int_{r_\mrm{ADIOS}}^{r_\mrm{H}} \frac{R_\mrm{S}}{v_\mrm{r,ADIOS}} \,\mathrm{d}r\,.  
\end{equation}
We use the radial velocity formula from \citet{narayan_advection-dominated_1995} in the set-ups with and without wind loss, since $v_\mrm{r}$ is not (strongly) dependent on density and therefore not very sensitive to the wind \citep{yuan_numerical_2012}, to obtain: 
\begin{equation}
    \tau_\mrm{visc,ADIOS} =  0.0196 \left(\frac{M_{\bullet}}{10 \ \Msun}\right) \left( \frac{0.25}{\alpha_\mrm{ADIOS}}\right) \left( \frac{r_\mrm{ADIOS}}{10}\right)^{3/2} \ \mathrm{s}\,,
\end{equation}
where $\alpha_\mrm{ADIOS}$ is the viscosity parameter for the ADIOS flow. Note that there are significant uncertainties in the value of this parameter \citep[see e.g.][]{mckinney_general_2012}, which depends on the accumulated magnetic flux, disc thickness, and SMBH spin. Therefore, we calibrate this value to achieve a smooth transition in accretion rates between the truncated disc regime and the pure ADIOS regime, choosing $\alpha_\mrm{ADIOS}$ such that the mass flow rate through the pure ADIOS flow for $r_\mrm{tr} \xrightarrow{} r_\mrm{ThD}$ tends to the mass flow rate through the outer thin disc for the same limit (see Section~\ref{subsubsec:MassFlowTruncDisc}).
Following \citet{blandford_fate_1999}, we can then use the accretion rate at $r_\mrm{ADIOS}$ to calculate the mass accretion rate at the event horizon ($r_\mrm{H}$):
\begin{equation}
    \dot{M}_\mrm{BH,0} = \dot{M}(r_\mrm{ADIOS}) \left( \frac{r_\mrm{H}}{r_\mrm{ADIOS}} \right)^{s} = \frac{M_\mrm{d}}{\tau_\mrm{visc,ADIOS}} \left( \frac{r_\mrm{H}}{r_\mrm{ADIOS}} \right)^{s}\,,
\end{equation}
where $s$ characterises the importance of the outflow. This parameter cannot exceed $s=1$ for energetic reasons and $s=0$ corresponds to no outflow. For our ADIOS-model-based simulations, we set $s=0.4$ for consistency with \citet{xie_radiative_2012} who assume this value in their numerical determination of the radiative efficiency of the outflow. We also run a reference simulation without wind loss setting $s=0.0$.

\subsubsection{Truncated disc model} \label{subsubsec:MassFlowTruncDisc}
In the truncated disc state, the inner hot flow is being fed by the outer thin disc, which limits the SMBH mass growth rate. The variability of the inner hot flow could in principle differ from the outer thin disc. Thick discs tend to exhibit higher variability \citep[e.g.][]{lalakos_jets_2023,liska_magnetic_2024}, which may be explained by a variable $\alpha$ parameter \citep[e.g.][]{turner_new_2023}, and have shorter viscous timescales. In our model, we set the mass flow rate through the inner hot flow (modulo wind mass loss) equal to the feeding rate from the outer thin disc, for simplicity. This will not affect the overall SMBH mass growth rate but may impact electromagnetic counterpart predictions and we discuss this in more detail in Section~\ref{subsec:results_accrate_lum}.

The derivation of the thin disc mass accretion rate in \citet{fiacconi_galactic_2018} relies on the inner disc radius being much smaller than the outer disc radius. This is generally true when the inner radius is at the ISCO, however, this assumption may no longer be valid for the truncated disc model when the truncation radius can reach a significant fraction of the thin disc radius. For this reason, we only use the formula for $\dot{M}_\mrm{ThD}$ from \citet{fiacconi_galactic_2018} if $r_\mrm{tr}/r_\mrm{ThD} < 10^{-2}$. In this case, the thin disc accretion rate is given by:
\begin{equation}
    \dot{M}(r_\mrm{tr})= f_\mrm{Edd,ThD} \times \dot{M}_\mrm{Edd}\,.
\end{equation}
If $r_\mrm{tr}/r_\mrm{ThD} \geq 10^{-2}$, we estimate the thin disc accretion rate via the viscous timescale $\tau_\mrm{visc}$:
\begin{equation}
    \dot{M}(r_\mrm{tr}) = \frac{M_\mathrm{ThD}}{\tau_\mrm{visc}}\,.
\end{equation}
Here we evaluate the viscous timescale based on the radial velocity for the thin $\alpha-$disc 
\begin{equation}
    v_\mrm{r, ThD} = - \frac{3}{2} \frac{\nu}{R} = - \frac{3}{2} \frac{\alpha c_\mrm{s} H}{R}\,,
\end{equation}
where $\nu= \alpha c_\mathrm{s} H$ is the kinematic viscosity. Integrating this expression between the truncation radius and the outer thin disc radius yields:
\begin{equation}
    \tau_\mrm{visc} = 1.39 \times 10^{-3} \left(\frac{M_{\bullet}}{10 \ \Msun}\right) \left( \frac{0.1}{\alpha}\right) \left( \frac{H}{R}\right)^{-2} \left( r_\mrm{ThD}^{3/2} - r_\mrm{tr}^{3/2}\right) \mrm{s}\,.
\end{equation}
The thin disc mass $M_\mathrm{ThD}$, adjusted for the truncation radius, is given by \citep{fiacconi_galactic_2018}:
\begin{equation}
    M_\mrm{ThD} = M_\mrm{d} \left(1 - \left( \frac{r_\mathrm{tr}}{r_\mathrm{ThD}} \right)^{5/4} \right)\,.
\end{equation}
The derivation of the (unresolved) scale height for the thin disc model is significantly modified in the presence of an inner hot flow \citep[e.g.][]{rozanska_vertical_1999} and will likely not apply for the heavily truncated discs where we employ the viscous timescale formula -- therefore we treat $H/R$ as a free parameter in this regime (though we enforce $H/R < \alpha$). We then calibrate the scale height such that $\frac{M_\mathrm{ThD}}{\tau_\mrm{visc}}$ matches $\dot{M}_\mrm{ThD}$ for $r_\mrm{tr}=0.01r_\mrm{ThD}$. 

To calculate the feeding rate at the truncation radius, the radiative losses in the thin disc need to be estimated. Note that for the `pure' thin disc model the radiative efficiency is given by:
\begin{equation}
    \eta_\mrm{ThD}(a) = 1 - \sqrt{1- \frac{1}{3 r_\mrm{ISCO}(a)}}\,,
\end{equation}
as $r_\mrm{ISCO}$ defines an effective surface for the maximum feasible energy extraction from infalling particles. In the thin disc case, we can only extract energy from the thin disc component down to $r_\mrm{tr}$, so the expression for $\eta_\mrm{TrD}$ then becomes:
\begin{equation}
    \eta_\mrm{TrD} = 1 - \sqrt{1- \frac{1}{3 r_\mrm{tr}}}\,.
\end{equation}
The ADIOS flow feeding rate at the truncation radius is therefore set by the thin disc mass accretion rate $\dot{M}_\mrm{ThD}$ and the radiative efficiency $\eta_\mrm{TrD}$. In analogy to the previous section, we can then use this feeding rate to estimate the accretion rate at the event horizon (ADIOS model):
\begin{equation}
    \dot{M}_\mrm{BH,0} = \dot{M}(r_\mrm{tr}) (1 - \eta_\mrm{TrD}) \left( \frac{r_\mrm{H}}{r_\mrm{tr}} \right)^{s}\,,
\end{equation}
where we assume $s=0.4$ for the simulations with wind loss and set $s=0.0$ for the comparison simulations without wind loss.

Figure~\ref{fig:fedd_disc_states} shows the Eddington fraction of the unified accretion disc model $f_\mathrm{Edd}$ as a function of the Eddington fraction in the thin disc model $f_\mathrm{Edd,ThD}$ for $M_{\bullet} = 10^{6} \ \Msun$. The dark purple and pink lines correspond to the pure ADIOS flow model without and with wind loss, respectively. The steeper slope in this regime is introduced by the dependence of the truncation radius on the thin disc Eddington fraction to the second power, which leads to a rapid increase in the viscous timescale as $f_\mathrm{Edd,ThD}$ decreases. The gradient shifts at $f_\mathrm{Edd,ThD} \lesssim 10^{-5}$, where the ADIOS radius becomes limited by the Bondi radius. The red and orange lines show the truncated disc Eddington fractions without and with wind loss, respectively. Without wind loss, this closely corresponds to the thin disc rate (indicated by the grey dashed line) as the inner hot flow rate is set by the outer feeding rate. Note the slight gradient change at $f_\mathrm{Edd,ThD} \lesssim 10^{-3}$ as the thin disc accretion rate prescription switches to being estimated by the viscous timescale. Finally, for $f_\mathrm{Edd,ThD} \gtrsim 10^{-2}$, the system switches to the pure thin disc state (blue line).

\subsection{Angular momentum transfer} \label{subsec:AngTransferADIOS}

In this Section, we describe how to model the angular momentum transfer for each disc state both due to accretion (see Section~\ref{subsubsec:AngMomAcc}) and due to Lense-Thirring precession (see Section~\ref{subsubsec:AngMomLT}). Note that, as for the mass transfer, if we are in the thin disc regime, we evolve the angular momenta of the disc and the SMBH according to the equations from \citet{fiacconi_galactic_2018}. Below we outline our scheme for the pure ADIOS flow and truncated disc cases.

\subsubsection{Accretion} \label{subsubsec:AngMomAcc}
For both the pure ADIOS flow and the truncated disc model with an inner ADIOS flow, we model the evolution of the SMBH angular momentum, $\mbf{J}_\mrm{\bullet}$, due to accretion as: 
\begin{equation}
\label{eq:AngularMomentumTransfer}
    \left.\frac{\mrm{d}\mbf{J}_\mrm{\bullet}}{\mrm{d}t}\right|_{\substack{\mrm{acc}}}=\dot{M}_\mrm{BH,0} L_\mrm{ADIOS}\;\mrm{sign}(\mbf{j}_\mrm{\bullet}\cdot \mbf{j}_\mrm{d})\;\mbf{j}_\mrm{\bullet}\,,
\end{equation}
where $\mbf{j}_\mrm{\bullet}$ and $\mbf{j}_\mrm{d}$ are the angular momentum unit vectors (`versors')  of the SMBH and the disc, respectively, and $L_\mrm{ADIOS}$ is the specific angular momentum of the ADIOS flow. The latter is evaluated at the event horizon because for thick discs the stress is non-zero down to the horizon \citep[see e.g.][]{sadowski_energy_2016}. To evaluate this expression, we need to know the value of $L_\mrm{ADIOS}$, which we would expect to be below the thin disc value due to additional (gas and magnetic) pressure support. We obtain $L_\mrm{ADIOS}$ from the analysis by \citet{lowell_rapid_2024} where they estimate the contributions to the spin-up function from both electromagnetic (jet) and hydrodynamic (disc) components using the GRMHD simulations of an ADIOS flow in the MAD state by \citet{tchekhovskoy_general_2012}. They find that there is no significant dependence of the specific angular momentum $L_\mrm{ADIOS}$ on the spin parameter and obtain an approximately constant value $L_\mrm{ADIOS} \sim 0.86$ in geometrical units for the radiatively inefficient regime.

\subsubsection{Lense-Thirring precession} \label{subsubsec:AngMomLT}

\begin{figure}
    \centering
    \includegraphics[width=\columnwidth]{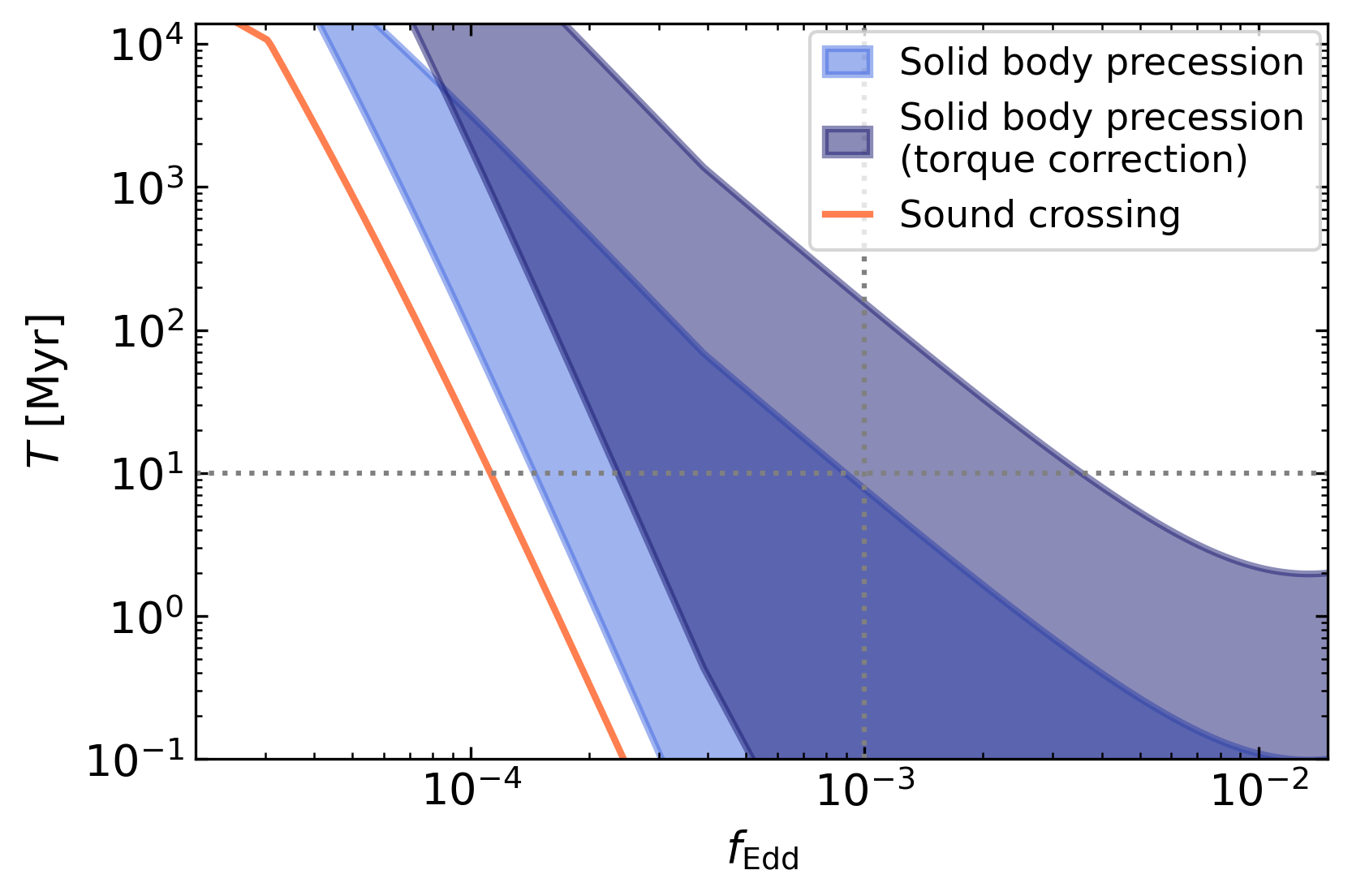}
    \caption{The Lense-Thirring precession timescales in the solid-body regime for $M_\mathrm{BH} = 10^{6} \ \Msun$, $M_\mathrm{d} = 10^{3} \ \Msun$ and $a=0.9$. The lower limit represents the classical solid body precession timescale for a pure ADIOS flow, whilst the upper limit of the shaded regions takes into account the fractional angular momentum of the inner hot flow in the truncated state (see text). The light blue shaded region does not take into account additional torques between the outer thin disc and inner hot flow, whilst the dark blue shaded region includes torque corrections due to the outer thin disc following \citet{bollimpalli_effect_2023}. Note that our simulations with $f_\mathrm{Edd} \sim 10^{-3}$ lead to typical precession rates of $\sim 10$~Myr, whilst the torque-corrected precession timescales correspond to $\gtrsim 100$~Myr.}
    \label{fig:solid_body_prec}
\end{figure}

\begin{table*}
\caption{Overview of the unified accretion disc model. We list the relevant equations for evolving the different disc states shown in the schematic Figure~\ref{fig:truncated_disc_sketch}: thin disc (top row), truncated disc (middle row) and ADIOS flow (bottom row). See Section~\ref{subsec:NumericalImplOverview} for an overview of the general equations governing the disc and SMBH evolution.}
\begin{center}
\begin{tabular}{@{}ll@{}}
\toprule
\textbf{Disc state} & \textbf{Relevant equations and quantities}  \\ 

\midrule

Thin & $\eta = \eta_\mrm{ThD}(a)$ [\citealt{bardeen_rotating_1972}] \\
Disc & \\
& $\dot{M} = \dot{M}_\mrm{BH,0} = \dot{M}_\mrm{ThD} = 0.76 \left( \frac{M_\mrm{d}}{10^{4} \ \Msun}\right)^{5}
    \left( \frac{M_{\bullet}}{10^{6} \ \Msun}\right)^{-47/7} \left( \frac{a J_\mrm{d}/J_\mrm{\bullet}}{3}\right)^{-25/7} \dot{M}_\mrm{Edd}$ \\
& [assuming no outflows, \citealt{fiacconi_galactic_2018}]  \\
  & \\  
 & $L_\mrm{inner}=L_\mrm{ISCO,ThD}(a)$ [\citealt{bardeen_rotating_1972}] \\    
 & \\
 & $\left.\frac{\mrm{d}\mbf{J}_\mrm{\bullet}}{\mrm{d}t}\right|_{\substack{\mrm{LT}}} = 
 \begin{cases}
      - J_{\bullet} \left\{\frac{\sin(\pi/7)}{\tau_{\rm align}} \left[\mathbf{j}_{\bullet} \times \mathbf{j}_{\rm d} \right] + \frac{\cos(\pi/7)}{\tau_{\rm align}} \left[ \mathbf{j}_{\bullet} \times \left( \mathbf{j}_{\bullet} \times \mathbf{j}_{\rm d} \right) \right]\right\} & \text{if $r_\mrm{warp} < r_\mrm{ThD}$}\\
      \text{[instantaneous alignment, \citealt{king_aligning_2005}]} & \text{if $r_\mrm{warp} \geq r_\mrm{ThD}$} 
    \end{cases}$  \\
    
  & [\citealt{martin_alignment_2007,perego_mass_2009,dotti_orientation_2013,fiacconi_galactic_2018}]   \\ 
  & \\ \midrule

Truncated & $\eta = \eta_\mrm{ADIOS}(f_\mrm{Edd})$  \\
Disc & [\citealt{xie_radiative_2012,ryan_radiative_2017,inayoshi_transition_2019}]  \\
& \\
& $\dot{M} = \dot{M}(r_\mrm{tr}) $  \\
& \\

& $\dot{M} (r_\mrm{tr}) =
    \begin{cases}
       \dot{M}_\mrm{ThD} \ \text{[\citealt{fiacconi_galactic_2018}]} & \text{if $r_\mrm{tr}/r_\mrm{ThD} < 10^{-2}$}\\
       M_\mathrm{ThD}/\tau_\mrm{visc} \ \text{[\citealt{pringle_accretion_1981,fiacconi_galactic_2018}]} & \text{if $r_\mrm{tr}/r_\mrm{ThD} \geq 10^{-2}$} 
    \end{cases}$   \\
    & \\
    
& $\dot{M}_\mrm{BH,0} = (1-\eta_\mrm{TrD}) \dot{M}(r_\mrm{tr}) \left( \frac{r_\mrm{H}}{r_\mrm{tr}} \right)^{s}$ [\citealt{blandford_fate_1999}]  \\
& \\
    
& $L_\mrm{inner}=L_\mrm{ADIOS}$ [\citealt{tchekhovskoy_general_2012,lowell_rapid_2024}]  \\  
& \\
 
& $\left.\frac{\mrm{d}\mbf{J}_\mrm{\bullet}}{\mrm{d}t}\right|_{\substack{\mrm{LT}}} = 
 \begin{cases}
      \text{[as thin disc case, \citealt{fiacconi_galactic_2018}]} & \text{if $r_\mrm{tr} < r_\mrm{warp}$}\\
      J_\mrm{\bullet}  \left\{ - \frac{J_\mrm{d,ADIOS}}{J_\mrm{\bullet}} \omega_\mrm{prec} \left[ \mbf{j}_\mrm{\bullet} \times \mbf{j}_\mrm{d} \right] - \frac{2 \pi}{t_\mrm{acc}} \left[ \mbf{j}_\mrm{\bullet} \times(\mbf{j}_\mrm{\bullet} \times \mbf{j}_\mrm{d}) \right]  \right\} \ \text{[\citealt{ingram_review_2019}]} & \text{if $r_\mrm{tr} \geq r_\mrm{warp}$} 
    \end{cases}$   \\
    & \\

     \midrule

ADIOS  & $\eta = \eta_\mrm{ADIOS}(f_\mrm{Edd})$  \\
Flow & [\citealt{xie_radiative_2012,ryan_radiative_2017,inayoshi_transition_2019}]  \\
& \\
& $\dot{M} = \frac{M_\mrm{d}}{\tau_\mrm{visc,ADIOS}}$ [\citealt{narayan_advection-dominated_1995,esin_advection-dominated_1997}]  \\
& \\
& $\dot{M}_\mrm{BH,0} = \frac{M_\mrm{d}}{\tau_\mrm{visc,ADIOS}} \left( \frac{r_\mrm{H}}{r_\mrm{ADIOS}} \right)^{s}$ [\citealt{blandford_fate_1999}]  \\
& \\
& $L_\mrm{inner}=L_\mrm{ADIOS}$ [\citealt{tchekhovskoy_general_2012,lowell_rapid_2024}]  \\  
& \\
& $\left.\frac{\mrm{d}\mbf{J}_\mrm{\bullet}}{\mrm{d}t}\right|_{\substack{\mrm{LT}}} = J_\mrm{\bullet}  \left\{ - \frac{J_\mrm{d,ADIOS}}{J_\mrm{\bullet}} \omega_\mrm{prec} \left[ \mbf{j}_\mrm{\bullet} \times \mbf{j}_\mrm{d} \right] - \frac{2 \pi}{t_\mrm{acc}} \left[ \mbf{j}_\mrm{\bullet} \times(\mbf{j}_\mrm{\bullet} \times \mbf{j}_\mrm{d}) \right]  \right\}$ [\citealt{ingram_review_2019}]  \\
& \\
\bottomrule
\end{tabular}
\end{center}
\label{tab:AccDiscEquations}
\end{table*}

Next, we focus on the angular momentum transfer between the SMBH and the disc due to the Lense-Thirring effect \citep{lense_uber_1918}. Lense-Thirring precession is a frame-dragging effect which manifests itself as nodal precession of orbits whose angular momentum is misaligned with the SMBH spin. In the weak-field limit, the precession frequency, $\omega_\mrm{LT}$, is given by \citep{lense_uber_1918}:
\begin{equation} \label{eq:weakfieldomegaLT}
    \omega_\mrm{LT} \approx \pm \frac{\mrm{c}}{R_\mrm{s}} \frac{2a}{2r^{3}}\,.
\end{equation}
Accretion discs are warped by the differential nature of the Lense-Thirring precession frequency. How these warps propagate depends on the type of accretion disc.

\begin{figure*}
    \centering
    \includegraphics[width=\textwidth]{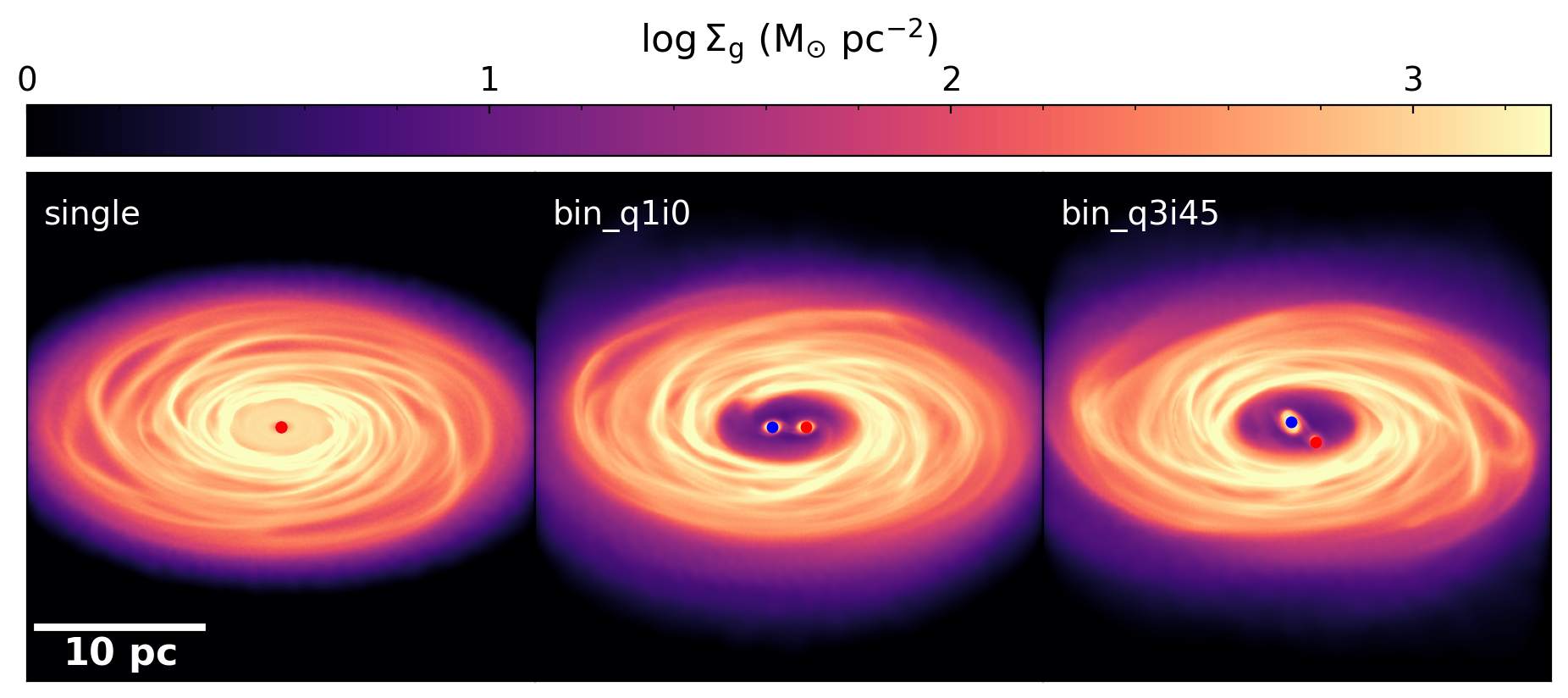} 
    \caption{Gas surface density maps of the three main simulation set-ups including the single SMBH simulation where the cavity is filled in with low density gas due to the absence of torques (\textit{left panel}), the equal-mass aligned binary simulation (\textit{middle panel}) and the non-equal mass (mass ratio $q=3$), misaligned (inclination angle $i=45\degree$) binary simulation (\textit{right panel}). Due to the super-Lagrangian refinement, we can resolve the binary system in detail (down to 0.01~pc), including streams in the cavity and the mini discs which feed the subgrid unified accretion disc.}
    \label{fig:proj_all}
\end{figure*}

In the diffusive regime ($\alpha > H/R$, thin disc case), the warps are communicated by viscosity resulting into the so-called Bardeen-Petterson configuration \citep{bardeen_lense-thirring_1975}, where the disc is aligned with the SMBH equatorial plane at small radii ($r < r_\mrm{warp}$) and retains its initial tilt at large radii ($r > r_\mrm{warp}$) with a smooth warp in between at $r_\mrm{warp}$ which is given by \citep[see][for derivation]{fiacconi_galactic_2018}:
\begin{equation}
    r_\mrm{warp} = 476 \, \xi^{-4/7} \left(\frac{\alpha}{0.1} \right)^{24/35} \left( \frac{M_{\bullet}}{10^{6} \ \Msun} \right)^{4/35} f_\mrm{Edd}^{-6/35} a^{4/7}\,,
\end{equation}
where $\xi$ is a parameter $\mathcal{O}(1)$ which is related to the ratio of the radial viscosity $\nu_{1}$ and vertical viscosity $\nu_{2}$ as $\nu_{1}/\nu_{2}=\xi/(2\alpha^{2})$ and can be determined numerically \citep{papaloizou_time-dependence_1983,lodato_warp_2007}.

In the wave-like regime ($\alpha < H/R$, ADIOS case), warps are communicated by pressure waves. These so-called bending waves result in a smooth warp far from the SMBH, whilst radial tilt oscillations are obtained close to the BH, as the wavelength of these pressure waves is a strong function of radius \citep[see][for details]{ingram_review_2019}.

For our pure ADIOS flow model, the system is in the wave-like regime. The whole flow, which is strongly coupled by the pressure waves (the local sound crossing timescale is much shorter than the precession timescale), then remains misaligned and precesses as a solid body \citep[see][]{ingram_low-frequency_2009,ingram_physical_2011,ingram_review_2019}. 

For our truncated disc model, the extent of the truncation radius determines whether we are in the diffusive or the wave-like regime.

If $r_\mrm{tr} < r_\mrm{warp}$, the thin disc feeds the ADIOS flow component aligned material, thereby twisting the ADIOS flow into alignment. Also see \citet{liska_formation_2018} who find in their numerical simulations that the angular momentum of the inner hot flow is negligible so that the thin disc can easily affect its alignment.

If $r_\mrm{tr} \geq r_\mrm{warp}$, the thin disc feeds misaligned material to the inner flow, so there is no alignment in the inner region and the inner ADIOS flow precesses as a solid body just as in the pure ADIOS flow case.

Following \citet{ingram_review_2019}, we can calculate the angular frequency of solid body precession, $\omega_\mrm{prec}$, as:
\begin{equation} \label{eq:omegaprec}
    \omega_\mrm{prec} = \frac{\int_{R_\mrm{in}}^{R_\mrm{out}} \omega_\mrm{LT}(R) \mcal{L}(R) R \mrm{d}R}{\int_{R_\mrm{in}}^{R_\mrm{out}}\mcal{L}(R) R \mrm{d}R}\,,
\end{equation}
where $\mcal{L}(R) = \Sigma R^{2} \Omega_{\phi}$ is the angular momentum per unit area. Note that since the $\mcal{L}(R)$ term appears in both the numerator and the denominator, we only need to know how this term scales with $R$, but not the absolute value. Using the scaling relations for the ADIOS model from \citet{yuan_hot_2014}, we have that the angular frequency scales as $\Omega_{\phi} \sim R^{3/2}$ and the volume density scales as $\rho \sim R^{-3/2+s}$. Under the assumption of a constant $H/R$ ratio, the surface density then scales as $\Sigma \sim R^{-1/2+s}$. Using the weak-field formula for $\omega_\mrm{LT}$ (Equation~(\ref{eq:weakfieldomegaLT})), the integrals in Equation~(\ref{eq:omegaprec}) can then be evaluated analytically \citep{fragile_global_2007,ingram_low-frequency_2009}. 

Note that with $s=0.4$, the surface density $\Sigma$ is nearly constant. However, GRMHD simulations predict that for misaligned discs, the density should drop off sharply in the region dominated by radial tilt oscillations. We therefore set the inner radius to be the bending wave radius $r_\mrm{bw} = 3.0 (H/R)^{-4/5} a^{2/5}$ and the outer radius is set to be equal to the extent of the ADIOS flow.

The torque due to Lense-Thirring precession is then given by:
\begin{equation} \label{eq:LTsolidbodyprec}
    \left.\frac{\mrm{d}\mbf{J}_\mrm{\bullet}}{\mrm{d}t}\right|_{\substack{\mrm{LT,prec}}} = \mbf{J}_\mrm{d,ADIOS} \times (\omega_\mrm{prec} \, \mbf{j}_\mrm{\bullet})\,,
\end{equation}
where in the pure ADIOS flow model, we simply have $\mbf{J}_\mrm{d,ADIOS} = \mbf{J}_\mrm{d}$. In the truncated disc case, we assume that the $\Sigma \sim R^{-3/4}$ scaling still holds for the thin disc component, so that we can scale the thin disc angular momentum as:
\begin{equation}
    \mbf{J}_\mrm{d,ThD} = \mbf{J}_\mrm{d} \left[1 - \left(R_\mrm{tr}/R_\mrm{ThD}\right)^{7/4} \right]\,,
\end{equation}
and consequently the ADIOS angular momentum is given by 
\begin{equation}
    \mbf{J}_\mrm{d,ADIOS} = \mbf{J}_\mrm{d} \left(R_\mrm{tr}/R_\mrm{ThD}\right)^{7/4}\,.
\end{equation}
In addition to disc precession, we also expect a disc alignment process -- much slower than the Bardeen-Petterson alignment in the thin disc case, with the thick disc aligning on the accretion timescale $t_\mrm{acc} = M_\mrm{BH}/\dot{M}_\mrm{BH,0}$ \citep[see e.g.][for analytical models of disc alignment]{king_aligning_2005,volonteri_distribution_2005} as well as the GRMHD simulations by \citet{liska_phase_2023-1}, which also find a global alignment mode for the disc-corona-jet system on the accretion timescale. The full Lense-Thirring torque due to disc precession and alignment is then given by:
\begin{equation} \label{eq:LT_solidbody}
    \left.\frac{\mrm{d}\mbf{J}_\mrm{\bullet}}{\mrm{d}t}\right|_{\substack{\mrm{LT}}} = J_\mrm{\bullet}  \left\{ - \frac{J_\mrm{d,ADIOS}}{J_\mrm{\bullet}} \omega_\mrm{prec} \left[ \mbf{j}_\mrm{\bullet} \times \mbf{j}_\mrm{d} \right] - \frac{2 \pi}{t_\mrm{acc}} \left[ \mbf{j}_\mrm{\bullet} \times(\mbf{j}_\mrm{\bullet} \times \mbf{j}_\mrm{d}) \right]  \right\}\,.
\end{equation}
In Figure~\ref{fig:solid_body_prec} we show the solid body precession timescales (blue-shaded regions) compared to the sound crossing timescale \citep[orange line, see][for derivation]{motta_different_2018}. From Equation~(\ref{eq:LTsolidbodyprec}), we can see that the precession timescales depends on $\omega_\mathrm{prec}$, the angle between the versors and the ratio of the ADIOS angular momentum to BH angular momentum. Indeed from this equation, we can derive that the rate of change of the azimuthal angle goes as $\frac{\mrm{d}\phi}{\mrm{d}t} = \omega_\mathrm{prec} \frac{J_\mathrm{d,ADIOS}}{J_\mathrm{d}}$ when the angular momentum of the BH dominates (recovering $\frac{\mrm{d}\phi}{\mrm{d}t} = \omega_\mathrm{prec}$ in the pure ADIOS regime). The $\frac{\mrm{d}\phi}{\mrm{d}t} = \omega_\mathrm{prec}$ case provides the lower limit on the precession timescales in Figure~\ref{fig:solid_body_prec} and the upper limit (used in our implementation) is given by the truncated-disc-corrected precession timescale, in all cases assuming $J_\mathrm{d}/J_\mathrm{BH}=0.9$, matching the typical angular momentum ratios in our binary SMBH simulations. The light-blue shaded region corresponds to the precession timescales without considering additional torques between the inner hot flow and outer truncated thin disc, whilst the dark-blue shaded region shows the timescales corrected by 95 per cent, following \citet{bollimpalli_effect_2023} who found that solid body precession may be slowed down by up to 95 per cent for truncated discs due to additional torques. In our model, we include this slowed down precession as an optional module to test the impact of reduced precession rates in this regime.

Finally, we note that in the MAD regime, jets can force the inner part of the ADIOS flow to align rapidly with the SMBH spin in GRMHD simulations \citep[e.g.][]{mckinney_alignment_2013}. However, \citet{liska_formation_2018} investigated a range of initial conditions for GRMHD simulations of tilted black hole discs and found that the importance of this effect crucially depends on the magnetic field configuration in the initial conditions. As we do not model the jet component, we do not include electromagnetic alignment due to jets here. However, we will consider this effect in future work when coupling our accretion disc model with an AGN jet subgrid prescription.

\subsection{Summary of the model} \label{subsec:ModelSummary}

As discussed, the evolution of the SMBH and its accretion disc within our unified model can be fully specified given the radiative efficiency, $\eta$, rest mass flow rate onto the SMBH, $\dot{M}_\mrm{BH,0}$, rest mass flow rate through the disc, $\dot{M}$, specific angular momentum at the inner boundary, $L_{\rm inner}$, and Lense-Thirring torque, $\left.\frac{\mrm{d}\mbf{J}_\mrm{\bullet}}{\mrm{d}t}\right|_{\substack{\mrm{LT}}}$. These quantities are listed for each state of the disc system in Table~\ref{tab:AccDiscEquations}. We also list the relevant references for all of the equations. 

Note that the derivations of the thin disc equations can be found in \citet{fiacconi_galactic_2018}, in particular the derivation of the Bardeen-Petterson torque for Lense-Thirring precession in the thin disc regime as listed in Table~\ref{tab:AccDiscEquations} (here $J_{\bullet}$ is the magnitude of the SMBH angular momentum and $\tau_\mrm{align}$ is the time-scale for the torque to modify the SMBH angular momentum). Note that if the warp radius exceeds the thin disc radius, instantaneous alignment of SMBH and disc angular momenta is assumed for the thin disc case, as listed in Table~\ref{tab:AccDiscEquations}.

\section{Simulations} \label{sec:sims}
\begin{table*}
\caption{Overview of single SMBH simulations. We list the simulation names (first column), accretion disc model employed (second column), wind loss (third column), initial mass of the subgrid accretion disc $M_\mathrm{d}$ (fourth column) and initial Eddington ratio for the subgrid accretion disc $f_\mathrm{Edd}$ (fifth column).}
\begin{center}
\begin{tabular}{@{}lcccc@{}}
\toprule
\textbf{Simulation name} & \textbf{Accretion disc model} & \textbf{Wind loss?}  &  \textbf{Initial disc mass} $M_\mathrm{d}$ [$\Msun$] & \textbf{Initial} $f_\mathrm{Edd}$  \\  \toprule
single\_uniadios & Unified & No & 2000 & $10^{-3}$	\\		
single\_uniadios\_winds&Unified	&Yes	&2000	&$10^{-3}$ \\
single\_thindisc	&Thin disc only	&No	&2000 &$10^{-3}$ \\
single\_uniadios\_light	&Unified	&No	&500	&$10^{-6}$ \\
single\_uniadios\_winds\_light	&Unified &Yes	&500	&$10^{-6}$ \\
single\_uniadios\_massive &Unified	&No&	7500	&0.78 \\		
\bottomrule
\end{tabular}
\end{center}
\label{tab:singleruns}
\end{table*}
To validate our new accretion disc implementation, we perform idealised simulations of a single or binary SMBH embedded in a gaseous disc, representing the inner region of a galaxy (see Figure~\ref{fig:proj_all}).

These test cases allow us to assess the sensitivity of our accretion disc subgrid model to different parameter choices and modelling assumptions, such as disc wind loss or reduced solid body precession, without being limited by the computational expense of full galaxy formation simulations. Ultimately, our accretion model has been developed with galaxy simulations in mind, however, exploring the accretion disc subgrid model in these set-ups is beyond the scope of this methodology-focused paper.

The single BH embedded in a circumnuclear disc (CND) represents a baseline case so that we can assess the impact of different modelling choices in a relatively steady environment. The binary SMBHs embedded in the circumbinary disc (CBD) allow us to study the behaviour of our new accretion disc subgrid model for strongly fluctuating gas inflow patterns, especially for the misaligned binary set-up. Furthermore, the binary simulations demonstrate a crucial science case for these types of SMBH accretion models, in particular with regards to electromagnetic counterpart and gravitational recoil predictions.

In this Section, we outline the different physical configurations and model variations explored in this paper. The set-up of the gaseous disc is based on the initial conditions (ICs) presented in \citet{bourne_dynamics_2023} and for completeness we provide a brief summary in Section~\ref{subsec:cndb_setup}. We then outline how these ICs are evolved for the single SMBH set-up (see Section~\ref{subsec:single_setup}) and the binary SMBH set-up (see Section~\ref{subsec:binary_setup}), including the subgrid model variations explored in both cases.
\begin{table*}
\caption{Overview of binary simulations. We list the simulation name (first column), mass ratio $q$ (second column), inclination angle with respect to the plane of the circumbinary disc $i$ (third column), wind loss (fourth column), and whether the reduced precession rate for truncated disc from \citet{bollimpalli_effect_2023} is employed (fifth column).}
\begin{center}
\begin{tabular}{@{}lccccc@{}}
\toprule
\textbf{Simulation name} & \textbf{Mass ratio q} & \textbf{Inclination angle i [\degree]}  &  \textbf{Accretion disc model} & \textbf{Wind loss?} & \textbf{Reduced precession?}  \\  \toprule
bin\_q1i0\_uniadios & 1 & 0 & Unified & No & No	\\	
bin\_q1i0\_uniadios\_slowprec &	1&	0&	Unified	&No	&Yes\\	
bin\_q1i0\_uniadios\_winds & 1 &0	&Unified	&Yes&	No	\\							bin\_q1i0\_uniadios\_winds\_slowprec&	1	&0	&Unified	&Yes	&Yes	\\			bin\_q1i0\_thindisc	& 1	&0	&Thin disc only&	No&	N/A	\\		
bin\_q3i45\_uniadios	&3	&45	&Unified &No	&No		\\		bin\_q3i45\_uniadios\_slowprec&	3&	45	&Unified	&No	&Yes	\\
bin\_q3i45\_uniadios\_winds	&3	&45	&Unified &Yes	&No	\\			bin\_q3i45\_uniadios\_winds\_slowprec	&3	&45	&Unified &Yes	&Yes	\\	bin\_q3i45\_thindisc	&3	&45&	Thin disc only	&No&	N/A	\\				
\bottomrule
\end{tabular}
\end{center}
\label{tab:binaryruns}
\end{table*}
\subsection{Gaseous disc set-up} \label{subsec:cndb_setup}

The gaseous disc for both the single and binary set-ups extends from $R_{\rm in}= 4$~pc to $R_{\rm out} = 14$~pc and follows a surface density profile of 
\begin{equation}
\Sigma(R) = \Sigma_{0} \left(\frac{R}{R_{\rm in}} \right)^{-\alpha}\,,
\end{equation}
where $\alpha = 2$ and the normalisation $\Sigma_{0}$ is set so that the total mass of the gaseous disc constitutes 10 per cent of the single/binary SMBH mass. As discussed in \citet{bourne_dynamics_2023}, this set-up represents an idealised version of the inner region of a post-merger galaxy, which offers ideal conditions for disc formation via the circularisation of infalling gas clouds.

The vertical structure is set up so that the aspect ratio of the gaseous disc, $H/R$, is approximately constant and the disc is in a stable configuration (initial Toomre parameter set to $Q=1.5$).

We then evolve this system as an ideal gas with adiabatic index $\gamma = 5/3$ and allow the gas to cool following a $\beta$-cooling prescription, where the local cooling timescale $T_\mathrm{cool}$ is proportional to the orbital timescale by a factor $\beta = 10$. This ensures that the disc can settle into a marginally stable configuration \citep[see][for details]{bourne_dynamics_2023}. 

The target gas mass resolution is set to $m_\mathrm{target}=0.2~\mathrm{M_{\odot}}$. To increase the spatial resolution in the gaseous disc cavity, and allow for accurate measurements of the mass and angular momentum inflow rates onto the subgrid accretion disc, we apply the super-Lagrangian refinement technique \citep{curtis_resolving_2015} to the gas cells in the cavity (within $R_\mathrm{ref}=3$~pc). To this end, we define an additional (de-)refinement criterion so that the maximum allowed cell size $R^\mathrm{cell}_\mathrm{max}$ decreases linearly within the cavity from $R^\mathrm{cell}_\mathrm{max}=0.2$~pc at $R_\mathrm{ref}$ (typical size of the mesh cell just inside the cavity) to $R^\mathrm{cell}_\mathrm{min}=0.01$~pc at the centre (chosen to match the gravitational softening of the BH). The cell sizes are allowed to be within a factor of two of these target radii and, to avoid excessive refinement, we set a minimum cell mass of $m_\mathrm{min}=10^{-5}m_\mathrm{target}$. Note that the super-Lagrangian refinement technique is not merely advantageous for the subgrid model, but also allows us to resolve important features in the low-density cavity (which is intrinsically difficult for Lagrangian codes); in particular we can resolve the formation of streams and mini discs around the binary SMBHs, and follow accurately how they torque the binary.

\subsection{Single SMBH simulations} \label{subsec:single_setup}
\subsubsection{Initial conditions and relaxation}
For all single SMBH simulation set-ups, we place a SMBH at the centre of the CND, with mass $M_\mrm{\bullet} = 2 \times 10^6 \ \Msun$. Following a relaxation time of $\sim 20$~Myr, we then activate the super-Lagrangian refinement and further relax the system for $\sim 6$~Myr. Note that we keep the overall relaxation time deliberately as short as possible. This allows for the disappearance of transient features, whilst testing our subgrid model \textit{before} the cavity has been refilled with gas due to the absence of binary torques, as we are mainly interested in the low-accretion-rate regime (see also the counter-rotating binaries in \citet{bourne_dynamics_2023}). Appendix~\ref{appsec:SingleBH} summarizes the main features of the single SMBH relaxation procedure.

\subsubsection{Subgrid model variations}

For the subgrid model, we need to specify the initial masses and angular momenta of the accretion disc and SMBHs.

The SMBH spin depends on the system's previous evolution with different accretion flows and feedback physics leading to either spin-up or spin-down \citep[see][for recent compilations]{narayan_jets_2022,lowell_rapid_2024}. Following \citet{bourne_dynamics_2023}, we assume that both SMBHs are initially highly spinning with a spin value of $a=0.9$, consistent with current observational constraints on $\sim 10^{6}~\Msun$ SMBHs \citep{reynolds_observing_2019}. The orientation is chosen randomly, however, to be able to make direct comparisons between different physics runs, the initially randomly chosen orientation for the first set-up, i.e. $\theta =62 \degree$, is kept the same for all accretion disc physics variations. The initial angular momentum versor of the subgrid accretion disc is aligned with the angular momentum of the surrounding gas inflows, i.e. with the $z$-axis.

For the subgrid  accretion disc evolution, we test the original thin disc model (\textit{single\_thindisc}) as well as the unified model with and without wind loss  (\textit{single\_uniadios\_winds} and \textit{single\_uniadios}). For these tests, we choose the initial disc mass as $M_\mrm{d} = 10^{-3} M_\mathrm{\bullet}=2000 \ \Msun$ and $f_\mathrm{Edd,initial,TD} = 0.003$, which leads to good agreement between the external inflow rate and the Eddington fraction for the ADIOS model, see Figure~\ref{fig:lum_acc_singleBH}.

Furthermore, we also test two variations of the initial disc mass set-up, rescaling the initial Eddington ratio according to Equation~(\ref{eq:fEddThD}). In the first case, we employ a significantly higher disc mass, $M_\mathrm{d}=7500 \ \Msun$, and the unified model without wind loss (\textit{single\_uniadios\_massive}), and in the second case, we employ a significantly lower disc mass, $M_\mathrm{d}=500 \ \Msun$, and the unified model both without and with wind loss (\textit{single\_uniadios\_light} and \textit{single\_uniadios\_winds\_light}). All of these runs and their parameters are listed in Table~\ref{tab:singleruns}.

The different disc mass choices here represent different previous evolution histories of the system and allow us to test the behaviour of our model in a low-inflow environment.

\subsection{Binary set-up} \label{subsec:binary_setup}

\subsubsection{Initial conditions and relaxation} \label{subsubsec:ICs_CBD}

The total binary mass is set to $M_{\rm bin} = 2 \times 10^{6}~\Msun$ and the binary is initially placed on a Keplerian orbit with semi-major axis $a_\mathrm{orbit} = 2$~pc, resulting in an orbital period of $T_\mathrm{bin}=0.187$~Myr. 

For the binary orbits, we consider two configurations. Firstly, we simulate a simple symmetric set-up with an equal mass ratio, $q=1$, zero eccentricity and with the binary orbit aligned with the plane of the gaseous disc (labelled as \textit{bin\_q1i0}). Secondly, we simulate binaries with a mass ratio $q=3$, zero eccentricity and with the binary orbit inclined by $i=45 \degree$ with respect to the CBD (labelled as \textit{bin\_q3i45}). These set-ups are equivalent to \textit{q01e00i00} and \textit{q03e00i45mod} in \citet{bourne_dynamics_2023}, respectively.

The initial conditions are evolved for 50 binary orbits corresponding to about $1.5T_\mathrm{cool}$ at the outer edge of the CBD without any additional refinement and without any accretion onto the SMBH -- accretion disc particle. Following the cooling relaxation period, the set-up is further relaxed with the super-Lagrangian refinement for another 50 binary orbits, so that the total relaxation period amounts to approximately three cooling timescales, allowing the CBD to reach a nearly steady state \citep[see][]{bourne_dynamics_2023}. The resulting snapshot is used as initial conditions to be further evolved for 500 binary orbits, testing the unified accretion disc model (varying the assumptions about wind loss and LT precession) as well as reference runs with the thin disc model from \citet{fiacconi_galactic_2018}, as described in Section \ref{subsubsec:model_var_bin}.

\subsubsection{Subgrid model variations} \label{subsubsec:model_var_bin}
For the subgrid modelling, we again need to specify the initial masses and angular momenta of the accretion disc and SMBHs. Following the single BH set-up and \citet{bourne_dynamics_2023}, we assume that both SMBHs are initially highly spinning with a spin value of $a=0.9$. The orientation is chosen randomly, however, to be able to make direct comparisons between different physics runs, the initially randomly chosen orientation for the first set-up, i.e. $\theta_\mathrm{primary}=31 \degree$ and $\theta_\mathrm{secondary}=62 \degree$, is kept the same for all accretion disc physics variations. Note, however, that this initial random orientation is different from the initial SMBH versor orientations in \citet{bourne_dynamics_2023}, so that our thin disc model runs display some small differences compared to the \citet{bourne_dynamics_2023} set-ups.

For the subgrid accretion disc, the initial mass is set to $M_\mrm{d} = 10^{-3} M_\mathrm{\bullet}$. We align the initial angular momentum versor with the angular momentum of the surrounding mini disc and we choose the ratio between the magnitudes of the disc and SMBH angular momenta such that the initial Eddington ratio, as set by Equation~(\ref{eq:fEddThD}), approximately matches the initial inflow rate from the mini disc (with $f_\mrm{Edd,initial1,2} = 10^{-3}$ for \textit{bin\_q1i0} and  $f_\mrm{Edd,initial1} = 2 \times 10^{-3}$ and $f_\mrm{Edd,initial2} = 6 \times 10^{-3}$ for \textit{bin\_q3i45}).

For both binary configurations, we run five different simulations. Firstly, we repeat the thin-disc only set-up from \citet{bourne_dynamics_2023}, though as noted above, we have a different initial orientation of the SMBH versors. Secondly, we rerun both binary configurations using the ADIOS model without wind loss (\textit{uniadios}) with efficient solid body precession for truncated discs following \citet{ingram_low-frequency_2009} as well as reduced solid body precession (\textit{slowprec}) following \citet{bollimpalli_effect_2023}. Finally, we repeat the same two precession cases, where we also account for wind loss in the hot flow (\textit{uniadios\_winds}). All of these simulations are listed in Table~\ref{tab:binaryruns}.
 
\section{Results} \label{sec:results}

The unified accretion disc model determines the evolution of the SMBH accretion rates, luminosities and spins across different radiative regimes. In this section, we analyse all of these aspects in detail for both the single SMBH set-ups, which allow us to test the impact of our subgrid model choices in a simpler setting without the binary torques, as well as the binary SMBHs where we obtain more complex dependencies between the variations in the resolved inflows and the response of the subgrid model, in particular for the spin alignment.

Figure~\ref{fig:proj_all} shows gas surface density maps of our three main simulation set-ups: the single SMBH simulation (\textit{single}, left panel), the equal-mass, aligned binaries (\textit{bin\_q1i0}, middle panel) and the unequal-mass, misaligned binaries (\textit{bin\_q3i45}, right panel). We show the single BH simulation at $t=50$~Myr to demonstrate how the cavity is filled with relatively low-density gas in the single SMBH simulation. For the binary simulations, on the other hand, streams and mini discs form in the cavity that feed gas and angular momentum to the SMBHs, here shown at the beginning of the simulation before the CBD is torqued by the misaligned binary \citep[also see][]{bourne_dynamics_2023}.

First, we consider the disc state transitions throughout the simulations in Section~\ref{subsec:results_states} and then go on to analyse the gas accretion and luminosities in Section~\ref{subsec:results_accrate_lum}. Finally, we investigate the SMBH spin and disc angular momentum evolution due to gas inflows and Lense-Thirring precession in Section~\ref{subsec:results_angmom_evolv}.

\subsection{Disc state transitions} \label{subsec:results_states}

\begin{figure*}
    \centering
    \includegraphics[width=\textwidth]{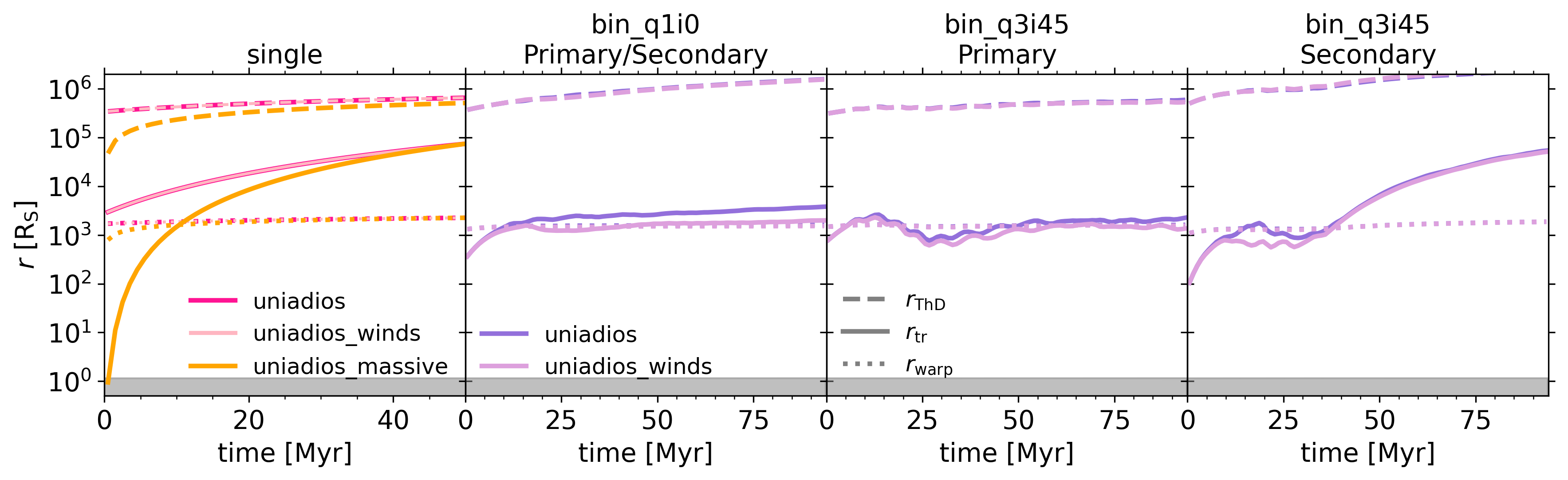} 
    \caption{Time evolution of the characteristic accretion disc radii, including the thin disc radius (dashed lines), truncation radius (solid lines) and warp radius (dotted lines). The grey-shaded region indicates the ISCO. The accretion disc models employed are given by the colour-coding as listed in the figure legend. See Tables~\ref{tab:binaryruns}~and~\ref{tab:singleruns} for details on the binary and single SMBH simulation runs, respectively. Overall, we find that the vast majority of our simulation set-ups are in the truncated disc state, with the Lense-Thirring regime switching between Bardeen-Petterson alignment and solid body precession as the truncation radius falls below and above the warp radius. Note that the models with and without wind loss share the same mass flow rate through the disc, assuming a constant SMBH mass and spin. Since there are negligible changes in SMBH properties in the single SMBH simulations, both models have nearly identical disc mass flow rates and, consequently, similar disc radii.}
    \label{fig:radii_allsims}
\end{figure*}

We begin our analysis of the simulation suite by considering the different characteristic disc radii which determine the disc state transitions. Figure~\ref{fig:radii_allsims} shows the thin disc radius $r_\mathrm{ThD}$ (dashed lines), the truncation radius $r_\mathrm{tr}$ (solid lines) and the warp radius $r_\mathrm{warp}$ (dotted lines) as a function of time. The grey-shaded region indicates the area inside the ISCO. The panel titles indicate the initial conditions and the colour-coding of the lines indicates the subgrid model employed as listed in the legend.

First, we consider the \textit{single} SMBH set-up. With the standard initial disc mass ($M_\mathrm{d} = 10^{-3} M_{\bullet} = 2000 \ \Msun$), the disc is in the truncated disc state for the whole simulation ($r_\mathrm{ISCO} < r_\mathrm{tr} < r_\mathrm{ThD}$), with the truncation radius larger than the warp radius. Hence the outer truncated thin disc is fully misaligned and feeds misaligned material to the inner hot flow leading to solid body precession (see Section~\ref{subsubsec:results_single_versors} for details on the precession behaviour in the single SMBH case). Note that the models with and without wind loss have the same mass flow rate through the disc for a given SMBH mass and spin (as the wind loss is at the expense of the SMBH growth and is assumed to occur at the ISCO). Given the insignificant changes in SMBH properties for the simulations considered here, these models hence have the same disc mass flow rates and therefore also virtually identical disc radii.

With the massive disc initialisation (\textit{uniadios\_massive}, $M_\mathrm{d} = 7500 \ \Msun$), however, the disc spends a short initial period of approximately 1~Myr in the thin disc state with the truncation radius inside the ISCO. Afterwards, the disc is rapidly depleted and changes to a truncated disc with steadily increasing truncation radius. At $t \sim 10$~Myr, the truncation radius starts exceeding the warp radius, so that the torque regime changes from Bardeen-Petterson alignment to solid body precession.\footnote{Note that we also performed a simulation with a light disc of initial disc mass $M_\mathrm{d} = 500 \ \Msun$, which is firmly in the pure ADAF/ADIOS regime (not shown here), with $r_\mathrm{tr}/r_\mathrm{ThD} \sim 10^{4}$ and does not significantly change its disc properties due to the low initial $f_\mathrm{Edd}=10^{-6}$, which means that both the disc mass and SMBH mass stay approximately constant throughout the simulation.}

Next, we consider the disc radii for the equal-mass, aligned binary (\textit{bin\_q1i0}, second panel). As the primary and secondary SMBH findings closely follow one another, we just show the disc radii evolution of the secondary here and comment on any (small) differences in the text. As with the standard single SMBH configuration, this simulation is in the truncated disc state for the whole duration of the simulation with $r_\mathrm{ISCO} < r_\mathrm{tr} < r_\mathrm{ThD}$. Initially the truncation radius is smaller than the warp radius, so that the system is in the Bardeen-Petterson configuration. As the truncation radius increases the \textit{uniadios} set-up (purple line) enters the solid body precession regime for the remainder of the simulation. The \textit{uniadios\_winds} set-up (pink line), however, briefly dips back into the Bardeen-Petterson regime as the truncation radius decreases due to the slower SMBH mass growth rate (which increases $f_\mathrm{Edd,ThD}$, see Equation~(\ref{eq:fEddThD})). Note that this temporary decrease does not occur for the primary SMBH due to small stochastic differences in their evolution.

For the unequal-mass ratio, misaligned binaries, we show the primary and secondary SMBH in separate panels. Due to the asymmetry of the set-up, there are significant quantitative differences in the disc radii evolution, although the qualitative behaviour is very similar. Both SMBHs are in the truncated disc regime with $r_\mathrm{ISCO} < r_\mathrm{tr} < r_\mathrm{ThD}$ for the whole simulation. For the primary SMBH, the system is initially in the Bardeen-Petterson configuration with $r_\mathrm{tr} < r_\mathrm{warp}$ and then switches from the solid-body precession regime back to Bardeen-Petterson and finally enters the solid-body regime again. This applies to both the \textit{uniadios} and \textit{uniadios\_winds} model, although the \textit{uniadios\_winds} model spends longer in the Bardeen-Petterson regime due to the smaller truncation radius. Similarly, the secondary SMBH switches between the different precession and alignment regimes, however with the \textit{uniadios\_winds} model, there is only one solid body precession phase at late times, with the early precession phase being suppressed by the slower growth in $r_\mathrm{tr}$. Note that the truncation radius of the secondary increases significantly towards late times (over one magnitude higher than the primary's truncation radius) as with our binary set-up, we have preferential accretion onto the primary \citep[see discussion in][]{bourne_dynamics_2023}.

Finally, we note that for the (primary) BHs in the binary simulations, the truncation radius closely tracks the warp radius. This is because our binary SMBHs are in the Eddington fraction regime of $f_\mrm{Edd} \gtrsim 10^{-3}$, which corresponds to the accretion regime where these two radii intersect for highly-spinning BHs (see Figure~\ref{fig:radii_adios}). The primary BHs closely maintain this accretion rate throughout the simulation as the external inflow rates are similar to the mass flow rate through the accretion disc so that the latter does not change its mass significantly. Whilst the single BH simulations also initially have $r_\mrm{tr} \sim r_\mrm{warp}$, these two radii rapidly diverge as the accretion rate decreases due to the significantly lower external inflows, with $r_\mrm{warp}$ having a much shallower dependency on $f_\mrm{Edd}$ than $r_\mrm{tr}$.

Having established the different disc regimes that govern the mass and angular momentum evolution of the SMBHs, we move on to a more detailed analysis of the gas accretion rates and luminosities in Section~\ref{subsec:results_accrate_lum} and of the disc angular momentum and SMBH spin evolution in Section~\ref{subsec:results_angmom_evolv}.

\subsection{Gas accretion rates and luminosities}
\label{subsec:results_accrate_lum}

\subsubsection{Single SMBH simulations} \label{subsubsec:results_accrate_lum_single}

\begin{figure}
    \centering
    \includegraphics[width=\columnwidth]{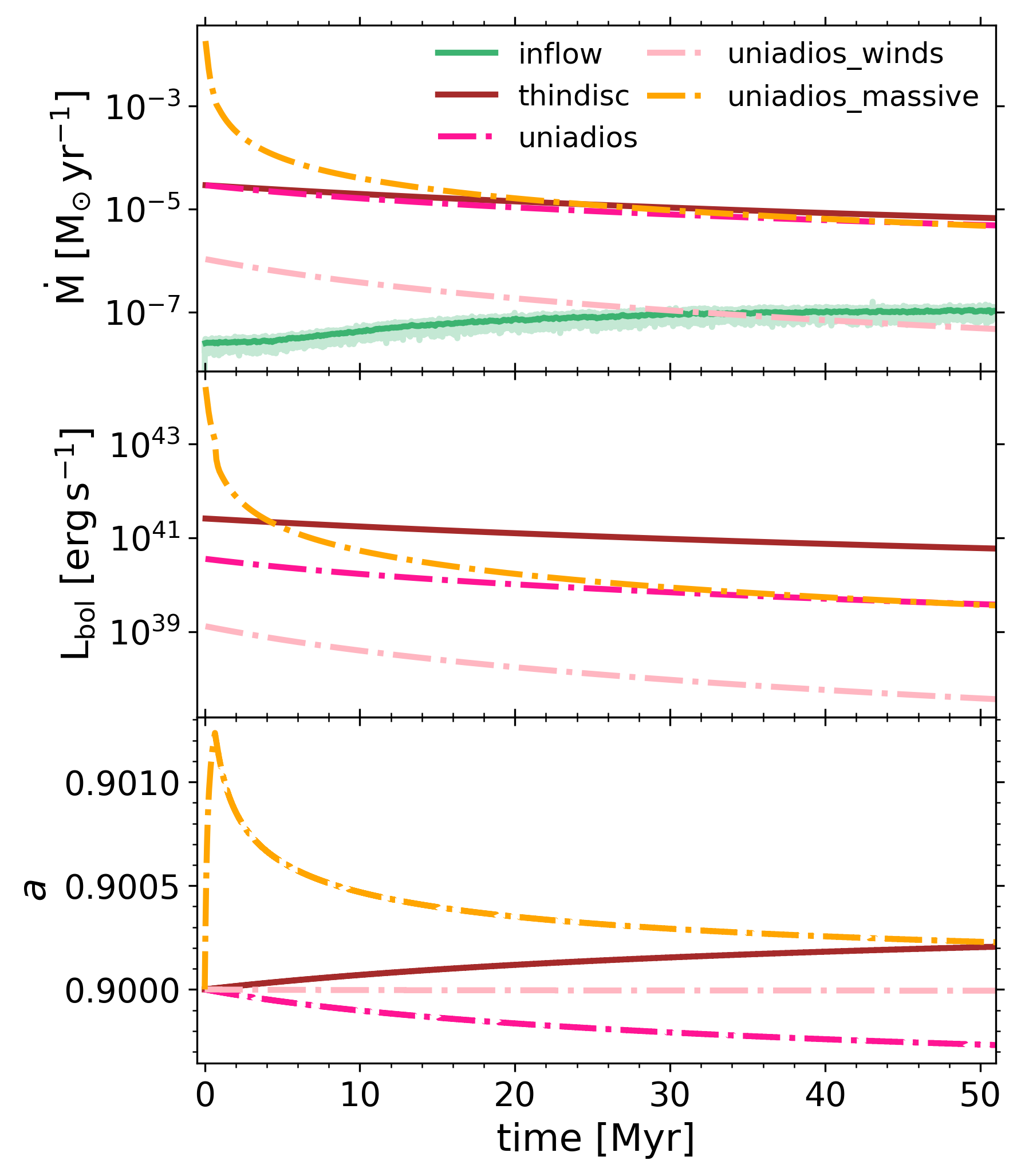} 
    \caption{Time evolution of the  gas mass accretion rate at ISCO $\dot{M}$ (\textit{top panel}), bolometric luminosity $L_\mathrm{bol}$ (\textit{middle panel}) and SMBH spin $a$ (\textit{bottom panel}). For comparison we also show the mass flow rate onto the subgrid accretion disc in light green (instantaneous) and dark green (binned). The unified model without wind loss predicts very similar mass accretion rates to the thin disc model, whilst the ADIOS mass accretion rates are significantly lower when accounting for wind loss. For the luminosities, both unified set-ups are significantly suppressed compared to the thin disc set-up due to the low radiative efficiencies in the truncated disc state. The SMBH spin only varies moderately in the simulations due to the low mass inflow rates, however, as encoded in the model, we can see how the thin disc model spins up the SMBH, whilst the ADIOS set-up leads to spin-down.}
    \label{fig:lum_acc_singleBH}
\end{figure}

We begin our analysis of the gas mass accretion rate by considering the single SMBH simulations. Figure~\ref{fig:lum_acc_singleBH} shows the time evolution of the gas mass accretion rate at the ISCO (top panel), bolometric luminosity (middle panel) and SMBH spin (bottom panel). The colour-coding indicates the subgrid model employed, as listed in the legend. For comparison, we also show the gas mass inflow rates onto the accretion disc in light green as well as the mean inflow rate in bins of 0.1~Myr in dark green.

We note that due to the low cooling rate, the cavity only fills in relatively slowly (also see Appendix~\ref{appsec:SingleBH}), so the initial disc mass plays an important role, leading to significant differences between the different disc initialisations (see Table~\ref{tab:singleruns}).

We first consider the reference thin-disc only run (solid brown line). Here the mass accretion rate and luminosity are relatively steady at $\sim 10^{-5} \ \Msun \, \mathrm{yr}^{-1}$ and $\sim 10^{41} \ \mathrm{erg \, s^{-1}}$, albeit slightly declining as the disc mass is depleted at a much faster rate than it is being replenished by the inflows, i.e. $ \dot{M}_\mathrm{inflow} \lesssim 10^{-7} \ \Msun \, \mathrm{yr}^{-1}$. The SMBH spin $a$ is very slowly increasing as the equilibrium spin for the co-rotating thin disc tends towards maximally spinning solution, i.e. $a_\mathrm{eq} \sim 1$.

With the unified accretion disc model, the mass accretion rate is virtually identical for the \textit{uniadios} set-up (dash-dotted dark pink line) as the disc is in the truncated state (see Section~\ref{subsec:results_states}) so that the mass flow rate is set by the outer thin disc. Note that this modelling choice results in the mass flow rates through the inner hot flow being significantly smoothed out. In principle, this boundary condition only needs to be true in a time averaged sense, and due to the significantly lower viscous timescale in the ADAF regime, the inner hot flow may display significant variability, as seen in GRMHD simulations \citep[e.g.][]{lalakos_jets_2023}. Furthermore, the thin disc may display significantly more variability than is captured in our prescription, for example due to opacity changes \citep[e.g.][]{jiang_opacity-driven_2020}. This has important implications for AGN bursts and multimessenger signatures, as discussed in more detail in Section~\ref{subsubsec:results_accrate_lum_binary}. 

The \textit{uniadios\_winds} model (dash-dotted light pink line) also takes into account wind loss in the inner hot flow, so that the accretion rate at ISCO is significantly suppressed. Both the \textit{uniadios} and the \textit{uniadios\_winds} model have suppressed luminosities compared to the \textit{thindisc} case due to the much lower radiative efficiency as well as wind loss for the ADIOS model.

Considering the SMBH spin evolution, we find that the truncated accretion disc model behaves in the opposite way from the \textit{thindisc} case, with the spin steadily decreasing. Again this decrease in the spin amplitude is extremely slow due to the low accretion rate (especially for the ADIOS case).

With the massive disc initialisation (\textit{uniadios\_massive}, dash-dotted yellow line), the disc is in the thin-disc state for $\sim 1$~Myr and then switches back to the truncated disc mode as the accretion rate rapidly decreases (also see Section~\ref{subsec:results_states}). This also leads to a very bright initial luminosity peak of $L_\mathrm{bol} \sim 10^{44} \ \mathrm{erg \, s^{-1}}$. The spin rapidly increases during the thin disc phase but then quickly decays again in the truncated disc phase.

Finally, we also tested a light disc set-up (not shown here), which is characterised by extremely low accretion rates and luminosities ($\dot{M} \sim 10^{-11} \ \Msun \, \mathrm{yr^{-1}}$ and $L_\mrm{bol} \sim 10^{33} \ \mrm{erg \, s^{-1}}$) so that, combined with the low inflow rates, there is virtually no change in the disc and SMBH properties.

\subsubsection{Binary SMBH simulations} \label{subsubsec:results_accrate_lum_binary}

\begin{figure*}
    \centering
    \includegraphics[width=\textwidth]{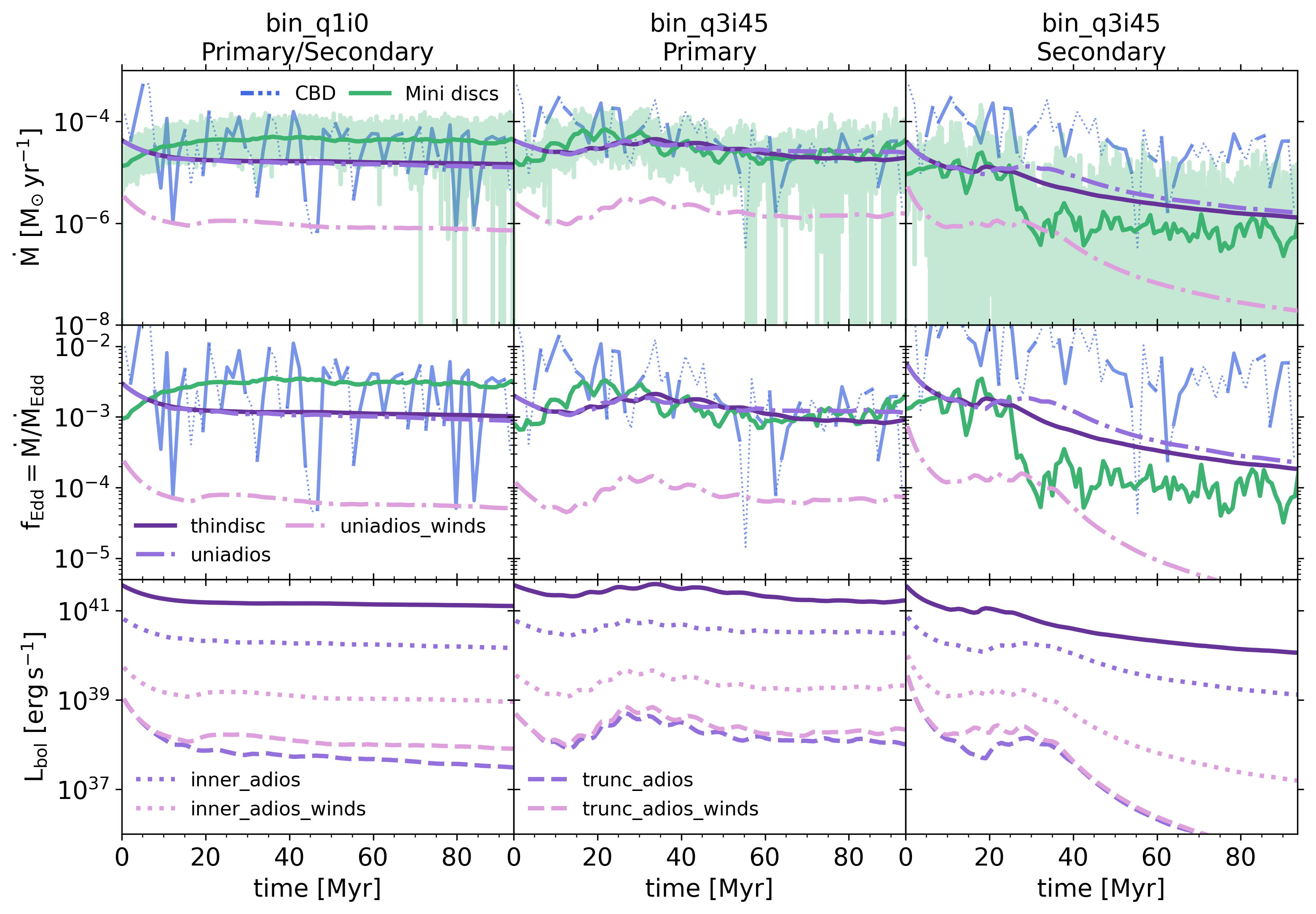} 
    \caption{Mass inflow rates (\textit{top row}), Eddington fractions (\textit{middle row}) and bolometric luminosities (\textit{bottom row}) for the binary simulations, with the subgrid model employed indicated by the colour-coding and line-styles as listed in the legend. In the top row, we also show the binned mass flow rates through the cavity, where solid blue lines indicate inflows and dashed blue lines indicate outflows. Furthermore, we plot the mass flow rates onto the accretion disc in light green (instantaneous) and dark green (binned). In the second row, we only include the binned inflow rates from the mini discs for clarity. In the third row, we split the bolometric luminosities into the contributions from the inner hot flow (dotted line) and outer truncated accretion disc (dashed lines) for the simulations performed with the unified accretion disc model. Overall, the ADIOS model substantially suppresses the mass accretion rates when accounting for wind loss, and the luminosities are significantly suppressed for all of the unified set-ups due to the low radiative efficiency in the truncated disc state.}
    \label{fig:inflow_binaries}
\end{figure*}

Having established the basic impact of the subgrid accretion disc models on the gas accretion rates and luminosities with the single SMBH simulations, we turn to the more complex and astrophysically interesting binary simulations. Figure~\ref{fig:inflow_binaries} shows the gas mass flow rates (top row), corresponding Eddington fractions (middle row), bolometric luminosities (bottom row). In the first column, we show the results from the equal-mass, aligned binary case, \textit{bin\_q1i0}, again just plotting the rates of the secondary SMBH which very closely follows the evolution of the primary SMBH due to the symmetric set-up. For the misaligned case, we separately plot the primary SMBH in the second column and the secondary SMBH in the third column.

We first focus on the gas mass flow rates in the top row. The blue lines indicate the mass flow rates from the CBD through the cavity calculated at $r=2a_\mathrm{orbit}=4$~pc, with the solid blue lines indicating net inflows for a given timebin and the dotted blue line showing net outflows, both averaged over $\sim 5$ binary orbits. We correct these CBD rates by a factor of $0.5$, to account for the average flow rate towards the primary/secondary SMBH (though note that preferential accretion means that for the unequal-mass binaries, the SMBHs will not receive an equal share of the CBD inflows). The CBD mass flow rates fluctuate between inflows and outflows, however, the net mass flow rate over the whole simulation time is inflowing. Furthermore, we show the gas mass flow rates from the mini discs onto the accretion disc as a light green shaded region with the mean inflow rate, in bins of $\sim 5$ binary orbits, plotted in dark green. We also show the mass flow rates through the subgrid disc for the different accretion disc models explored, including the thin-disc only reference run (\textit{thindisc}, solid dark purple lines) as well as the unified accretion disc model without wind loss (\textit{uniadios}, dash-dotted light purple lines) and with wind loss (\textit{uniadios\_winds}, dash-dotted pink lines). 

The mean feeding rate from the CBD and the gas mass flow rates through the mini disc onto the accretion disc are similar overall, in particular for the aligned set-up and the primary SMBH in the misaligned set-up. For the secondary SMBH, the CBD feeding rate somewhat exceeds the mini disc rate. The latter significantly decreases at $t \sim 30$~Myr and we have preferential accretion onto the primary BH \citep[also see][]{bourne_dynamics_2023}.

As noted in Section~\ref{subsubsec:results_accrate_lum_single}, the subgrid accretion disc model smooths out the inflow rates due to the relatively long viscous timescale of the thin disc, which also sets the accretion disc timescale in the truncated disc state \citep[see][]{hogg_dynamics_2018}. In principle, the inner hot flow could experience significant variability due to disc turbulence and much shorter viscous timescale. Capturing this variability is beyond the scope of this work, however, we note that there may be intermittent accretion bursts that our model does not capture.

Comparing the mass flow rates between the different subgrid models, we find similarly to the single SMBH case that the \textit{thindisc} and \textit{uniadios} models yield very similar mass flow rates. The slight positive offset of the \textit{uniadios} rates is due to the extremely low radiative efficiency in the ADIOS regime so that the disc mass is being drained more slowly leading to somewhat higher accretion rates at late times. However, this is vastly outweighed by the wind loss in the \textit{uniadios\_winds} set-up, which has much lower accretion rates. Note that this represents somewhat of a lower limit as we do not have feedback in our simulations so the lost wind mass is not re-injected. If we included this effect, we may have a fountain effect resupplying mass to the discs, especially in the inner region \citep[e.g.][]{sadowski_energy_2013}, though the winds may also act to reduce the mini disc mass flow rate.

The Eddington fractions of the inflow rates from the mini discs are generally $\lesssim 10^{-3}$, and with our set-ups approximately matching these initial inflow rates, the mini discs therefore set the truncated disc state. 

For the luminosities, we separately plot the contributions from the outer truncated thin disc (dashed lines) and inner hot flow (dotted lines). This demonstrates that the inner hot flow significantly dominates the disc luminosity in the truncated state, which means that the spectrum of the binary would also be shifted towards harder X-rays with the truncated disc model \citep{giustini_global_2019}. Energy extraction in the outer truncated thin disc is extremely inefficient for intermediate Eddington ratios as the gravitational potential energy released up to the truncation radius is only minimal, with most of the energy release in the full thin disc mode occurring at ISCO. Indeed even the inner hot flow in the \textit{uniadios} model is one order of magnitude fainter than the \textit{thindisc} model. For the \textit{uniadios\_winds} model, the low radiative efficiency and wind loss combined lead to a dramatic decrease in the inner hot flow luminosity.

\begin{figure*}
    \centering
    \includegraphics[width=\textwidth]{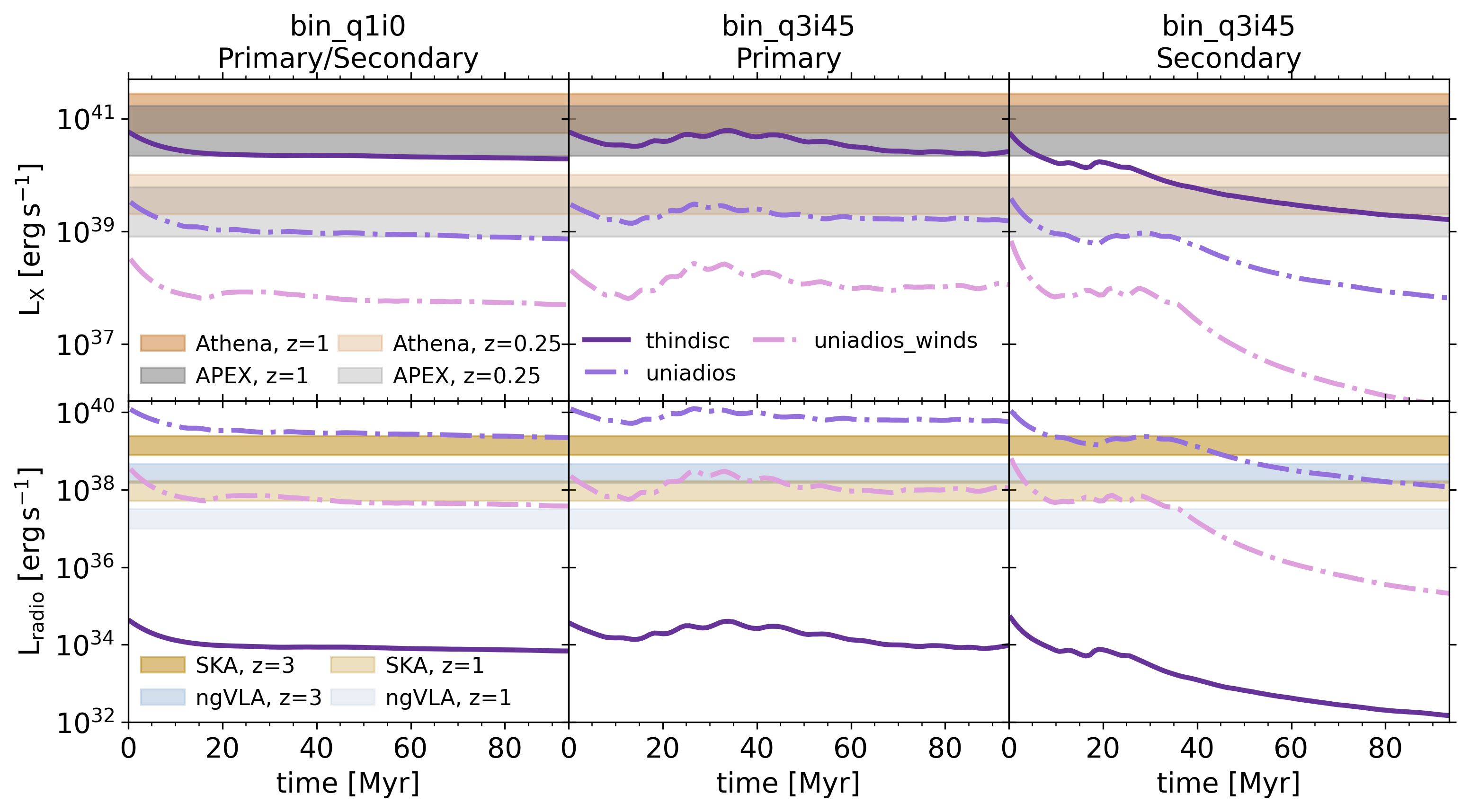} 
    \caption{X-ray luminosities (\textit{top row}) and radio luminosities (\textit{bottom row}) for the binary simulations, with the subgrid model employed indicated by the colour-coding and line-styles as listed in the legend. In the top row, we also plot the X-ray flux limits from \textit{NewAthena} and the proposed NASA X-ray probes for the Astrophysics Probe Explorer (\textit{APEX}) programme as brown- and grey-shaded regions, respectively, for $z=1$ and $z=0.25$. In the bottom row, we indicate typical GHz flux limits (where most of the jet emission is expected to occur) for SKA and ngVLA as gold- and blue-shade regions, respectively, for $z=3$ and $z=1$. The lower and upper limits correspond to proposed deep and wide surveys, respectively. The low radiative efficiency and wind loss in the ADIOS model lead to significantly lower X-ray luminosities than the thin disc predictions. For the radio regime, however, the situation is reversed as thick discs are associated with more efficient radio jet production.}
    \label{fig:inflow_binaries_luminosities}
\end{figure*}

This has important implications for multi-messenger signatures, and in particular electromagnetic counterpart predictions. To investigate this further, we show the X-ray and radio luminosities of the binaries in Figure~\ref{fig:inflow_binaries_luminosities}. 

To obtain the X-ray luminosities for the thin discs, we assume the bolometric correction from \citet{shen_bolometric_2020}, appropriate for a quasar template. For the ADIOS X-ray luminosities, we use the Eddington-ratio dependent bolometric correction from \citet{vasudevan_piecing_2007}. For comparison, we also show the detectable luminosities for upcoming X-ray missions, with \textit{NewAthena} plotted as brown-shaded regions and the proposed NASA APEX missions as grey-shaded regions. In both cases, the light-shaded regions indicate the detectable luminosities at $z=0.25$, whilst the dark-shaded regions indicate detectable luminosities at $z=1.0$. The \textit{NewAthena} lower luminosity limit is based on the flux limit for a deep survey with the maximum proposed angular resolution ($5''$) \footnote{These are taken from a recent \textit{NewAthena} status update presentation which can be found at \url{https://api.cloud.ifca.es:8080/swift/v1/ACO/Presentations/20230616NewAthena_XRU.pdf}.}. Similarly, as a representative example of the NASA X-ray Probe-class missions that are currently under review, we plot the flux limits for the proposed AXIS deep and wide surveys \citep{reynolds_overview_2023}. Though note that (in addition to infrared missions), there are several other proposed X-ray probes for the APEX program that will target dual SMBH and close pairs, e.g. HEX-P which would have the ability to spatially resolve dual AGN targeting the hard X-rays \citep{civano_high_2024,pfeifle_high_2024} and LEM which could spectrally resolve dual AGN at high redshifts \citep{pfeifle_spectrally_2023}.

This demonstrates the crucial need for accurate accretion disc modelling as the different disc models make significantly different predictions for the detectability of the binaries. In particular, at $z=1$, even with the state-of-the-art NASA probe concepts, the binaries would not be detectable according to our unified accretion disc model -- regardless of whether wind loss is included. For nearby galaxies within $z=0.25$, which may harbour a significant SMBH binary population as revealed by the population modelling based on gravitational wave background constraints from nanoGRAV \citep[see][though note that strongest constraints are for more massive SMBHs than the ones considered here]{agazie_nanograv_2023-1}, X-ray detection prospects are more favourable, although the proposed probes may struggle to detect these objects if wind loss is significant. With the thin disc model, the predicted X-ray luminosities are much higher and may be detectable out to $z=1$, with the proposed NASA X-ray Probe-class missions. 

Radio interferometers will play a particularly important role in constraining wide binaries, such as the ones simulated in this work, as very long baseline interferometry (VLBI) allows to directly resolve these systems, whilst future X-ray missions will still be limited to kpc scales \citep{de_rosa_quest_2019}. Current radio interferometers have already identified parsec scale binaries at $z \sim 0.1$ and future facilities will push this to much higher redshifts with SKA resolution sufficient to at least $z \sim 1$ and the ngVLA resolving wide SMBH binaries out to redshifts $z \gtrsim 3$ \citep{burke-spolaor_next-generation_2018,bogdanovic_electromagnetic_2022}.

To this end, we plot radio luminosity predictions for the different disc states in the bottom row of Figure~\ref{fig:inflow_binaries_luminosities}. Both analytical theory \citep{meier_association_2001} and GRMHD simulations \citep{tchekhovskoy_black_2010} predict that geometrically thick accretion discs should be more efficient at jet production and may even explain the radio loud/quiet AGN dichotomy. The spin-driven jet model from \citet{meier_association_2001} is also applied in \citet{barausse_evolution_2012} and more recently in \citet{mangiagli_massive_2022}, and we follow their disc-state-dependent formalism here for our binary electromagnetic counterpart predictions. We then convert the jet luminosities into radio luminosities using the scaling relationship between jet power and synchrotron luminosity from \citet{cavagnolo_relationship_2010}. Following \citet{mangiagli_massive_2022}, we also take into account beaming effects assuming a typical Lorentz factor of $\Gamma=10$ for AGN.

For comparison, we also again indicate detectable luminosities for future radio facilities based on typical sensitivities in the GHz band (where most of the jet emission is expected to occur) for SKA \citep[$\sim 200$--$2000$~nJy, see][]{braun_anticipated_2019} and the ngVLA \citep[$\sim 40$--$400$~nJy, see][]{plotkin_science_2018}, as collated by \citet{latif_radio_2024}. As a note of comparison, the upper limit for detectable luminosities with SKA corresponds to the lower limit for LOFAR.

Based on current SMBH spin constraints \citep{reynolds_observing_2019}, we have initialised our SMBHs as highly spinning ($a=0.9$) which favours strong jet production \citep{blandford_electromagnetic_1977,blandford_hydromagnetic_1982}. In the ADIOS regime, the jets are further boosted due to the smaller solid angle subtended by the base of the jet, increasing the jet power \citep{tchekhovskoy_black_2010}.

Hence we obtain very high radio jet luminosities in the ADIOS regime, even when considering wind loss, which exceed the thin disc predictions by several orders of magnitude. This has important implications for VLBI SMBH binary searches, which will mainly be limited by angular resolution for highly-spinning SMBHs in the truncated or ADIOS state. For thin discs, however, these searches will also be fundamentally limited by sensitivity. In our case, whilst the ngVLA may be expected to resolve a parsec-scale SMBH binary out to $z\sim3$, with the thin disc model we would not expect to detect the jet luminosities of this system. 

One important caveat is the role of magnetic fields in jet production. Indeed recent GRMHD simulations have demonstrated that thin discs can have much more efficient jets if the inner region of the disc is magnetically saturated (`MAD state') which may provide a potential explanation for radio-loud quasars \citep{liska_bardeen-petterson_2019,ricarte_recipes_2023}. Incorporating these effects is beyond the scope of this work though we note that a stronger magnetic field would increase the spin-driven jet power for both disc types following the Blandford-Znajek mechanism \citep{blandford_electromagnetic_1977}, so that the qualitative statement of stronger jets in the ADIOS regime would likely still hold.

As a final caveat, we note that these strong jets would spin down the SMBH, decreasing the jet power. As we are only estimating the jet power in post-processing we do not include this effect here, however, in the future we plan to couple our unified accretion disc model to a resolved jet feedback implementation \citep[also see][]{talbot_blandford-znajek_2021,talbot_blandford-znajek_2022,talbot_simulations_2024}
 and include the jet spin down self-consistently.

Overall, we emphasise that the thin disc model and truncated disc model for the same inflow rate predict markedly different counterparts. For the former, we obtain relatively bright X-ray AGN but weak radio jets, whilst the latter produces strong radio jets and weak X-ray emission. Hence it is of critical importance to incorporate as realistic AGN accretion disc models as possible when making electromagnetic counterpart predictions for SMBH binaries.

\subsection{Evolution of SMBH spin and accretion disc angular momentum}
\label{subsec:results_angmom_evolv}
Here, we explore the evolution of the SMBH and accretion disc angular momentum, which is governed by the external inflows from the mini discs, the accretion of gas angular momentum from the accretion disc onto the SMBH and the mutual Lense-Thirring torques. The latter two are set by the assumed subgrid model for the accretion disc. In particular the Lense-Thirring torques, which arise from a misalignment between the SMBH and disc angular momentum versors, operate in two very different physically regimes depending on the disc model assumed. In the thin disc case, these differential torques are communicated via viscosity, leading to the so-called Bardeen-Petterson configuration, where the inner part of the thin disc is aligned with the SMBH versor up to a warp radius $r_\mathrm{warp}$. In the ADIOS regime, the whole flow is strongly coupled via pressure waves and precesses as a solid body, only aligning very slowly with the total angular momentum versor. See Section~\ref{subsubsec:AngMomLT} for a detailed description of the subgrid treatment of Lense-Thirring precession in the different disc states.

\subsubsection{Single SMBH simulations: angular momentum versor evolution} \label{subsubsec:results_single_versors}

\begin{figure}
    \centering
    \includegraphics[width=\columnwidth]{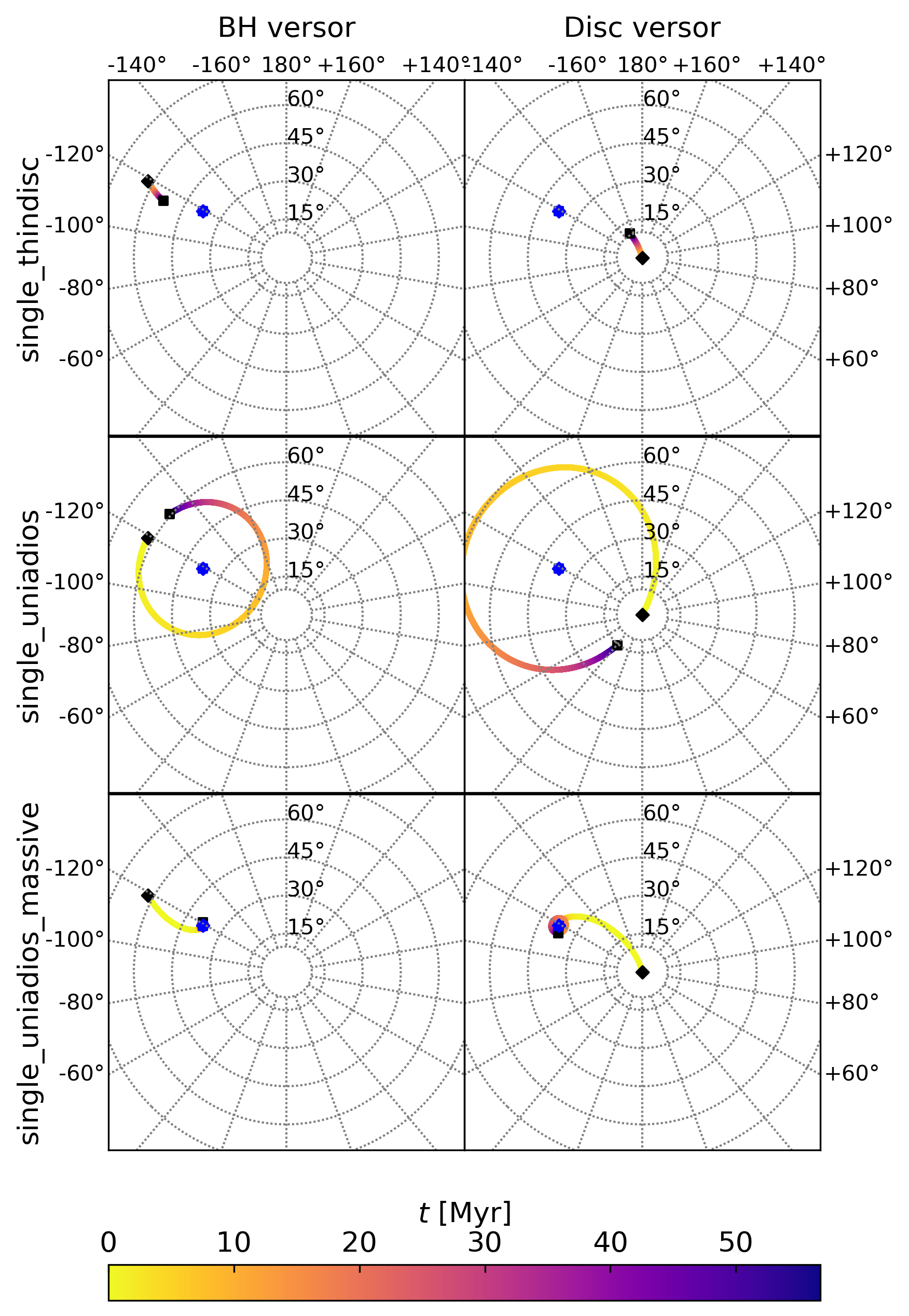}
    \caption{Polar projections of the evolutionary tracks of the SMBH (\textit{left panel}) and accretion disc (\textit{right panel}) angular momentum versors (unit vectors co-directional with angular momentum vector) for the single SMBH simulations, where the colour coding indicates the simulation time. The black diamond marker indicates the initial orientation, whilst the black square marker shows the final versor orientation. We also plot the versor position of the total angular momentum versor as a blue dot. For the standard thin disc case (\textit{single\_thindisc}, top row) we obtain slow Bardeen-Petterson alignment, whilst for the standard unified accretion disc model (\textit{single\_uniadios}, middle row), the system is in the solid-body precession mode covering a significant solid angle. With the massive disc initialisation (\textit{single\_uniadios\_massive}, bottom row), the system aligns rapidly with the total angular momentum vector and, as it switches to the truncated disc mode, goes through small-scale solid body precession.}
    \label{fig:versor_evolv_single}
\end{figure}

We first consider the angular momentum evolution in the single SMBH simulations. Here the spin alignment is mainly driven by the Lense-Thirring torques as both the SMBH accretion rates and the external gas inflow rates are negligible (see Section~\ref{subsubsec:results_accrate_lum_single}). Figure~\ref{fig:versor_evolv_single} shows polar projections of the SMBH and accretion disc versors' time evolution for the thin disc model (\textit{single\_thindisc}, top row), unified accretion disc model without wind loss (\textit{single\_uniadios}, middle row) and unified accretion disc model with a massive initial disc and without wind loss (\textit{single\_uniadios\_massive}, bottom row). The colour coding of the tracks indicates the time evolution and the black diamond and square markers indicate the initial and final versor orientation, respectively. The blue filled circle indicates the location of the total angular momentum versor $\mathbf{j}_\mathrm{tot}$. Note that the total versor does not evolve throughout the simulation due the extremely low gas inflow levels. 

For the thin disc model, \textit{single\_thindisc}, both the SMBH and the accretion disc versor are slowly aligning with the total angular momentum versor due to the Bardeen-Petterson alignment. For the unified accretion disc model, \textit{single\_uniadios}, we instead obtain efficient precession of the inner hot flow. Note that the wind loss model (\textit{single\_uniadios\_winds}, not shown), results in an identical versor evolution as the precession frequency only depends on the disc and SMBH state variables (i.e. the masses and angular momenta) and not the current accretion rate (see equation~\ref{eq:omegaprec}). Since the state variables are only changing negligibly due to the low accretion rates, the solid body precession pattern is therefore identical with or without wind loss in these simulations. Finally, we also consider the versor evolution for different initial disc masses. With the massive disc set-up, \textit{single\_uniadios\_massive}, the Eddington fraction is initially much higher (see Section~\ref{subsubsec:results_accrate_lum_single}), so that we obtain extremely rapid realignment of the SMBH and disc versor with the total versor in the Bardeen-Petterson configuration. As the accretion rate drops, the disc switches to the truncated disc mode and starts precessing as a solid body, albeit with an extremely small precession angle due to the previous realignment. With the light disc model (\textit{single\_uniadios\_light}, not shown) the precession timescales are much longer than the simulation time and the alignment timescale exceeds the Hubble time, so that for the duration of the simulation the versor orientation remains constant.

\subsubsection{Single SMBH simulations: precession and alignment analysis}

\begin{figure*}
    \centering
    \includegraphics[width=\textwidth]{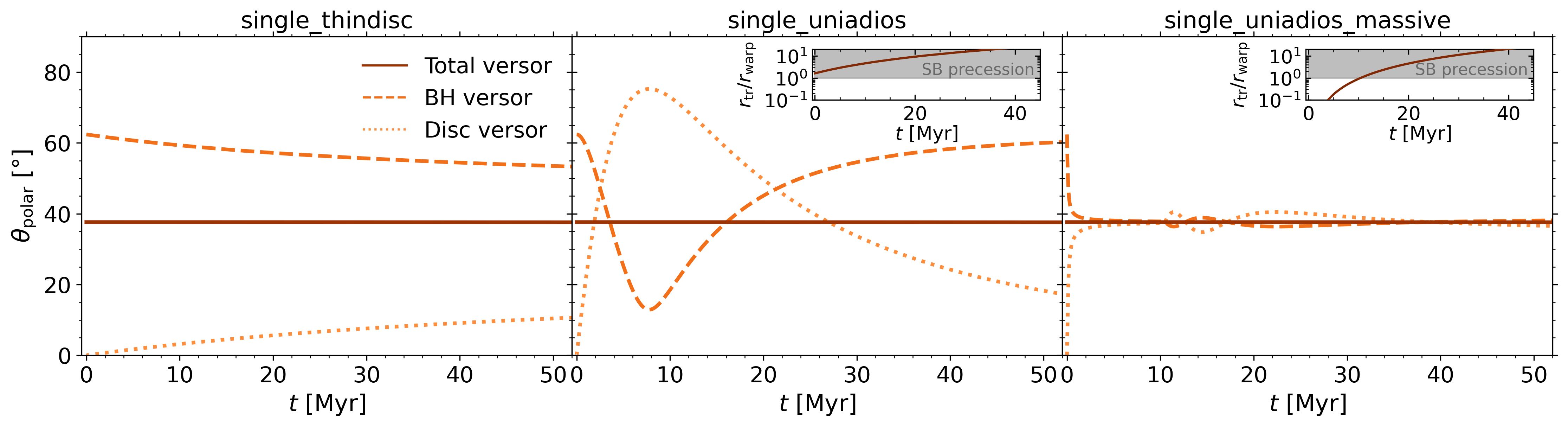} 
    \caption{Polar angle of the total (solid line), SMBH (dashed line) and disc (dotted line) angular momentum versors for the \textit{single\_thindisc} (left panel), \textit{single\_uniadios} (middle panel) and \textit{single\_uniadios\_massive} (right panel) set-ups. For the unified accretion disc model, we also show inset plots with the ratio between the truncation and warp radius to indicate when the system enters the solid body precession regime. This demonstrates the alignment towards the total versor in the Bardeen-Petterson regime as well as the solid body precession evident in the oscillations of the polar angle, with the disc and SMBH precessing in anti-phase.}
    \label{fig:precess_analysis_single}
\end{figure*}

\begin{figure}
    \centering
    \includegraphics[width=\columnwidth]{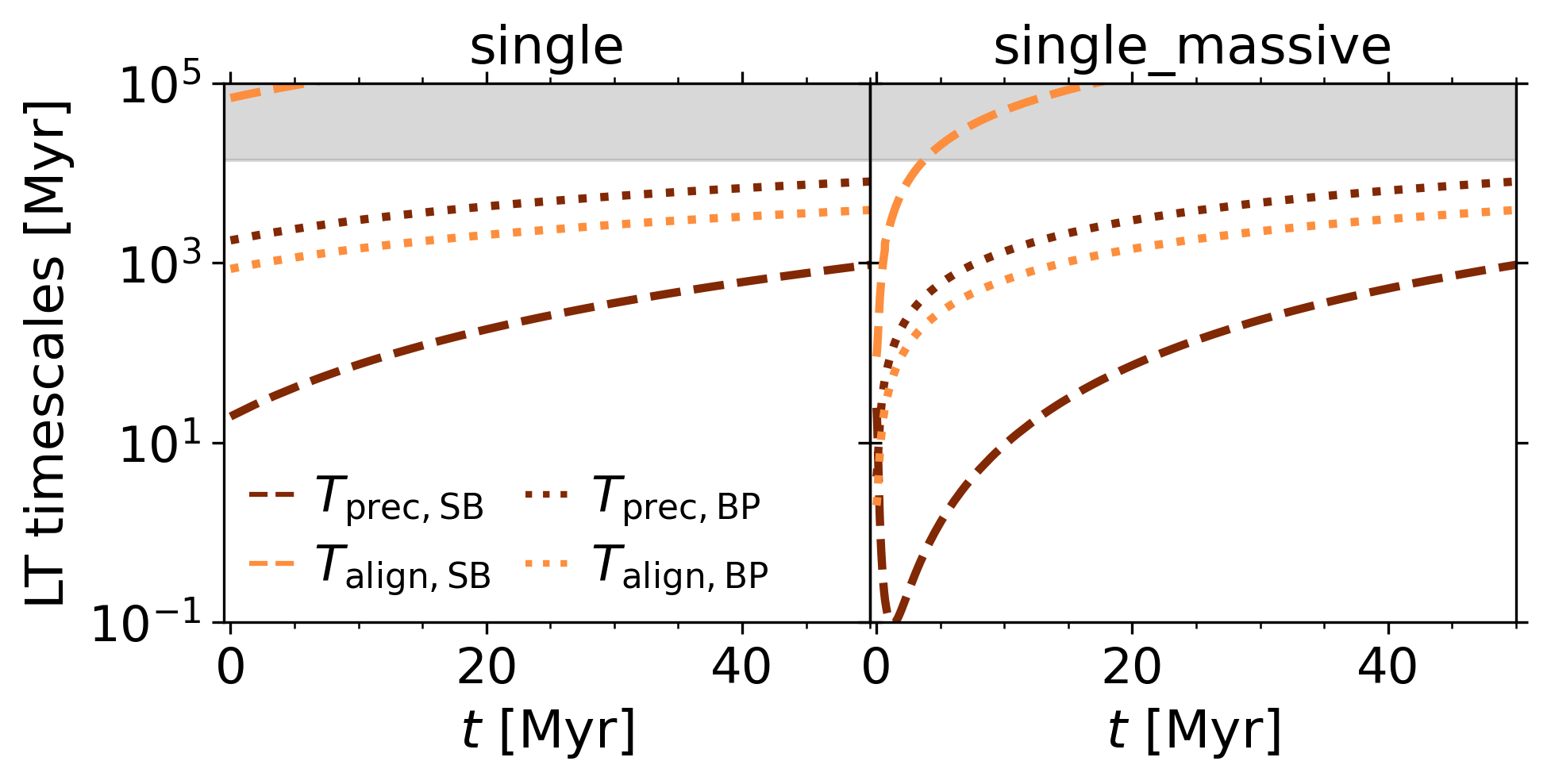} 
    \caption{Lense-Thirring timescales for the \textit{single} and \textit{single\_massive} set-up in the left and right panels, respectively. We plot the alignment timescales (orange lines) and precession timescales (brown lines) for the solid body precession regime (dashed lines) and Bardeen-Petterson regime (dotted lines). The grey-shaded region indicates timescales exceeding the Hubble time. Precession is significantly more efficient for thick discs, whilst alignment is extremely inefficient in the solid body precession regime.}
    \label{fig:LT_timescales_single}
\end{figure}

We investigate these different evolution scenarios more quantitatively in Figure~\ref{fig:precess_analysis_single}, which shows the polar angle of the total, SMBH and accretion disc versors as a function of time and Figure~\ref{fig:LT_timescales_single}, which shows the time evolution of the precession and alignment timescales in the Bardeen-Petterson and solid body precession regimes, respectively.

The left-hand panel of Figure~\ref{fig:precess_analysis_single} shows how in the \textit{single\_thindisc} simulation, the disc and the SMBH versor are torqued towards the total versor by the Bardeen-Petterson effect. However, this alignment process is very slow with timescales $T_\mathrm{align,BP} \gtrsim 10^{3}$~Myr (see left-hand panel of Figure~\ref{fig:LT_timescales_single}). Meanwhile the precession timescales are even longer, so that the versor evolution in the \textit{single\_thindisc} set-up is dominated by the alignment process, with precession only playing a negligible role.

For the \textit{single\_uniadios} model, the disc and SMBH versors are precessing in anti-phase to one another in the solid body regime (see middle panel of Figure~\ref{fig:precess_analysis_single}), with the initial precession timescale $T_\mathrm{prec,SB} \gtrsim 20$~Myr, which then increases to $T_\mathrm{prec,SB} \sim 10^{3}$~Myr at the end of the simulation (see Figure~\ref{fig:LT_timescales_single}), so that the BH-disc-system never completes a full precession period.

Finally, with the \textit{single\_uniadios\_massive} model, we have a phase change where the disc goes through rapid Bardeen-Petterson alignment in the first Myr with $T_\mathrm{align,BP} \sim 1$~Myr and then enters the solid-body precession regime at $t \sim 10$~Myr, undergoing small-scale precession due to previous efficient realignment with the period increasing from $T_\mathrm{prec,SB} \lesssim 10$~Myr to $T_\mathrm{prec,SB} \sim 10^{3}$~Myr.

\subsubsection{Binary SMBH simulations: angular momentum versor evolution}

Following on from the simpler single SMBH case, we move on to analyse the angular momentum evolution of the SMBH and accretion disc in the binary simulations. Here the inflow rates from the mini discs can at times be significant, so that the versor evolution is not just driven by the Lense-Thirring torques but also by external inflows of gas angular momentum.

\begin{figure*}
    \centering
    \includegraphics[width=\columnwidth]{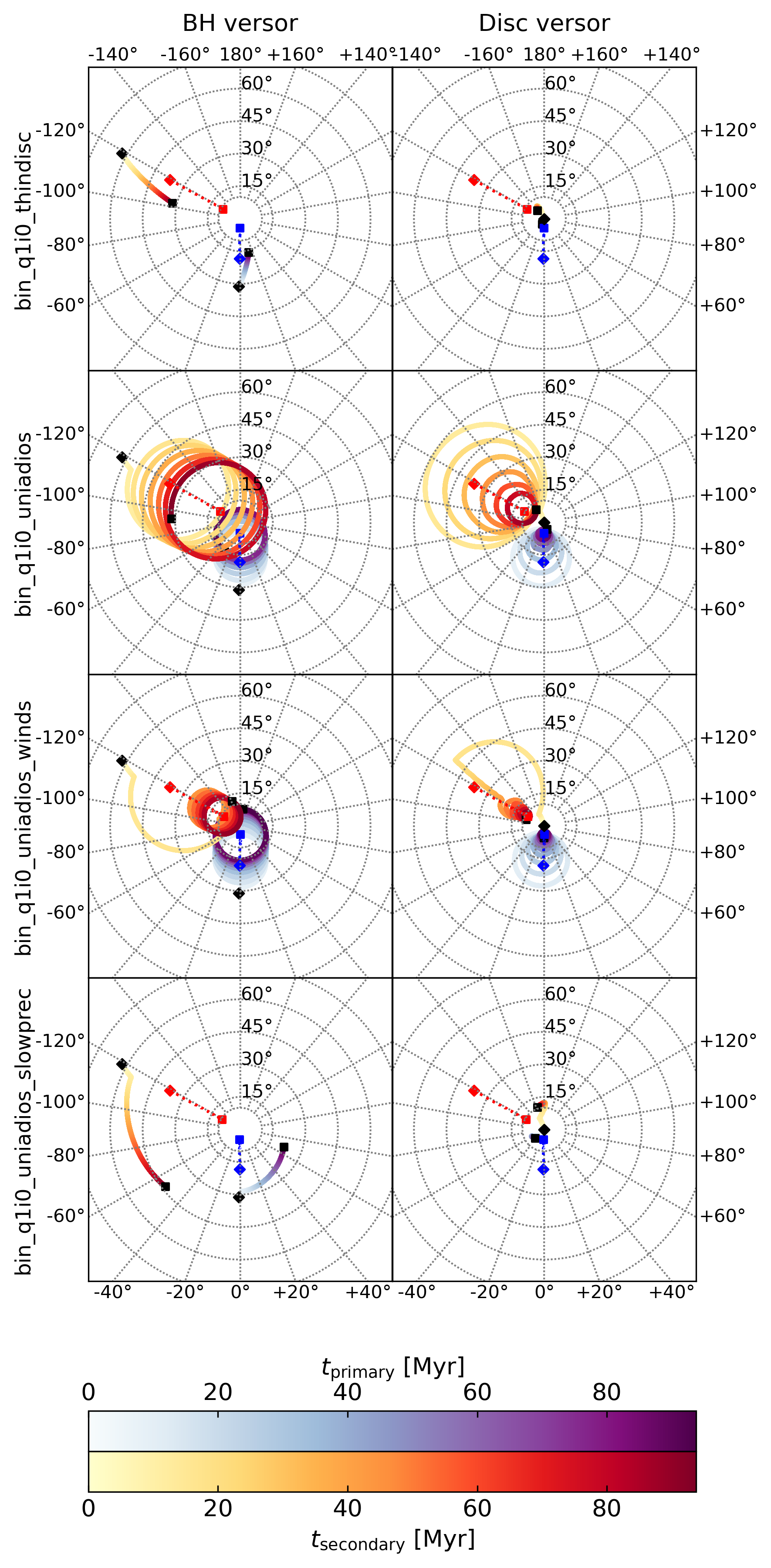}
    \includegraphics[width=\columnwidth]{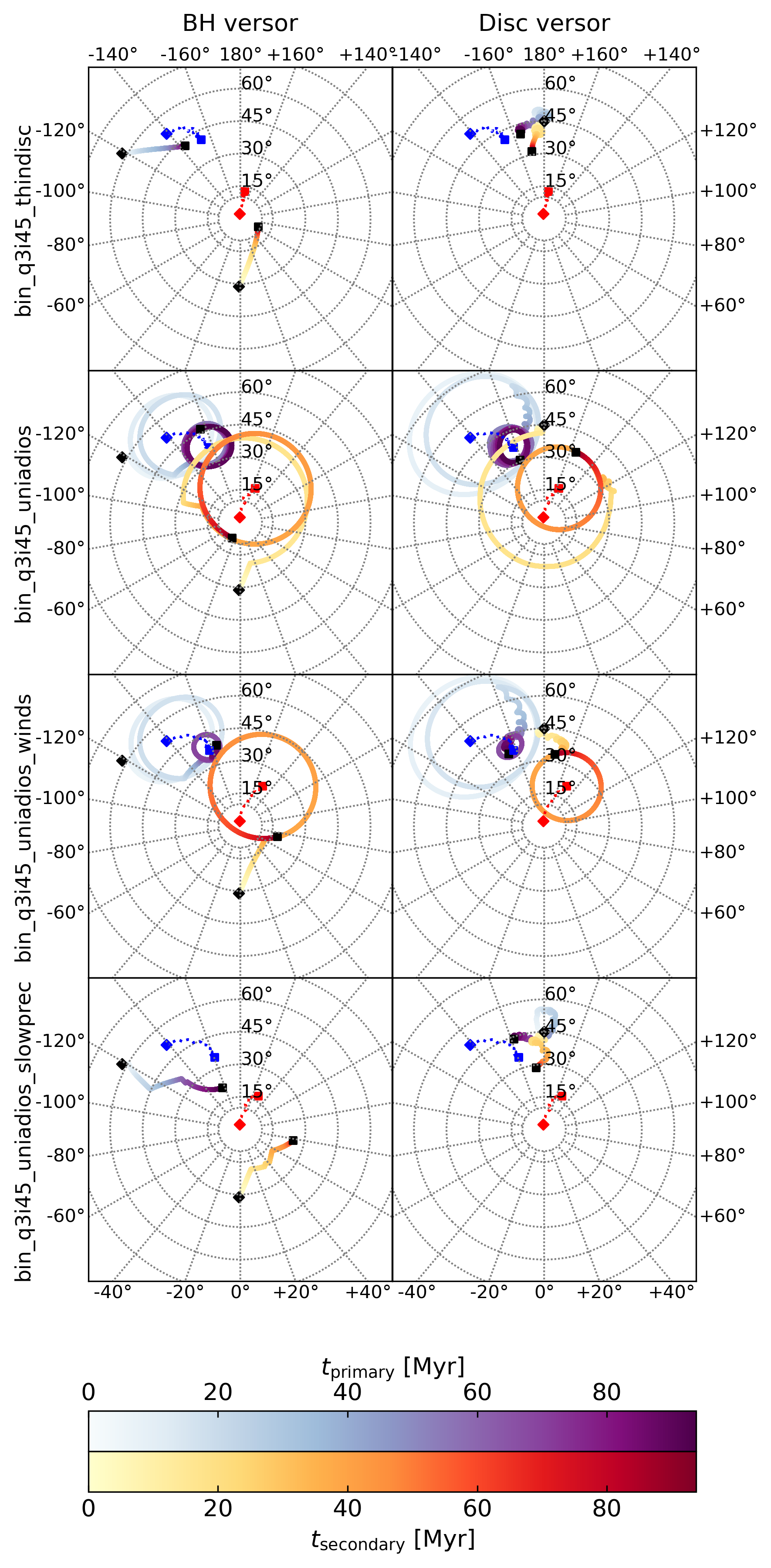}
    \caption{Polar projections of the evolutionary tracks of the SMBH (left columns) and accretion disc (right columns) angular momentum versors for the aligned binaries (\textit{bin\_q1i0}, left panel) and misaligned binaries (\textit{bin\_q3i45}, right panel). We show the time evolution of the primary SMBH as blue-purple colour-coded tracks and the time evolution of the secondary SMBH as yellow-red colour-coded tracks. For comparison, we also plot the evolution of the total angular momentum versors as blue and red tracks for the primary and secondary SMBH, respectively. The rows correspond to different subgrid model set-ups as indicated by the labels. Again with the thin-disc only model, we simply get slow alignment due to the Bardeen-Petterson effect. For the unified accretion disc model, the system goes through alternating periods of solid body precession and Bardeen-Petterson alignment, covering a significant solid angle. However, if the precession rate for the truncated disc mode is reduced following \citet{bollimpalli_effect_2023} then the system neither efficiently precesses nor efficiently aligns.}
    \label{fig:versor_evolv_binary}
\end{figure*}

Figure~\ref{fig:versor_evolv_binary} shows polar projections of the SMBH and accretion disc versors' time evolution for the aligned \textit{q1i0} (left panel) and misaligned \textit{q3i45} binary runs (right panel). The blue-purple tracks represent the versor evolution of the primary, whilst the orange-red tracks show the versor evolution of the secondary SMBH. The red and blue dotted lines show the evolution of the total angular momentum versor $\mathbf{j}_\mathrm{tot}$. 

The evolution of the total angular momentum versors for a given binary configuration is virtually identical for all of the subgrid models explored as it is governed by the external inflows from the mini discs. For the aligned binary, \textit{bin\_q1i0}, we find efficient realignment of the total versors with the poles as we have coherent accretion from the mini discs, whilst for the misaligned binary, \textit{bin\_q3i45}, the accretion is less coherent as the binary realigns with the outer CBD \citep[see][for details]{bourne_dynamics_2023}.

The first row shows the runs with the thin disc model. As with the single SMBH set-up, we have Bardeen-Petterson alignment of the SMBH and disc angular momentum versors with the total angular momentum versors. However, as the total angular momentum versors are being reoriented towards the mini disc angular momentum and the subgrid disc is initially assumed to be aligned with the mini discs (see Section~\ref{subsubsec:model_var_bin}), the alignment between the disc and total versors is accelerated by the external inflows. For the SMBH versors, on the other hand, the external inflows lead to less efficient alignment as the total versor recedes from the SMBH versor. 

The second row shows the simulations performed with the unified accretion disc model without wind loss and with standard precession. For the \textit{bin\_q1i0} set-up, after a short Bardeen-Petterson alignment period, the SMBH and disc versors are rapidly precessing. The SMBH versors' precession angles remain roughly constant, whilst for the disc versor the precession angle is steadily decreasing as the disc and total angular momentum versor align due to accretion from the mini discs. For the \textit{bin\_q3i45} simulation, the picture is more complex as the system alternates between Bardeen-Petterson alignment and solid-body precession, so that the precession angle decreases for both the SMBH and disc versors.

The simulations shown in the third row were also performed with the unified accretion disc model, but account for wind loss with the ADIOS model. Note that this slows down the SMBH mass growth rate, which decreases the truncation radius so that the inner hot flow is more likely to be smaller than the aligned region of the warped outer thin disc, so that the inner hot flow is also twisted into alignment. Indeed for both binary configurations, the SMBH -- accretion disc system tends to spend more time in the Bardeen-Petterson alignment phase with the ADIOS wind loss being taken into account. The impact is particularly dramatic for the secondary SMBH in the \textit{bin\_q1i0} set-up. Here the combination of solid body precession followed by Bardeen-Petterson alignment realigns the SMBH and disc versors even more efficiently than the thin disc only case.

The simulations presented in the fourth row were performed with a reduced solid body precession rate for the truncated disc case, following the results from \citet{bollimpalli_effect_2023}, who find that the precession rate of the inner hot flow is reduced by 95 per cent due to torques from the outer thin disc (see Section~\ref{subsubsec:AngMomLT} for details on the solid body precession model). With this model, the system never completes a full precession cycle within the simulation time and there is only very slow alignment from the short Bardeen-Petterson phases. The equivalent set-ups with wind loss (\textit{uniadios\_winds\_slowprec}, not shown here) are very similar. Since the precession proceeds very slowly, the duration of the Bardeen-Petterson alignment has a much less significant impact on the overall evolution.

\subsubsection{Binary SMBH simulations: precession and alignment analysis}

\begin{figure*}
    \centering
    \includegraphics[width=0.495\textwidth]{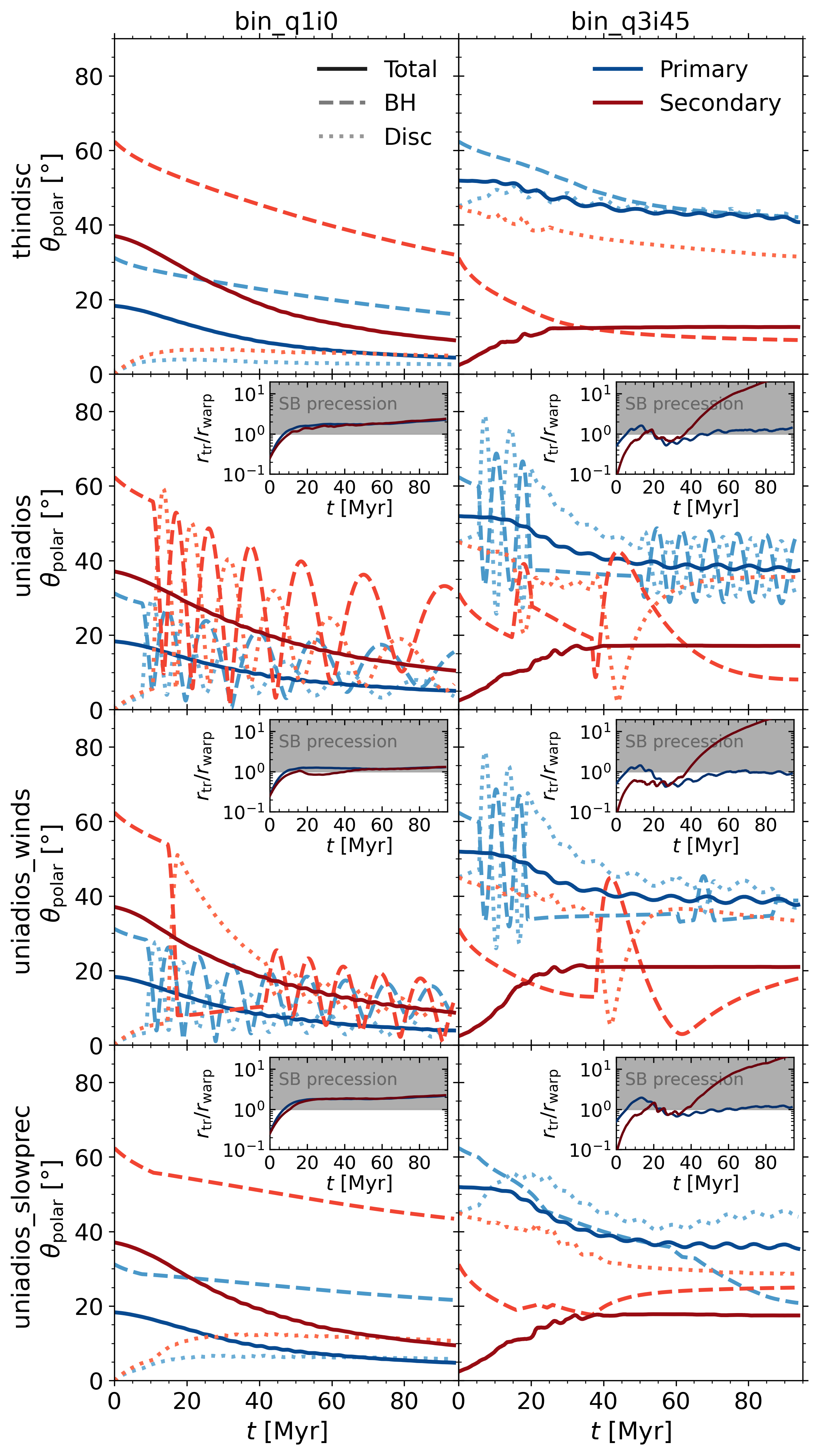} 
    \includegraphics[width=0.495\textwidth]{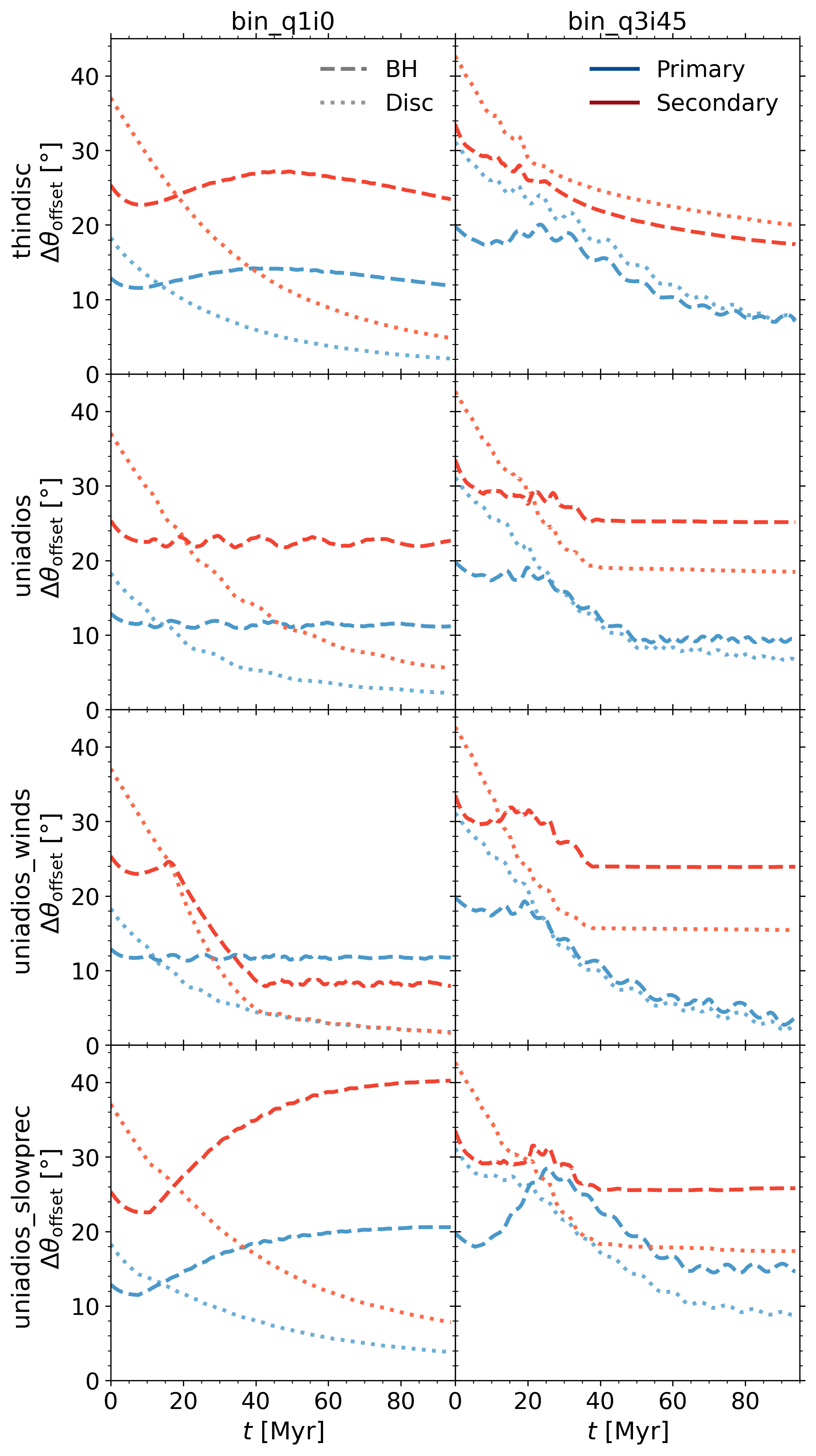} 
    \caption{\textit{Left panel:} Time evolution of the polar angles of the total (solid lines), SMBH (dashed lines) and accretion disc (dotted lines) angular momentum versor for the primary (blue lines) and secondary SMBHs (red lines). The insets show the ratio between the truncation and warp radius, indicating the precession regime of the system. \textit{Right panel:} To better illustrate the evolution of the versor alignment and precession, we also plot the offset angle between the total versor and SMBH/disc versors. For the \textit{bin\_q1i0} set-up (left columns), the accretion disc alignment is mainly driven by angular momentum inflows from the mini discs, whilst the SMBH alignment depends on the subgrid model and interplay between the different Lense-Thirring regimes. For the \textit{bin\_q3i45} set-up (right columns), the inflows from the mini discs are less coherent, so both disc and SMBH alignment are sensitive to the accretion disc model.}
    \label{fig:precess_analysis_binary}
\end{figure*}

\begin{figure*}
    \centering
    \includegraphics[width=\textwidth]{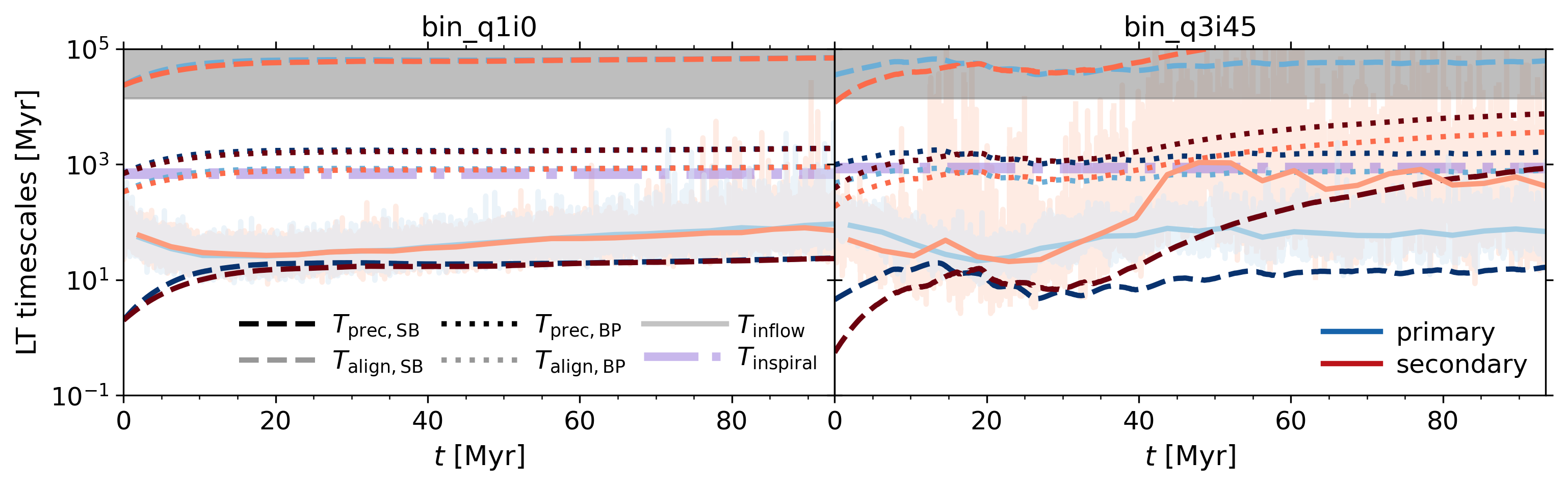}
    \caption{Lense-Thirring timescales for the \textit{bin\_q1i0} and \textit{bin\_q3i45} simulations in the left and right panels, respectively. We plot the alignment timescales (dark lines) and precession timescales (light lines) for the solid body precession regime (dashed lines) and Bardeen-Petterson regime (dotted lines). The timescales for the primary are colour-coded in blue, whilst the timescales for the secondary are colour-coded in red. The grey-shaded region indicates timescales exceeding the Hubble time. For reference, we also plot the inflow timescales from the mini discs as shaded regions (instantaneous) and solid lines (binned). Furthermore, we show the binary inspiral timescale as a thick light purple dash-dotted line. The interplay of external inflows and solid body precession as well as Bardeen-Petterson alignment all influence the spin alignment in our simulations, as all of the relevant timescales for these processes are in a similar range to the simulation runtime.}
    \label{fig:LT_timescales}
\end{figure*}

We now turn to investigate the complex precession behaviour of the binary SMBHs more quantitatively. Figure~\ref{fig:precess_analysis_binary} shows the time evolution of the versors' polar angle in the left-hand panels. Due to efficient angular momentum accretion from the mini disc, the total angular momentum versor evolves significantly in the binary simulations. To control for this effect, we also consider separately the time evolution of the `offset angle' between the BH/disc versors and the total angular momentum versor in the right-hand panels. Moreover, Figure~\ref{fig:LT_timescales} shows the alignment and precession timescales in the Bardeen-Petterson and solid body regimes as well as the characteristic external inflow timescale (defined as $T_\mathrm{inflow} = M_\mathrm{d} / \dot{M}_\mathrm{inflow}$) and the binary inspiral timescale.

We first focus on the angular momentum evolution in the equal-mass, aligned binary case (\textit{bin\_q1i0}). Due to the inflows from the mini discs, the angular momentum versor $\mathbf{j}_\mathrm{d}$ (and therefore the total versor $\mathbf{j}_\mathrm{tot}$) is driven towards alignment with the pole. As discussed in the previous section, $\mathbf{j}_\mathrm{tot}$ may only change due to external inflows, and therefore the $\mathbf{j}_\mathrm{tot}$ evolution follows the same pattern in all of the \textit{bin\_q1i0} runs. From Figure~\ref{fig:LT_timescales}, we can see that the mean inflow timescale varies between $\sim 20 - 100$~Myr (light-blue and light-red solid lines) and hence $\mathbf{j}_\mathrm{tot}$ aligns with the poles during the duration of the simulation.

The alignment and precession of the accretion disc and SMBH angular momentum versors, however, depends significantly on the accretion disc model assumed. With the thin disc model (first row), the disc is in the Bardeen-Petterson configuration, where the SMBH and spin versors may efficiently realign, both being driven towards $\mathbf{j}_\mathrm{tot}$ on an alignment timescale of a few 100~Myr (see light-blue and light-red dotted lines in Figure~\ref{fig:LT_timescales}). Considering the evolution of the angle between $\mathbf{j}_\mathrm{tot}$ and the disc and SMBH versors in the right-hand panel of Figure~\ref{fig:precess_analysis_binary}, we can see that the disc efficiently aligns with $\mathbf{j}_\mathrm{tot}$ as both the external inflows and the Bardeen-Petterson alignment work in tandem. For the SMBH, on the other hand, the inflows drive $\mathbf{j}_\mathrm{tot}$ away from the SMBH versor so that there is a phase where the misalignment angle increases. Though eventually, as the inflow timescale increases, Bardeen-Petterson alignment becomes more efficient. Overall the SMBH spin evolution is dominated by the alignment process, as the precession timescale in the Bardeen-Petterson configuration is much longer ($T_\mathrm{prec,SB} \sim 10^{3}$~Myr), exceeding the runtime of our idealised set-up.

With the unified accretion disc model (second to fourth rows), we get significantly different behaviour. As the disc is in the truncated state, the relevant Lense-Thirring precession regime is determined by the ratio between the truncation radius and the warp radius from the Bardeen-Petterson alignment process of the outer thin disc (shown as inset panels in Figure~\ref{fig:precess_analysis_binary}). If the truncation radius is larger than the warp radius, then the outer disc is feeding misaligned material to the inner hot flow which therefore enters the solid-body-precession mode, characteristic of misaligned thick discs (see Section~\ref{subsubsec:AngMomLT} for details). 

With the \textit{uniadios} model (second row), we have a short period of Bardeen-Petterson alignment followed by solid-body precession for the remainder of the simulation. In the case of the disc the precession angle is steadily decreasing as the total angular momentum versor is being driven towards the poles. For the SMBH versor, on the other hand, as the precession is in anti-phase (close to the pole), the total angular momentum versor is moving towards the SMBH versor and the precession angle is decreasing. When the SMBH precession has completed a full cycle (far from the pole), the total angular momentum versor is moving away from the SMBH versor and the precession angle increases. These small oscillations can be seen in the third panel of the second row in Figure~\ref{fig:precess_analysis_binary}, which as expected, are shifted from the precession motion by half a phase. Overall this means that the mean precession angle is approximately conserved, with the SMBH versor following the translational motion of the total versor, thereby also aligning with the poles and the disc angular momentum.

With the \textit{uniadios\_winds} model (third row), we also take wind-loss into account which leads to lower SMBH masses and therefore lower (auxiliary) Eddington fractions and smaller truncation radii. For the \textit{bin\_q1i0} simulation, the truncation radius to warp radius ratio is relatively close to one (see insets in left-hand panels of Figure~\ref{fig:precess_analysis_binary}), so that the system is highly sensitive to small changes in the truncation radius and the secondary SMBH dips back into the Bardeen-Petterson regime at $t=20$~Myr for the \textit{uniadios\_winds} set-up. This change-over occurs when the SMBH is in anti-phase (close to the poles), so that now the SMBH and the total versor are being driven towards one another by both the external inflows and the Bardeen-Petterson alignment. Meanwhile, whilst the disc is far away from the pole, the disc angular momentum is directly affected by the external inflows (with a higher fractional change than the total versor) and therefore also realigns rapidly with the poles. At $t=40$~Myr, when the system re-enters the solid-body precession regime, both the disc and the SMBH versor are within $10\degree$ of the total versor, so that we only have small-scale precession for the remainder of the simulation. Remarkably, the interplay between the Bardeen-Petterson regime and solid-body precession leads to even more efficient alignment than the thin disc model.

In the fourth row of Figure~\ref{fig:precess_analysis_binary}, we also show the time evolution of the polar and precession angles for the slow precession model for truncated discs following \citet{bollimpalli_effect_2023}. Due to the longer precession period, the SMBH versor remains in the phase space where the total versor is receding and therefore the offset angle for the SMBH is steadily increasing. Hence with this model, the final SMBH versor is the least aligned out of all set-ups considered. Note that including wind loss (\textit{bin\_q1i0\_uniadios\_winds\_slowprec}, not shown here) leads to very similar results as in neither case a significant re-orientation of the SMBH versor is achieved.

Next, we analyse the precession and alignment patterns in the non-equal-mass, misaligned binary simulations \textit{bin\_q3i45} shown in the second columns of the two panels in Figure~\ref{fig:precess_analysis_binary}. We find that in this misaligned scenario the primary total versor is initially already in close alignment with the disc, leading to efficient realignment of the primary disc and SMBH versors with $\Delta \theta_\mathrm{offset} < 20 \degree$ for all accretion disc models explored (see right-hand panel of Figure~\ref{fig:precess_analysis_binary}). For the secondary SMBH, the initial offset angle is much larger and the inflow timescale increases dramatically at $t=40$~Myr (see Figure~\ref{fig:LT_timescales}), so that alignment is much less efficient. We also note that the accretion from the mini discs for the misaligned simulations is highly stochastic and incoherent \citep{bourne_dynamics_2023}, so that the evolution of the total versors (see left-hand panel of Figure~\ref{fig:precess_analysis_binary}) slightly differs between the different simulations performed.

With the thin-disc model, the initial Bardeen-Petterson alignment timescale for both SMBHs is a few 100~Myr. This remains relatively constant for the primary, however, for the secondary SMBH the alignment timescale steadily increases from $t\sim40$~Myr onwards, significantly slowing down the disc alignment. Note that since the inflows for the misaligned binary case are much less coherent we do not observe the slowing down of the SMBH versor alignment that we saw with the equal-mass binary. As discussed in \citet{bourne_dynamics_2023}, there is some small precessional motion for both the disc and the total versor due to the incoherent inflows from the mini discs which are realigning with the outer CBD. It is worth noting that the realignment process slightly differs in our simulation from the set-up presented in \citet{bourne_dynamics_2023} as they have a different randomly chosen initial SMBH versor orientation, which is much more significantly offset for the primary SMBH leading to less efficient alignment with the mini discs.

With the unified accretion disc model, we obtain several changeovers between the Bardeen-Petterson and solid-body regime due to the stronger variation of the Eddington fraction in the misaligned case (see Figure~\ref{fig:inflow_binaries}). Firstly, we focus on the set-up without wind loss, \textit{uniadios}, in the second row of Figure~\ref{fig:precess_analysis_binary}. For the primary and secondary SMBH, we initially have a period of Bardeen-Petterson alignment and then transfer into the solid-body precession at $t=5$~Myr and $t=15$~Myr, respectively. During the solid-body precession period, we again have the precession angle decreasing for the disc versor as the external inflows provide additional realignment, whilst the SMBH precession angle remains constant. The primary and secondary SMBH then switch back to the Bardeen-Petterson configuration at $t \sim 20$~Myr, further aligning with the total versor until $t=40$~Myr and $t=50$~Myr, respectively. Having further realigned, the SMBHs then carry out solid-body precession at a reduced angle for the remainder of the simulation. Note that the disc precession angle only marginally decays in this second phase of solid body precession as the angular momentum inflow rates are less coherent. 

With wind loss, \textit{uniadios\_winds} (third row), we obtain qualitatively similar behaviour for the primary SMBH, although the second Bardeen-Petterson alignment phase is significantly prolonged as the SMBH mass grows more slowly, leading to a smaller truncation radius. This leads to a reduced precession angle in the second solid-body precession phase. For the secondary SMBH, on the other hand, the smaller truncation radius means that the BH-disc system remains in the Bardeen-Petterson configuration until $t \sim 40$~Myr and then transfers to the solid body precession regime with a significantly reduced precession angle.

Finally, we also consider the modified set-up, where the precession rates are reduced by 95 per cent in the truncated disc mode, \textit{uniadios\_slowprec}. As with the \textit{bin\_q1i0} set-up, this means that the SMBH versors are lagging behind the total versor alignment due to external inflows throughout, so that the precession angle increases, leading to a larger final offset angle for the primary SMBH. The secondary SMBH experiences a similar albeit much smaller effect as it spends significantly less time in the solid body precession phase.

\subsubsection{Summary and implications of angular momentum evolution}
Figures~\ref{fig:versor_evolv_single}~and~\ref{fig:versor_evolv_binary} clearly show that the accretion disc subgrid model has a significant impact on the spin alignment and precession process. In the truncated disc regime, precession is much more prevalent with the SMBH spin versors covering a large solid angle. This has crucial implications for distributed feedback, in the particular in the context of spin-driven jets (see Section~\ref{subsec:discu_feedback} for a more detailed discussion of implications for AGN feedback). Also note that VLA and Merlin observations have uncovered precessing jets with inferred precession periods of $1-10$~Myr \citep{krause_how_2019}. Whilst this could be attributed to geodetic precession, it would also be consistent with solid-body precession in the truncated disc regime within our framework and LOFAR is expected to shed more light on the origin of these variability patterns \citep[e.g.][]{horton_new_2023}.

Furthermore, it is interesting to note that despite the alignment torques between the accretion disc and SMBH being extremely weak in the ADAF/ADIOS regime, the SMBH may still realign with the disc if the external inflow timescale matches the precession timescale. Furthermore, Bardeen-Petterson alignment may be accelerated if the solid body precession reorientates the SMBH versor towards the mini discs. On the other hand, if the precession process is slowed down significantly due to additional torques in the truncated disc state \citep{bollimpalli_effect_2023}, then the spins will not align within the inspiral timescale, with crucial implications for recoil predictions, influencing merger rates and the gravitational wave background.

\section{Discussion} \label{sec:discussion}

We have developed a unified accretion disc model suitable for galaxy formation simulations that can resolve the outer edge of the accretion disc. Our work builds on the thin disc model prescription presented in \citet{fiacconi_galactic_2018} and extends this scheme by harnessing the results from high-resolution GR(R)MHD simulations of accretion discs. Our model spans a range of radiative regimes from the optically thin, geometrically thick discs to the optically thick, geometrically thin discs via a truncated disc configuration. We also track the SMBH spin evolution with a disc-state-dependent Lense-Thirring precession model. Our predictions for the SMBH spin alignment and precession as well as SMBH luminosities are highly sensitive to the disc state with important implications for future multi-messenger observational campaigns which we outline in Section~\ref{subsec:obs_implications}. We also discuss the caveats of our model and important future extensions in Section~\ref{subsec:caveats_future}.

\subsection{Implications for observations} \label{subsec:obs_implications}

Parsec-scale SMBH binaries have proved extremely challenging to observe, with only a handful of detections so far \citep{rodriguez_compact_2006,bansal_constraining_2017,kharb_candidate_2017}, as these are in the regime where extremely high angular resolution is necessary to resolve the binary, however, the SMBHs are not close enough to display significant variability on observable timescales. This makes radio interferometers the tool of choice for searching for these binaries and future facilities such as SKA or ngVLA are expected to significantly increase sample sizes \citep[e.g.][]{burke-spolaor_next-generation_2018}, though see a cautionary note from \citet{dorazio_observational_2023}.

To estimate radio luminosities, it is crucial to have reliable estimates for the mass accretion rates, spin orientations and solid angles \citep[e.g.][]{mangiagli_massive_2022}. With our unified accretion disc model, we can estimate all of the relevant quantities and make predictions for the jet luminosities depending on the disc state which reveals a much more promising avenue for detecting SMBH binaries accreting at intermediate luminosities than the thin disc modelling (see Figure~\ref{fig:inflow_binaries_luminosities}).

According to analytical calculations, our binary system should have an inspiral timescale of $\lesssim 10^{3}$~Myr (see Figure~\ref{fig:LT_timescales}), so this represents a precursor to the gravitational-wave-emitting binaries. Nevertheless it is crucial to make predictions for these precursor luminosities, in particular for X-ray confirmations \citep{piro_chasing_2023}.

Furthermore, accurate AGN modelling will be crucial for dual AGN (and close pairs) searches with JWST where recent tantalizing observations have revealed hints at a large fraction of candidate dual AGN \citep{maiolino_jades_2023,ubler_ga-nifs_2023-1}, highlighting the importance of imaging spectroscopy in tracking SMBH mergers throughout cosmic time. Combined with results from gravitational wave interferometers, these JWST constraints can provide crucial constraints on the merging SMBH population \citep[e.g.][]{padmanabhan_constraints_2024}.

The AGN lifetime and dynamical evolution of the binary in our simulations also provide important information for astrometric searches for SMBH binaries with \textit{Gaia} \citep[e.g.][]{dorazio_detecting_2019}.

Finally, our spin evolution predictions play an important role in estimating the gravitational wave background signal as they determine the recoil velocity of the SMBH remnant -- in particular large and misaligned SMBH spins may lead to recoil velocities of $\sim 5000 \ \mathrm{km \, s^{-1}}$ \citep{lousto_hangup_2011,lousto_nonlinear_2013,lousto_kicking_2019} which would have a significant impact on SMBH occupation fractions and merger rates.

\subsection{Caveats and future work} \label{subsec:caveats_future}

\subsubsection{Accretion modelling}
We have extended the thin disc model from \citet{fiacconi_galactic_2018} into the heavily sub-Eddington regime based on the ADAF/ADIOS models. However, we note that there are several other disc models that have been proposed for the radiatively inefficient accretion regime, including convection-dominated accretion flows \citep[CDAFs;][]{narayan_self-similar_2000,quataert_convection-dominated_2000} or more recently simple convective flows \citep{xu_simple_2023} and strongly magnetised Bondi models \citep{cho_bridging_2023-1}. \citet{li_rotating_2013} carried out an extensive simulation suite, capturing the accretion flows from Bondi scale to the small scale and demonstrated how the inflow and outflow patterns will change as a function of cooling. Similarly, they show how strong outflows suppress the Bondi rate, however, they also demonstrate a strong dependency on cooling. Note that in principle, we could incorporate these accretion flow scenarios within our framework by varying the $s$ parameter from the ADIOS wind loss model, which effectively sets the flow density profile. However, this is beyond the scope of this paper and we plan to investigate the impact of $s$ parameter choices more systematically in the future.

Whilst this work has focused on the radiatively-inefficient regime, in the future it will also be important to further consider the disc modelling in the radiatively efficient regime. Several recent studies have questioned and improved the standard thin $\alpha$-disc model \citep[e.g.][]{jiang_global_2019} and this will require careful consideration to improve the predictions of our model in the quasar regime.

Misaligned discs experience radius-dependent Lense-Thirring torques. How these differential torques are communicated depends on the accretion disc regime. In our unified accretion disc model, we evolve the SMBH spin according to the Bardeen-Petterson alignment \citep{bardeen_lense-thirring_1975} in the thin disc state and following the solid body precession model from \citet{fragile_global_2007} in the thick disc state. We also include the possibility of slowed down solid body precession in the truncated disc regime \citep{bollimpalli_effect_2023}. However, GR(R)MHD simulations paint an even more complex picture, including disc tearing \citep[e.g.][]{liska_radiation_2023} which could suppress Bardeen-Petterson alignment \citep[e.g.][]{steinle_bardeen-petterson_2023} and induce phase lags between the inner hot flow and outer thin disc \citep[e.g.][]{liska_phase_2023-1}. Furthermore, if the thick disc is in the magnetically arrested state, strong magneto-spin alignment could significantly suppress solid body precession \citep[e.g.][]{mckinney_alignment_2013,chatterjee_misaligned_2023}.

Finally, another crucial future extension to our model will be to incorporate the super-Eddington regime. This accretion disc regime may be particularly important in the early Universe, where JWST is starting to uncover a population of overmassive AGN (compared to the local scaling relations) which could stem from sustained super-Eddington accretion episodes \citep[e.g.][]{goulding_uncover_2023,maiolino_jades_2023}. Furthermore, cosmological simulations show that there is likely enough gas present around the early SMBHs for super-Eddington accretion \citep[e.g.][]{rennehan_three_2023,bennett_growth_2024,juodzbalis_dormant_2024}. Note that many of the key physical properties of our ADIOS modelling in the sub-Eddington regime will also apply to the super-Eddington regime. The latter is advection-dominated due to the long radiative diffusion times \citep[rather than long cooling times, see][]{yuan_hot_2014}, so that the radiative efficiency drops in the super-Eddington regime and accretion (as well as jets) cause the SMBH to spin down \citep[also see][]{lowell_rapid_2024,ricarte_recipes_2023}. 

\subsubsection{Feedback modelling} \label{subsec:discu_feedback}

Within our current framework we only model gas accretion through a unified accretion disc model. For some model variations we take into account mass loss through a wind, however, we do not yet re-inject the wind mass into the simulation. The obvious next step is to incorporate AGN feedback mechanisms into our model. The most straightforward extension here would be to incorporate ADIOS winds with the mass loading directly informed by the wind mass loss in our model. The momentum and energy loading of the wind could then be inferred from the kinetic wind efficiencies obtained from GRMHD simulations \citep[e.g.][]{tchekhovskoy_general_2012,sadowski_energy_2013,sadowski_kinetic_2017}. We note that the thin disc component is also expected to drive winds, in particular at high Eddington fractions, and including these winds would be an important extension for future work \citep[also see][]{sala_non-isotropic_2021}, representing the `quasar mode' feedback employed in cosmological simulations which is especially crucial in the early Universe. In addition to wind feedback, radiative feedback would be another important channel to consider. Here our unified accretion disc model provides the ideal starting point as we could base the AGN emission on the subgrid accretion disc state.

Moreover, AGN jets are an ubiquitous phenomenon and are likely powered by the SMBH spin energy extraction via the Blandford-Znajek (BZ) mechanism \citep{blandford_electromagnetic_1977}. \citet{talbot_blandford-znajek_2021} have coupled spin-driven BZ jets self-consistently with the thin disc model from \citet{fiacconi_galactic_2018} and considered their evolution in Seyfert galaxies \citep{talbot_blandford-znajek_2022} and a galaxy merger \citep{talbot_simulations_2024}. In future work, it would be timely to extend this scheme to include spin-driven jets across different accretion disc types \citep[also see][]{husko_spin-driven_2022,husko_winds_2024}. In particular, the solid body precession model could have a crucial impact here leading to more `distributed' jet feedback. Note that the precession periods in the solid body regime are highly uncertain (see Figure~\ref{fig:solid_body_prec}), and there are hints at Lense-Thirring precession in AGN on the scale of $\sim 10$~yr \citep{cui_precessing_2023-1,von_fellenberg_radio_2023}. In terms of AGN subgrid models, this would imply that for timescales relevant for galaxy formation the AGN feedback would be effectively distributed over a cone, which may significantly increase its impact and potentially lead to more direct interaction with the interstellar medium as demonstrated in idealised simulations of precessing jets by \citet{su_which_2021}.

\subsubsection{Porting the unified accretion disc model to cosmological simulations}

In addition to improving the subgrid SMBH modelling, it will also be crucial to move towards more realistic environments and test different accretion regimes. We found that more chaotic accretion, such as in the misaligned binary case, can have a significant impact on luminosities and spin evolution so a natural next step would be to test this framework within galaxy merger simulations \citep[also see][]{liao_modelling_2023,talbot_simulations_2024}. Furthermore, extending our model to cosmological zoom-in simulations will allow us to track the mass and angular momentum flows from the cosmological environment down to the scale of the accretion disc. For the zoom-in simulations, the super-Lagrangian refinement technique will permit us to reach the required resolution \citep[also see][]{curtis_powerful_2016,bourne_agn_2019,bourne_agn_2021,angles-alcazar_cosmological_2021,hopkins_forged_2024,hopkins_forged_2024-1}. For large cosmological volumes, applying the super-Lagrangian refinement technique to the central regions of tens of thousands of galaxies, however, is computationally infeasible. One avenue to circumvent these issues could involve using machine-learning techniques such as generative adversarial networks or learned coarse models using convolutional neural networks \citep[e.g.][]{li_ai-assisted_2021,stachenfeld_learned_2021}. This would require appropriate training data sets covering the vast range of relevant scales and different environments and could build on recent pilot projects pioneering this approach for subgrid models in planet formation \citep[e.g.][]{pfeil_neural_2022}.

Finally, our unified accretion disc model could be modified for lower-resolution applications, in particular for the SMBH mass evolution. Here more coarse-grained prescriptions such as the Bondi model or the torque-based accretion model could provide an estimate of the inflow rate onto the disc. The accretion disc could then act as a gas reservoir that gets drained on the viscous timescale which gets calculated in a disc-state dependent fashion. Similar approaches have already been introduced in the literature \citep[e.g.][]{power_accretion_2011,wellons_exploring_2023}, however, these employ either a constant viscous timescale or a viscous timescale based on the thin $\alpha$-disc model, whereas our approach would also be directly applicable in the radiatively inefficient regime and therefore appropriate for both the quasar and radio mode in cosmological simulations.

\section{Conclusions} \label{sec:conclusions}

We have developed a novel unified accretion disc model designed to be embedded in galaxy formation simulations. Drawing inspiration from high-resolution GR(R)MHD simulations, our model tracks mass flows through three distinct disc states and evolves the SMBH spin based on disc-state-dependent Lense-Thirring precession models. This innovative accretion disc framework has been integrated into the moving mesh code \textsc{arepo}, and our work includes a comprehensive suite of validation simulations. These simulations encompass idealized scenarios featuring single SMBHs and binary SMBHs embedded in gaseous discs, with variations in mass ratio, inclination angle, and subgrid accretion disc models. Our key findings are summarized below.

\begin{itemize}
\item The ADIOS model predicts substantial wind loss in the thick disc regime, leading to a significant suppression of gas mass accretion rates onto SMBHs by an order of magnitude or more. Additionally, the decrease in radiative efficiency within the advection-dominated accretion regime results in luminosity estimates that are decreased by another order of magnitude. Consequently, \textit{predictions for the luminosities of electromagnetic counterparts of SMBH binaries exhibit significant sensitivity to the disc state}.
\item Hence, it is imperative to incorporate these realistic accretion disc models when forecasting outcomes for future missions both for single and binary SMBHs. Specifically, our unified accretion disc model reveals a drastic reduction in the redshift range for accretion disc luminosities compared to predictions based on the thin $\alpha$-disc model, regardless of whether winds are accounted for. This has important implications for SMBH binary searches in the X-ray, optical and infrared. Furthermore, the predicted spectrum undergoes significant shifts, with the luminosity being dominated by the hard inner hot flow.
\item For the radio jet luminosities, on the other hand, the situation is reversed with the unified accretion disc model predictions much more favourable than the thin disc predictions, since the geometrically thick ADIOS flow leads to significantly more efficient jet production. We predict that future VLBI searches with SKA and the ngVLA should significantly increase the known samples of parsec-scale binaries.
\item The evolution of the spin magnitude depends only mildly on the disc state, primarily due to relatively low external inflow rates in all cases considered here. We note however, that depending on the disc state adopted we predict for the {\it same} large-scale gas inflow, the SMBH spin may either increase or decrease. Clearly, in case of higher and more variable accretion rates, significant differences in the spin magnitude evolution are expected.
\item Importantly, \textit{the SMBH spin direction's evolution is markedly influenced by the assumed disc model}, ranging from slow alignment in the Bardeen-Petterson configuration to fast solid-body precession in the truncated disc regime. This has important implications for the impact of jet-mode feedback on galaxies and the circumgalactic medium, as more rapid precession would result in distributing the jet energy over a larger volume. 
\item For binary systems, the variations in spin evolution could significantly impact recoil velocities, consequently influencing BH merger rates and the gravitational wave background. 
\end{itemize}

This model provides a crucial advancement in AGN accretion modelling, and there are numerous promising avenues for extensions, including spatially resolved feedback. In the near future, next generation cosmological simulations, incorporating this type of subgrid physics, will be able to give us unprecedented insights into the complex interplay of SMBHs and their host galaxies, and provide the scientific community with much needed robust predictions for the era of multi-messenger science. 

\section*{Acknowledgements}
The authors would like to thank the referee for a thorough review which improved the quality of this manuscript. Furthermore, the authors are grateful to Alexander Tchekhovskoy, Beverly Lowell, Matthew Liska, Jenny Greene, Shy Genel, Angelo Ricarte, Doug Rennehan and Samuel Turner for useful discussions during the development of this work. The authors would also like to thank Lucy Reading-Ikkanda at the Simons Foundation for creating the illustration of our model in Figure~\ref{fig:truncated_disc_sketch}. S.K. is supported by a Flatiron Research Fellowship at the Flatiron Institute, a division of the Simons Foundation and a Junior Research Fellowship from St Catharine's College, Cambridge. S.K, M.A.B., and D.S. acknowledge support by European Research Council Starting Grant 638707 ``Black holes and their host galaxies: coevolution across cosmic time''. M.A.B and D.S. additionally acknowledge support from the Science and Technology Facilities Council (STFC). K.P. is supported by the Simons-NSBP Scholars Program. This work was supported by the Simons Collaboration on ``Learning the Universe''. The computations in this work were, in part, run at facilities supported by the Scientific Computing Core at the Flatiron Institute, a division of the Simons Foundation. This work used the DiRAC Data Intensive service (CSD3) at the University of Cambridge, managed by the University of Cambridge University Information Services on behalf of the STFC DiRAC HPC Facility (\url{www.dirac.ac.uk}). The DiRAC component of CSD3 at Cambridge was funded by BEIS, UKRI and STFC capital funding and STFC operations grants. This work used the DiRAC Memory Intensive service (Cosma8) at Durham University, managed by the Institute for Computational Cosmology on behalf of the STFC DiRAC HPC Facility (\url{www.dirac.ac.uk}). The DiRAC service at Durham was funded by BEIS, UKRI and STFC capital funding, Durham University and STFC operations grants. DiRAC is part of the UKRI Digital Research Infrastructure.

\section*{Data Availability}
The data underlying this article will be shared on reasonable request
to the corresponding author.


\bibliographystyle{mnras}
\bibliography{UnifiedAccDiscPaper} 




\appendix

\section{CND relaxation for single SMBH simulation} \label{appsec:SingleBH}
\begin{figure*}
    \centering
    \includegraphics[width=\textwidth]{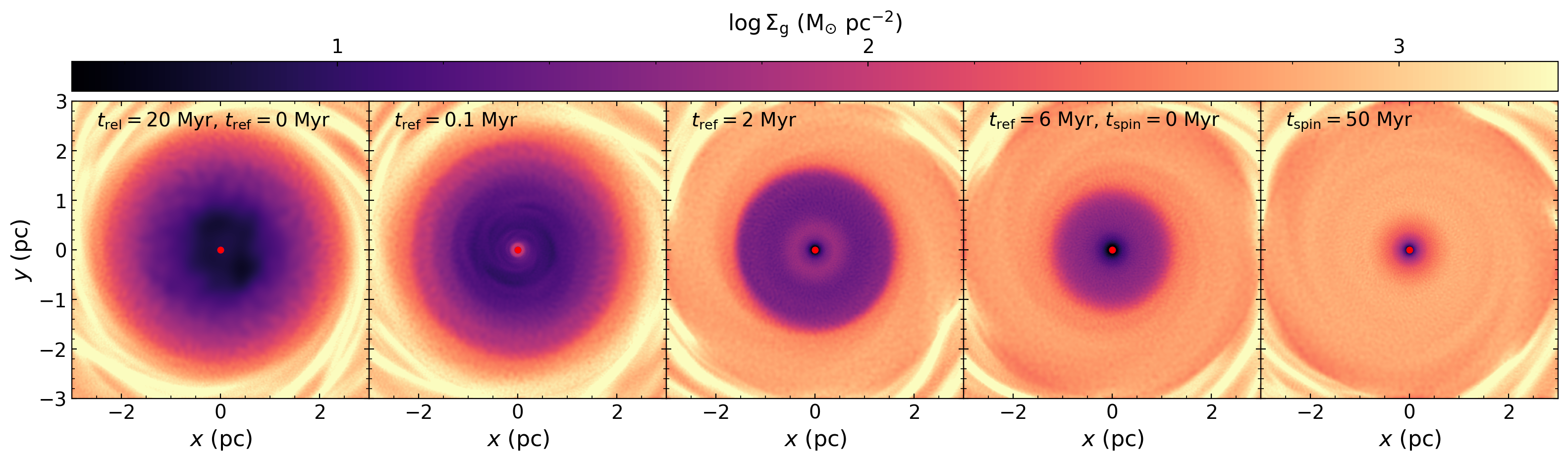} 
    \caption{Gas surface density maps during the refinement relaxation (panels 1 -- 4) and the production run with the subgrid disc model (panels 4 -- 5). This illustrates the creation and dissipation of overdensities and spiral arms during the refinement relaxation phase.}
    \label{fig:proj_single_relaxation}
\end{figure*}
Here we present more detailed plots illustrating the evolution of the cavity for the single SMBH simulations during the refinement relaxation and once the accretion disc subgrid model is switched on. Note that we deliberately do not let the cavity fully fill with gas before activating our model as we are mainly interested in the low-accretion-rate regime. 

Figure~\ref{fig:proj_single_relaxation} shows gas density projections of the cavity during the refinement relaxation (panels 1 -- 4) and the production run with the subgrid disc model (panels 4 -- 5). After the refinement model is activated, several instabilities develop, amplified by cooling from the CND onto the central region \citep[also see][]{bennett_resolving_2020,angles-alcazar_cosmological_2021}. This leads to the formation of spiral arms and a dense core which then propagates outwards (see third panel). By the time that the subgrid model is activated (fourth panel), this overdensity has dissipated. However, during the simulation, the cavity continues to fill with gas leading to a mild increase in inflow rates.

\begin{figure*}
    \centering
    \includegraphics[width=\textwidth]{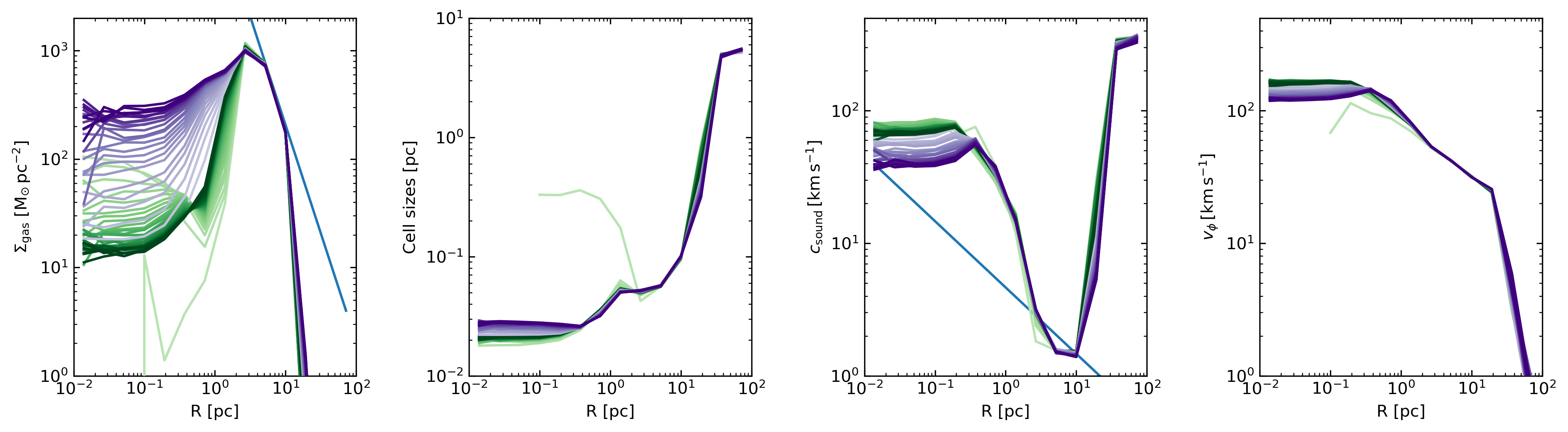} 
    \caption{Gas surface density, cell sizes, sound speed and azimuthal velocity as a function of radius during the refinement relaxation phase (green lines) and the subgrid model simulation phase (purple lines).}
    \label{fig:prop_single_relaxation}
\end{figure*}

Figure~\ref{fig:prop_single_relaxation} shows this more quantitatively, plotting the gas density, cell sizes, sound speed and azimuthal velocity as a function of radius during the refinement relaxation phase (green lines) and the subgrid model simulation phase (purple lines).


\bsp	
\label{lastpage}
\end{document}